\documentclass[12pt]{article}
\pdfoutput=1 
\usepackage{amssymb, amsthm,amsmath,amsfonts}
\usepackage{graphicx}
\usepackage{xcolor, colortbl}
\usepackage{empheq}
\usepackage[pdftex, bookmarks=true,colorlinks,linkcolor=red,urlcolor=blue, citecolor=blue]{hyperref}
\usepackage{cite}
\usepackage{slashed}
\usepackage{caption}
\usepackage{amsthm}
\usepackage{array}
\usepackage{mathrsfs}

\usepackage{soul}
\usepackage{float}
\usepackage{bm}

\makeatletter\@addtoreset{equation}{section}\makeatother

\newcommand{\abs}[1]{\left| #1 \right|}

\newcommand{\dT}{\Delta T}
\newcommand{\dF}{\Delta \mathcal{F}}
\newcommand{\OO}{\langle\hat{O}\rangle}
\newcommand{\dpt}{\Delta p_t}
\newcommand{\epsO}[1]{\mathcal{O}(\epsilon^{#1})}

\newcommand{\ptsch}{p_t^{\text{Sch}}}
\newcommand{\pxsch}{p_x^{\text{Sch}}}
\newcommand{\pphisch}{p_{\phi}^{\text{Sch}}}
\def\bea{\begin{equation}\begin{aligned}}
\def\eea{\end{aligned}\end{equation}}

\newcommand{\preprint}[1]{\begin{table}[t]  
             \begin{flushright}               
             {#1}                             
             \end{flushright}                 
             \end{table}}                     
\renewcommand{\title}[1]{\vbox{\center\LARGE{#1}}\vspace{5mm}}
\renewcommand{\author}[1]{\vbox{\center#1}\vspace{5mm}}
\newcommand{\address}[1]{\vbox{\center\em#1}}

\newcommand{\be}{\begin{equation}}
\newcommand{\ee}{\end{equation}}
\newcommand{\bse}{\begin{subequations}}
\newcommand{\ese}{\end{subequations}}
\newcommand{\beqa}{\begin{eqnarray}}
\newcommand{\eeqa}{\end{eqnarray}}
\newcommand{\beqar}{\begin{eqnarray*}}
\newcommand{\eeqar}{\end{eqnarray*}}
\newcommand{\bi}{\begin{itemize}}
\newcommand{\ei}{\end{itemize}}
\newcommand{\bn}{\begin{enumerate}}
\newcommand{\en}{\end{enumerate}}

\newcommand{\ba}{\begin{array}}
\newcommand{\ea}{\end{array}}
\newcommand{\bc}{\begin{center}}
\newcommand{\ec}{\end{center}}

\newcommand{\e}{\epsilon}

\newcommand{\al}{\alpha}

\definecolor{darkgreen}{rgb}{0,0.3,0}
\definecolor{darkblue}{rgb}{0,0,0.3}
\definecolor{darkred}{rgb}{0.7,0,0}

\def\lae{\mathrel{\mathop{\smash{\lower .5 ex \hbox{$\stackrel<\sim$}}}}}
\def\lae{\mathrel{\mathop{\smash{\lower .5 ex \hbox{$\stackrel>\sim$}}}}}

\def\arXiv#1{\href{http://arxiv.org/abs/#1}{arXiv:#1}}
\def\arXiv#1#2{\href{http://arxiv.org/abs/#1}{arXiv:#1}}

\def\doi#1#2{\href{http://doi.org/#1}{#2}}

\newcommand{\comment}[1]{{\bf {\textcolor{blue}{ [#1]}}}}

\begin{document}

\unitlength = .8mm

\begin{titlepage}
\vspace{.5cm}
\preprint{}

\date{\today}
\begin{center}
\hfill \\
\vskip 0cm

\title{\boldmath Critical and multicritical Kasner scaling 
\\ in holographic phase transitions}

\vskip 0.5cm
{Ling-Long Gao$^{\,a}$}\footnote{Email: {\tt linglonggao@buaa.edu.cn}},
{Yan Liu$^{\,a}$}\footnote{Email: {\tt yanliu@buaa.edu.cn}} and
{Hong-Da Lyu$^{\,b}$}\footnote{Email: {\tt hongdalyu@sdu.edu.cn}} 

\address{${}^{a}$Department of Space Science, \\ and Peng Huanwu Collaborative Center for Research and Education, \\Beihang University, Beijing 100191, China}


\address{${}^{b}$Key Laboratory of Particle Physics and Particle Irradiation (MOE),\\
Institute of Frontier and Interdisciplinary Science,\\
Shandong University, Qingdao, Shandong 266237, China}

\end{center}
\vskip 1.1cm

\abstract{
We study how holographic phase transitions imprint their scaling laws on the local Kasner geometry inside Einstein-scalar black holes. At fixed double-trace coupling, we derive the near-critical scaling of the deviation of the first-epoch Kasner exponent from the Schwarzschild value: $p_t^{(1)}+\frac{1}{3} \propto O^2 \propto (T_c-T)^{1/r}$, where $O$ denotes the boundary condensate and $r$ labels ordinary criticality ($r=1$),  tricriticality ($r=2$), and higher multicriticality ($r\geq3$). 
A super-exponential scalar potential generates a sequence of Kasner epochs separated by scalar-field bounces. We further find that any later epoch that can be tracked continuously across the transition inherits the same temperature exponent, while its coefficient depends on the epoch. Numerical solutions confirm these predictions for the ordinary and tricritical cases. Finally, we show analytically that the leading irrelevant exponent of the infrared fixed point governs the low-temperature Kasner scaling, and verify the resulting scaling numerically for the first Kasner exponent. 
}

\end{titlepage}

\begingroup 
\hypersetup{linkcolor=black}
\tableofcontents
\endgroup

\section{Introduction}
\label{sec:intro}

A black-hole horizon hides the interior from an exterior observer, but it does not make the interior an independent system. 
In holography, suitably defined boundary observables can retain information about regions behind the horizon and even about the approach to the singularity, e.g. analytically continued thermal correlation functions \cite{Fidkowski:2003nf, Festuccia:2005pi, Ceplak:2024bja}. Since the interior geometry belongs to the same bulk solution that determines the thermodynamics measured at infinity, changes in the phase structure of the boundary state may leave an imprint not only outside the horizon, but also on the approach to the singularity. Understanding this connection offers a concrete way to explore how strongly coupled quantum matter organizes spacetime in a region that is otherwise difficult to probe. 

A canonical asymptotic description of spacelike singularities is provided by Kasner geometry. In homogeneous Einstein-scalar systems, the scalar kinetic energy can dominate the interior evolution and produce a Kasner regime characterized by a set of scaling exponents \cite{Frenkel:2020ysx}. These exponents provide a compact measure of the anisotropic geometry near the singularity, and depend on the exterior black-hole solution. 
Thermal phase transitions therefore offer a natural setting for studying how changes in the boundary state are transmitted across the horizon.

This idea was first explored in Ref.~\cite{Liu:2021hap} for Einstein-scalar black holes with a double-trace deformation and a polynomial scalar potential. It was found that the Kasner exponents are continuous but non-differentiable across an ordinary second-order transition, whereas they jump across a first-order transition. This established a direct relation between the order of a boundary phase transition and the behavior of a black-hole singularity. It also raises two broader questions. How are different orders of phase transitions and their scaling laws encoded in the interior? And what replaces this picture when the scalar potential prevents the interior from settling into a single Kasner regime?

To address these questions, we consider four-dimensional Einstein-scalar gravity with the scalar potential consisting of polynomial and super-exponential terms, 
together with alternative quantization and a negative double-trace deformation that allows scalar condensation \cite{Witten:2001ua, Faulkner:2010gj, She:2011cm}. 
For convenience, we choose the super-exponential sector such that, after absorbing its constant contribution into the cosmological term, its expansion around $\phi=0$ begins at order $\phi^8$. It therefore leaves the scalar mass and the independent quartic and sextic interactions unchanged, allowing the boundary phase structure to be controlled by the polynomial sector. This separation is not essential: more general small-field expansions can be accommodated by appropriately redefining the lower-order polynomial couplings. At large $\lvert\phi\rvert$, however, the super-exponential interaction dominates and acts as a reflecting wall for the scalar in the black-hole interior. 
As a result, the interior no longer settles into a single asymptotic Kasner regime, but instead passes through a sequence of local epochs separated by scalar bounces \cite{Hartnoll:2022snh,Gao:2025otg}. Each epoch is characterized by its own scalar velocity $v_n$ and a corresponding set of Kasner exponents ${p_t^{(n)},p_x^{(n)},p_\phi^{(n)}}$. A boundary phase transition may therefore affect not only the value of one Kasner exponent, but also the organization and evolution of the entire sequence. This richer structure makes it possible to ask how boundary critical behavior is transmitted between different interior epochs and how it is modified by the large-field dynamics of the scalar potential. 

We establish an analytic connection between boundary critical behavior and the multiple local Kasner epochs of the black-hole interior. We use a perturbation method to relate the boundary condensate to the first local Kasner epoch, and  further track this critical scaling across later epochs that remain continuously identifiable.  We then use IR matching to show that the leading irrelevant exponent of the IR fixed point governs the low-temperature approach of the Kasner exponents.  We numerically test these predictions for the ordinary and tricritical transitions and verify the low-temperature scaling in the first-order case as well.

The paper is organized as follows. We begin in Sec.~\ref{sec:setup} by introducing the Einstein-scalar model, outlining the local Kasner epochs of the black hole interior, and discussing the role of the super-exponential potential. In Sec.~\ref{sec:pt} we perform a perturbative analysis of the solution and the interior Kasner exponent near the critical point and at low temperature, focusing on the scaling behaviors of Kasner exponents. In Sec. \ref{sec:num} we perform numerical studies and compare them with the analytic predictions. We conclude and discuss possible extensions in Sec. \ref{sec:cd}. 
Further technical details are provided in Appendices~\ref{app:ther-bh}--\ref{app:screened-kasner}.

\section{Holographic setup and black hole interior}
\label{sec:setup}

In this section, we introduce the Einstein-scalar black hole model 
and collect the essential ingredients needed to connect boundary critical phenomena with the evolution of the black-hole interior.

We consider a 2+1-dimensional boundary field theory dual to the following
3+1-dimensional Einstein-scalar gravity,
\be
\label{EinScalarAction}
S = \int d^4x\;\sqrt{-g}\,\Big[R -
\frac{1}{2}(\nabla \phi)^2 - V(\phi)\Big]\,.
\ee
We set $16\pi G=1$ and the AdS radius $L=1$. 
The scalar potential is chosen to be\footnote{This choice is convenient for the discussion of the phase transition up to the fourth-order multicritical point. The generalization of our discussion to other combinations of polynomial and super-exponential terms is straightforward.}
\be
\label{eq:potential}
V(\phi) = (-6-\lambda)+ \frac{1}{2}m^2 \phi^2 + \lambda_4\phi^4
+\lambda_6 \phi^6+\lambda\,e^{\lambda_8 \phi^8}\,.
\ee
We take $m^2=-2$, so that the scalar lies in the window allowing
both standard and alternative quantizations \cite{Klebanov:1999tb}.  The super-exponential term is
irrelevant for the ultraviolet expansion around $\phi=0$. However, it can become dominant in the black-hole interior, where the scalar may reach large values.

To describe homogeneous states at finite temperature, we consider the planar black-hole ansatz,
\be
\label{field-ansatz0}
 ds^2 = \frac{1}{z^2}
 \left(-f(z)e^{-\chi(z)}dt^2 +\frac{dz^2}{f(z)} + dx^2 + dy^2\right)\,,
 \qquad \phi=\phi(z)\,.
\ee
The AdS$_4$ boundary is located at $z=0$, the event horizon at $z=z_h$, while the interior corresponds to $z>z_h$. The planar Schwarzschild-AdS black hole is recovered by setting $f=1-z^3/z_h^3$ and $\chi=\phi=0$.  Substituting the ansatz \eqref{field-ansatz0} into the Einstein-scalar equations yields
\bea 
\label{eq:HoloEOM}
\chi' &= \frac{1}{2}z\phi'^2\,, \\
\left( \frac{e^{-\chi/2} f}{z^3} \right)'&=
\frac{e^{-\chi/2} V}{2z^4}\,,\\
\left( \frac{e^{-\chi/2}f\phi'}{z^2} \right)'&=
\frac{e^{-\chi/2}\,\partial_{\phi}V}{z^4}\,,
\eea
where primes denote derivatives with respect to $z$.

At the horizon $z=z_h$, regularity gives
\bea
\label{eq:nhexpansion}
 f&=-\frac{V(\phi_h)}{2z_h}(z_h-z)+\cdots\,,\\
 \chi&=\chi_h-\frac{2}{z_h}
 \left(\frac{V'(\phi_h)}{V(\phi_h)}\right)^2(z_h-z)+\cdots\,,\\
 \phi&=\phi_h-\frac{2V'(\phi_h)}{z_hV(\phi_h)}(z_h-z)+\cdots\,.
\eea
Since $f>0$ outside the horizon, we have $f'(z_h)<0$ for a regular finite-temperature horizon.
The Hawking temperature is therefore
\be
T=\frac{|f'(z_h)|e^{-\chi_h/2}}{4\pi}
=-\frac{V(\phi_h)e^{-\chi_h/2}}{8\pi z_h}\,.
\ee
The limiting case $T=0$ makes the above non-extremal expansion \eqref{eq:nhexpansion} degenerate, and must be treated separately.

There are two useful scaling symmetries. The first is a time rescaling,
\be
\label{scalsym1}
 t\rightarrow bt\,,\qquad \chi\rightarrow\chi+2\log b\,,
\ee
which is used to set $\chi=0$ at the boundary. It also gives the radially conserved quantity
\be
\label{conserved-charge}
 Q=\frac{e^{-\chi/2}}{z^2}\left(f'-f\chi'\right)\,,
 \qquad Q'=0\,.
\ee
This quantity serves as a bridge between boundary thermodynamics, horizon data and interior Kasner data. It can also be used to check the numerical accuracy.
The second symmetry is the planar scaling
\begin{align}
\label{scalsym2}
 (t,z,x,y)\rightarrow b(t,z,x,y)\,,
\end{align}
which indicates that the system is characterized by dimensionless combinations. Note that the Noether charge associated with 
\eqref{scalsym2} reduces to the same expression as \eqref{conserved-charge}, after using the equations of motion \eqref{eq:HoloEOM}.

\subsection{Dual field theory}
We deform the boundary theory by a double-trace operator,
\be
S \rightarrow S-\frac{\kappa}{2}\int d^3x\;O^2\,,
\ee
where $\kappa$ is the double-trace coupling. We adopt the alternative quantization, in which the dual operator $O$ has dimension $\Delta=\Delta_-=1$.
The fields have the expansions near the conformal AdS boundary,
\bea 
\label{eq:nbexp}
 f&=1+\frac{\alpha^2}{4}z^2+
 m_T z^3+\cdots\,,\\
 \chi&=\chi_b+\frac{\alpha^2}{4}z^2+
 \frac{2\alpha\beta}{3}z^3+\cdots\,,\\
 \phi&=\alpha z+\beta z^2 +\left(\frac{1}{8}+2\lambda_4\right)\alpha^3 z^3+
 \cdots\,.
\eea
In the presence of this deformation, the source is given by $J=\kappa\alpha-\beta$, while $\alpha$ is the expectation value \cite{Witten:2001ua, Papadimitriou:2007sj, Vecchi:2010dd}. The sourceless condition is therefore
\be
\kappa\alpha-\beta=0\,.
\ee
We take $\kappa<0$ so that the double-trace deformation can trigger scalar condensation and a finite-temperature phase transition. 

The additional higher-order terms in the potential \eqref{eq:potential} compared with the potential in \cite{Liu:2021hap}, such as $\lambda_6 \phi^6$ and $\lambda \exp{(\lambda_8 \phi^8)}$, do not affect the holographic renormalization. Therefore, the renormalized free energy density remains the same as that obtained in \cite{Liu:2021hap},
\be
\label{eq:free-energy}
 \mathcal{F}\equiv \frac{F}{V_2}
 =m_T-\frac{\kappa\alpha^2}{2}
 =\frac{\kappa \alpha^2}{6}-\frac{4\pi T}{3 z_h^2}
 \,,
\ee
where we have used the conserved charge \eqref{conserved-charge} in the second equality. This result is consistent with the thermodynamic identity,
\be
\label{fmT-thermo-relation}
 \mathcal{F}=\mathcal{E}-Ts\,,
\ee
where $\mathcal{E}$ and $s$ are the energy density and entropy density, respectively.
For the Schwarzschild black hole, $\alpha=0$ and $m_T=-1/z_h^3$, so
\begin{equation}
 \mathcal{F}_{\rm Sch}=-\frac{1}{z_h^3}
 =-\left(\frac{4\pi}{3}\right)^3T^3\,.
\end{equation}
The details of these results are given in Appendix \ref{app:ther-bh}. 

\subsection{Interior with super-exponential potential}

We now turn to the black hole interior, which corresponds to $z>z_h$. In any radial interval where $V$ and
$V_{,\phi}$ are negligible compared with the scalar kinetic terms,
\eqref{eq:HoloEOM} admits the local Kasner-like solution \cite{Kasner:1921zz, BKL, BKL-book},
\begin{align}
\label{eq:rel1}
\phi=v\log z+\cdots\,,
\qquad
\chi=\frac{v^2}{2}\log z+\chi^{(s)}+\cdots\,,
\qquad
f=-f^{(s)} z^{3+\frac{v^2}{4}}+\cdots\,,
\end{align}
with $f^{(s)}>0$. Note that $v$ is the scalar field ``velocity'', while $\chi^{(s)}$ and $f^{(s)}$ are integration constants. For the Schwarzschild-AdS solution, the interior contains a single
Kasner epoch with $v=0$, giving the Kasner exponents 
\be
p_x^{\text{Sch}}=p_y^{\text{Sch}}=2/3\,,~~~~ p_t^{\text{Sch}}=-1/3\,,~~ {\rm and}~~ p_\phi^{\text{Sch}}=0\,.
\ee

The Kasner approximation \eqref{eq:rel1} breaks down when the super-exponential potential terms become comparable to the kinetic-dominated terms \cite{Hartnoll:2022snh, Gao:2025otg}. Beyond the transition region, the potential terms become negligible again, and the interior enters a new Kasner epoch.  
Consequently, the Kasner solution \eqref{eq:rel1} should not be assumed to be the final $z\to\infty$ asymptotic. It instead describes a local kinetic-dominated epoch over a finite interval of $z$. We label different Kasner epochs by an index $n$,
\be
\label{eq:nth-kasner-epoch}
\phi=v_n\log z+\cdots\,,
\qquad
\chi=\frac{v_n^2}{2}\log z+\chi_n^{(s)}+\cdots\,,
\qquad
f=-f_n^{(s)} z^{3+\frac{v_n^2}{4}}+\cdots\,.
\ee  
The corresponding Kasner exponents for the metric and scalar field are 
\begin{align}
\label{eq:rel2}
 p_x^{(n)}=p_y^{(n)}=\frac{8}{12+v_n^2}\,,
 \qquad
 p_t^{(n)}=\frac{v_n^2-4}{12+v_n^2}\,,
 \qquad
 p_\phi^{(n)}=
 \frac{4\sqrt2\,v_n}{12+v_n^2}\,.
\end{align}
It is sufficient to focus on the independent exponent $p_t^{(n)}$, which is equivalently characterized by the scalar velocity $v_n$.

\begin{figure}[h!]
\begin{center}
\includegraphics[
width=0.43\textwidth]{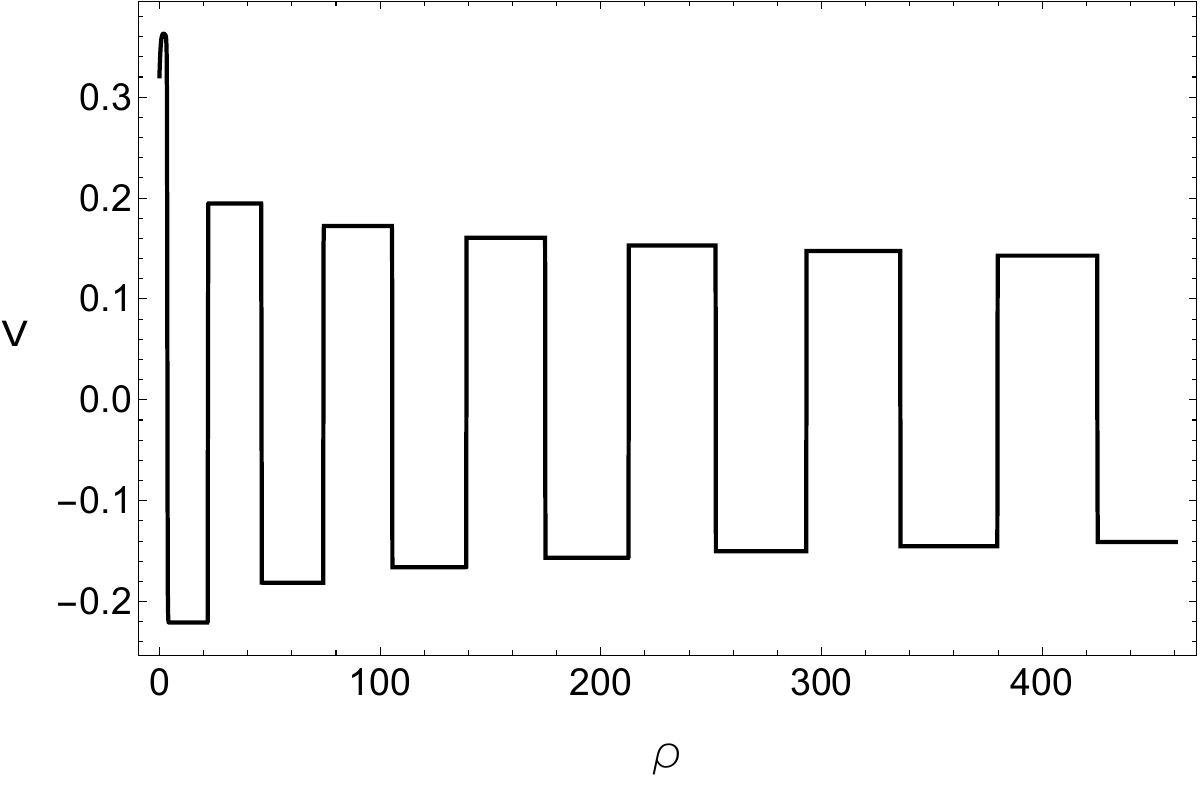}
~~~~
\includegraphics[
width=0.42\textwidth]{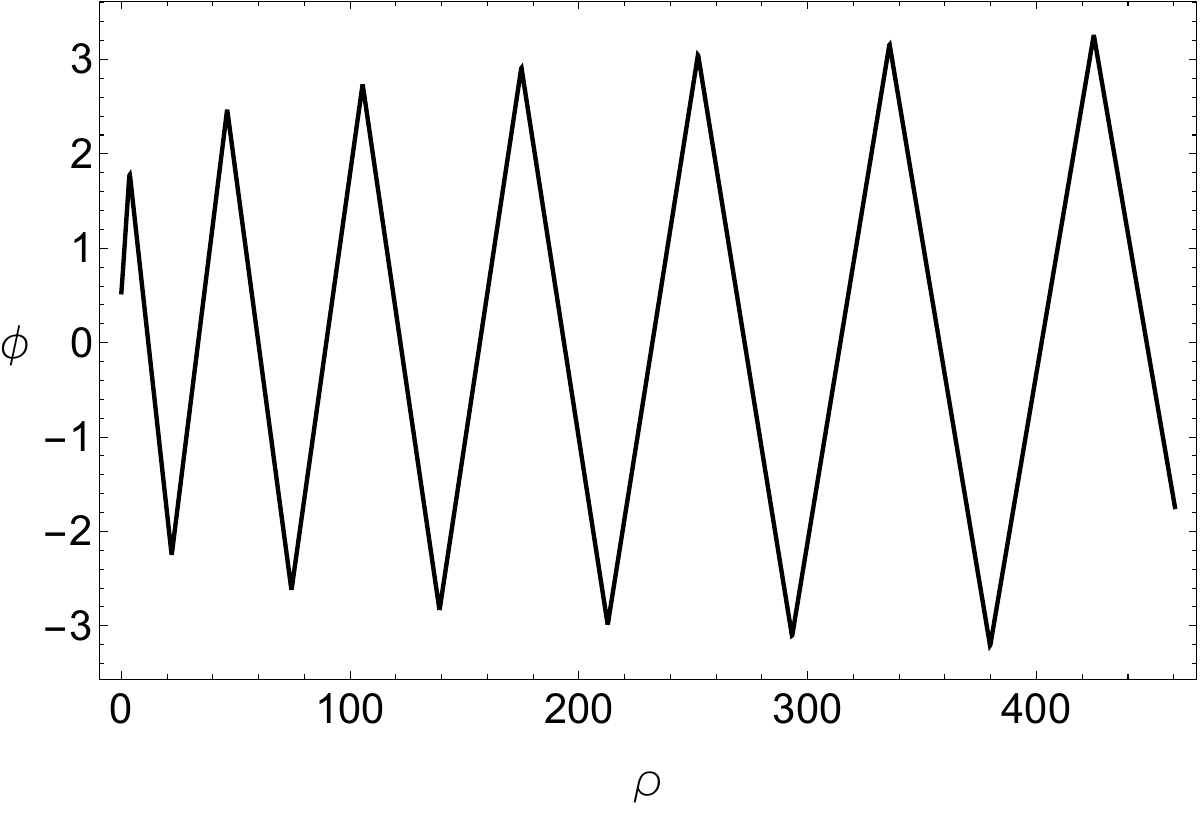}
\end{center}
\vspace{-0.6cm}
\caption{\small A typical interior evolution of a hairy black brane solution in Einstein-scalar theory with super-exponential potential \eqref{eq:potential}. Here $\rho=\log (z/z_h)$. Kasner epochs correspond to the plateau regime in the left panel and the sloped regime in the right panel, while bounces correspond to the jumps between two epochs in both panels.}
 \label{fig:schematic}
\end{figure}

Fig. \ref{fig:schematic} shows a schematic interior evolution with infinitely many Kasner epochs separated by bounces. All bounces overlap with their neighboring Kasner epochs. This allows us to obtain an analytical description of the bounces and further derive the evolution of \(\abs{v_n}\) as a function of $n$.

The sequence of local Kasner epochs and
scalar-field bounces can be viewed as a restricted realization of the
cosmological-billiard picture~\cite{Damour:2000hv,Damour:2000wm,Damour:2002et}.
Within the homogeneous ansatz~\eqref{field-ansatz0}, only a reduced set of dynamical
degrees of freedom is retained, while the super-exponential scalar potential
provides an effective reflecting wall. The resulting bounce dynamics may
therefore be regarded as a reduced counterpart of the generic BKL billiard
dynamics.

Compared with Ref.~\cite{Hartnoll:2022snh}, fewer approximations are employed in Ref.~\cite{Gao:2025otg} when solving the dynamical equations analytically, leading to more complete solutions. 
In Ref.~\cite{Gao:2025otg}, a critical value $v_c$, which is crucial for the evolution of $v_n$, is identified as
\be 
\label{eq:critical value}
v_c=2\sqrt{3}\,.
\ee 
Starting from a Kasner epoch with $|v|<v_c$, the sequence ${v_n}$ is monotonically decreasing in magnitude, i.e., $|v_{n+1}|<|v_n|$ for all subsequent $n$. While starting from an epoch with $|v|>v_c$, we have $|v_{n+1}|>|v_n|$. Consequently, this critical value $v_c$ is crucial for the evolution of all Kasner exponents \eqref{eq:rel2}.

There are two possible types of Kasner transitions, the decreasing type with $\abs{v_n}>\abs{v_{n+1}}$ and the increasing one with $\abs{v_n}<\abs{v_{n+1}}$. These two types of transitions correspond to two qualitatively different fates of the interior evolution. However, it was numerically found in Ref.~\cite{Hartnoll:2022snh} that only decreasing transitions exist. Here we give a direct proof.  
Evaluating the conserved charge $Q$ in \eqref{conserved-charge} both at the horizon and in the $n$-th Kasner epoch gives
\be
\label{eq:ccre}
 -\frac{4\pi T}{z_h^2}
 =f_n^{(s)} e^{-\chi_n^{(s)}/2}\,\left(\frac{v_n^2}{4}-3\right)\,.
\ee
Since $T>0$ and $f_n^{(s)}>0$, any local Kasner epoch obeys $\abs {v_n}<2\sqrt{3}$, where the upper bound is exactly the critical value $v_c$ in \eqref{eq:critical value}. Therefore, all Kasner epochs satisfy $\abs{v_{n+1}}<\abs{v_n}<v_c$\,, or equivalently $p_t^{(n+1)}<p_t^{(n)}<1/3$. 
The conserved charge \eqref{conserved-charge} and critical value \eqref{eq:critical value} provide powerful constraints for the field velocity $v_n$, ruling out the increasing transitions with $p_t^{(n+1)}>p_t^{(n)}>1/3$.

Note that throughout this paper, we only find the bounces satisfying $v_n v_{n+1}<0$,\footnote{ A possible reason is as follows. The case $v_n v_{n+1}>0$ 
indicates that the magnitude of the field value $\abs{\phi_n}$ continues to grow, and thus the super-exponential potential can no longer be neglected as the system evolves deep in the interior. As a result, there could no longer be any subsequent Kasner epoch. } i.e., the field velocity changes sign before and after a bounce.
Since only decreasing transitions exist, the analysis in Ref.~\cite{Hartnoll:2022snh, Gao:2025otg} directly applies and gives the asymptotic interior evolution as $n\to \infty$,
\bea
\label{eq:metric-exp-approach}
p_t^{(n)}-p_t^{\text{Sch}} &\simeq \frac{v_n^2}{9}\propto n^{-1/3}\,,\\
p_x^{(n)}-p_x^{\text{Sch}}& \simeq -\frac{v_n^2}{18}\propto -n^{-1/3}\,,\\
\abs{p_\phi^{(n)}-p_{\phi}^{\text{Sch}}}&\simeq \frac{\sqrt2}{3}|v_n|\propto n^{-1/6}\,.
\eea
The final singularity is therefore approached through a sequence of local
Kasner epochs that become asymptotically Schwarzschild-like. In the phase-transition analysis below, a quoted Kasner
exponent should always be understood as the exponent extracted from a
specified local epoch.

\section{Analytical perturbative analysis}
\label{sec:pt}

In this section we explain how second-order critical, tricritical  and higher-order multicritical scalings arise from analytic perturbative analysis, and show how the same perturbative parameter is transmitted to the later 
interior Kasner exponents. We also point out the existence of an intermediate temperature at which the first Kasner epoch is lost because of the super-exponential potential. Finally we analyze the low-temperature limit. These analytic results will be verified using the numerical solutions in the next section.

\subsection{Critical phase transitions and scaling relations}
\label{sec:analy}

In this subsection we first develop a unified perturbative description of how the phase structure of the hairy black-brane branch is encoded in its critical scaling near the Schwarzschild bifurcation point, and then discuss the Kasner scaling law. We work at fixed double-trace coupling $\kappa=-1$ and impose the vanishing-source condition. Since the phase structure is governed by the dimensionless ratio $T/(-\kappa)$, varying the temperature at fixed $\kappa$ is sufficient to explore the entire phase diagram. 

We analyze the neighborhood of the critical point by expanding the hairy black-hole solution around the critical Schwarzschild background. At the critical temperature, the Schwarzschild solution admits a normalizable static scalar mode satisfying the fixed double-trace boundary condition. We use the leading scalar amplitude \(\epsilon\) as an expansion parameter and solve the scalar and metric equations order by order. At each order, regularity at the horizon and the boundary condition in UV AdS$_4$ determine the allowed corrections 
and fix the horizon location and, consequently, the temperature. The first nonvanishing temperature correction determines the critical exponent. By successively tuning the quartic and sextic couplings, the same construction describes ordinary criticality, tricriticality, and fourth-order multicriticality. The expansion also determines the free-energy difference and the leading deviations of the first-epoch Kasner exponents from their critical values, thereby establishing a direct analytic connection between the boundary phase transition and the black-hole interior.

\subsubsection{Framework of perturbative expansions}

Note that all dimensionful quantities are measured in units of \(-\kappa\), where we set \(\kappa=-1\).  The source-free boundary condition is then
\begin{equation}
 \beta=-\alpha .
 \label{eq:fixed-kappa-bc}
\end{equation} 
It is convenient to introduce
\begin{equation}
 u\equiv\frac{z}{z_h},
\end{equation}
so that the AdS boundary and the horizon remain at \(u=0\) and \(u=1\),
respectively.  This is only a computational change of coordinates; no scaling
transformation is used to change the physical coupling. 

Regarding \(f\), \(\chi\), and \(\phi\) as functions of \(u\), the
background equations of motion become 
\begin{align}
 \chi'
 &=\frac{u}{2}\phi'^2\,,
 \label{eq:eom-u-chi-expanded}\\
 f'-\frac{3f}{u}
 &=\frac{V(\phi)}{2u}+\frac{uf}{4}\phi'^2\,,
 \label{eq:eom-u-f-expanded}\\
 \phi''
 +\left(\frac{1}{u}+\frac{V(\phi)}{2uf}\right)\phi'
 -\frac{V_{,\phi}(\phi)}{u^2f}
 &=0\,.
 \label{eq:eom-u-phi-compact}
\end{align}
Here and below in this subsection, primes denote derivatives with respect to \(u\). All explicit
factors of \(z_h\) cancel from the differential equations.
We define the horizon-unit temperature by 
\begin{equation}
 \tau\equiv z_h T
 =-\frac{V(\phi_h)e^{-\chi_h/2}}{8\pi}\,,
 \qquad \phi_h\equiv\phi(1)\,,
 \qquad \chi_h\equiv\chi(1)\,,
 \label{eq:horizon-unit-temperature}
\end{equation}
and the physical temperature is \(T=\tau/z_h\).  The quantity \(\tau\) is
the dimensionless Hawking temperature of the solution with its horizon fixed
at \(u=1\); it is not a local temperature measured at the horizon.

We now choose the perturbatively small parameter. Near the AdS boundary,
\begin{equation}
 \label{eq:UV-expansion-u}
 \phi(u)=A u+B u^2+\cdots\,.
\end{equation}
At the critical point the hairy branch merges continuously with the 
Schwarzschild solution, so that \(A\to0\). We use this coefficient as the expansion parameter for nearby hairy branch,
\begin{equation}
 \epsilon\equiv A\,.
 \label{eq:epsilon-definition}
\end{equation}
The parameter $\epsilon$ merely labels the family of solutions:
the physical coupling remains fixed at
\(\kappa=-1\) and the source vanishes for every
\(\epsilon\). 

The small-field expansion of the potential \eqref{eq:potential} is taken to be
\begin{equation}
 V(\phi)=-6+\frac{m^2}{2}\phi^2
+\lambda_4\phi^4+\lambda_6\phi^6+\lambda\lambda_8\phi^8+\mathcal O(\phi^{16}),
 \qquad m^2=-2\,.
 \label{eq:potential-small-field}
\end{equation}
Because the scalar potential is even, the equations are invariant under
\(\phi\rightarrow-\phi\). 
The two symmetry-related hairy branches correspond
to \(\epsilon\rightarrow-\epsilon\). Assuming that the branch is
analytic near \(\epsilon=0\), the scalar field and its boundary
coefficients are odd functions of \(\epsilon\), whereas the metric,
horizon radius, and temperature are even functions. We therefore write the perturbative expansion near the critical point as
\begin{align}
 \phi(u;\epsilon)
 &=\phi_0(u)+\epsilon\phi_1(u)
 +\epsilon^3\phi_3(u)
 +\epsilon^5\phi_5(u)
 +\mathcal O(\epsilon^7),\\
 f(u;\epsilon)
 &=f_0(u)
 +\epsilon^2f_2(u)
 +\epsilon^4f_4(u)
 +\mathcal O(\epsilon^6),\\
 \chi(u;\epsilon)
 &=\chi_0(u)+\epsilon^2\chi_2(u)
 +\epsilon^4\chi_4(u)
 +\mathcal O(\epsilon^6)\,,
 \label{eq:epsilon-expansions}
\end{align}
where the planar Schwarzschild background is
\begin{equation}
 f_0(u)=1-u^3,
 \qquad \chi_0(u)=0\,,
 \qquad \phi_0(u)=0\,.
 \label{eq:Sch-background-u}
\end{equation}
These expansions lead to the expansion of $B$ in \eqref{eq:UV-expansion-u}, and that of $z_h$,
\bea
\label{eq:exp-Bzh}
 B(\epsilon)
 &=B_1\epsilon+B_3\epsilon^3
 +B_5\epsilon^5+\mathcal O(\epsilon^7),\\
 z_h(\epsilon)
 &=z_{h0}+z_{h2}\epsilon^2
 +z_{h4}\epsilon^4+\mathcal O(\epsilon^6).
\eea
Note that the horizon radius is not fixed independently: it changes along the hairy branch so that the physical coupling and the vanishing-source boundary condition remain unchanged.
A similar perturbative method was used to study the analytical holographic superconductor \cite{Herzog:2010vz}. 

We now impose the boundary conditions for the perturbative functions. Since the choice \(A=\epsilon\) fixes the freedom to redefine the branch
parameter, the higher-order scalar corrections contain no term proportional to \(u\). This gives the boundary conditions for the perturbative scalars as $u\to0$,
\begin{align}
 \phi_1(u)=u+B_1u^2+\cdots\,,~~~
 \phi_3(u)=B_3u^2+\cdots\,,~~~
 \phi_5(u)=B_5u^2+\cdots\,.
 \label{eq:phiu-bc}
\end{align}
These scalars also satisfy the regularity conditions at the horizon. The boundary conditions for the perturbative metric functions are
\be 
\label{eq:bc-fchi}
f_2(1)=f_4(1)=0\,,\qquad \chi_2(0)=\chi_4(0)=0\,.
\ee 
Unless explicitly stated otherwise, the above boundary conditions are imposed throughout the remainder of this section.

Under the above boundary conditions, the expansion coefficients in \eqref{eq:exp-Bzh} are not independent of each other. 
Comparing the UV expansion in $u$ \eqref{eq:UV-expansion-u} with that in $z$ \eqref{eq:nbexp}, the coefficients are related by
\begin{equation}
 \alpha=\frac{A}{z_h}\,,\qquad
 \beta=\frac{B}{z_h^2}\,.
\end{equation}
The source-free boundary condition \eqref{eq:fixed-kappa-bc} then becomes
\begin{equation}
 B(\epsilon)+\epsilon\, z_h(\epsilon)=0\,.
 \label{eq:fixed-source-epsilon}
\end{equation}
It follows order by order that 
\begin{equation}
 z_{h0}=-B_1\,,\qquad
 z_{h2}=-B_3\,,\qquad
 z_{h4}=-B_5\,.
 \label{eq:zh-from-B}
\end{equation}
These relations are important for determining the relation between $\Delta T$ and $\epsilon$.

\subsubsection{The scalar onset mode and Kasner-condensate scaling}
At order \(\epsilon\), the scalar equation is
\be 
\phi_1''+\left(-\frac{2}{u}+\frac{f_0'}{f_0}\right)\phi_1'
+\frac{2\phi_1}{u^2 f_0}=0\,.
\ee  
The solution that is regular at the horizon and normalized by
\(\phi_1=u+B_1u^2+\cdots\) in \eqref{eq:phiu-bc} at the boundary is
\begin{align}
\label{eq:sol-phi1}
\phi_1(u)
&=u\,
\,{}_2F_1\!\left(
\frac{1}{3},\frac{1}{3};
\frac{2}{3};
u^3
\right)
-3\left[\frac{\Gamma(2/3)}{\Gamma(1/3)}\right]^3
u^2\,{}_2F_1\!\left(
\frac{2}{3},\frac{2}{3};
\frac{4}{3};
u^3
\right)\,. 
\end{align}
Hence
\begin{equation}
\label{eq:B1}
 B_1=-3\left[\frac{\Gamma(2/3)}{\Gamma(1/3)}\right]^3\,.
\end{equation}
The critical horizon radius and temperature are therefore 
\begin{equation}
 z_{h0}=-B_1\,,
 \qquad
 T_c=\frac{3}{4\pi z_{h0}}=\frac{1}{4\pi}\left[\frac{\Gamma(1/3)}{\Gamma(2/3)}\right]^3\,.
 \label{eq:tc-ana}
\end{equation}
The hairy black hole solution condenses only below this critical temperature, which matches the numerical result in the next section and has also been obtained in 
\cite{Mefford:2014gia, Liu:2021hap}.

After substituting $\alpha=A/ z_h$ and $\beta=B/z_h^2 $, 
the physical order parameter is  
\begin{equation}
\label{eq:O-scaling}
O=\alpha=\frac{A}{z_h}
 =\frac{\epsilon}{z_{h0}}+\mathcal O(\epsilon^3)
 =\frac{1}{3}\left[\frac{\Gamma(1/3)}{\Gamma(2/3)}\right]^3\,\epsilon+\mathcal O(\epsilon^3)\,.
\end{equation}
This indicates that the order parameter is always of order $\epsilon$. We do not need to calculate its higher order expansions. 

The continuation of the zero mode into the black-hole interior has the
large-\(u\) form
\begin{equation}
 \phi_1(u)=C_v\log u+C_0+\mathcal O(u^{-3}\log u)\,,
 \qquad
 C_v=\frac{3\,\Gamma(2/3)}{\Gamma(1/3)^2}\,.
 \label{eq:phi1-interior}
\end{equation}
This is a matched asymptotic statement, not the terminal
\(u\to\infty\) behavior at fixed nonzero \(\epsilon\).  In the overlap region
\begin{equation}
 1\ll\rho\equiv\log u\ll |\epsilon|^{-1}\,,
 \label{eq:Kasner-overlap}
\end{equation}
the zero mode has reached its logarithmic form while nonlinear corrections
remain parametrically small.  Denoting the scalar velocity in the first local
Kasner epoch by \(v_1\), the matching gives
\begin{equation}
 v_1=C_v\epsilon+\mathcal O(\epsilon^3)
 =C_v z_{h0} O
 +\mathcal O(O^3)\,.
 \label{eq:v-order-parameter}
\end{equation}

Using
\begin{equation}
\label{eq:pt1-v1}
 p_t^{(1)}=\frac{v_1^2-4}{v_1^2+12},
 \qquad p_t^c=-\frac13\,,
 \end{equation}
we obtain
\begin{align}
 \Delta p_t^{(1)}
 \equiv p_t^{(1)}-p_t^c
 =\frac{v_1^2}{9}+\mathcal O(v_1^4)
 &=\frac{C_v^2 z_{h0}^2}{9}O^2
 +\mathcal O(O^4)=\frac{9\,\Gamma\left(2/3\right)^8}{\Gamma\left(1/3\right)^{10}}\,O^2
 +\mathcal O(O^4).
 \label{eq:ptO}
\end{align}
This relation is universal to all continuous branches that emanate from
the same zero mode of the Schwarzschild solution, while higher-order corrections depend on the
potential couplings. The results here do not apply to a finite-condensate jump at a first-order coexistence point.

\subsubsection{Generic second-order criticality}
\label{sec:Generic second-order criticality}

Having identified the scalar onset mode, we now follow its nonlinear continuation into the generic continuous hairy branch. We first calculate the scaling of temperature with respect to $\epsilon$. Then we use the thermodynamic identity to determine the corresponding scaling of the free energy. Together with the scaling of condensate \eqref{eq:O-scaling}, and that of Kasner exponent \eqref{eq:ptO}, this allows us to obtain a complete characterization of the near-critical behavior, including the temperature dependence of the free energy, condensate, and Kasner exponent.

It is convenient to first discuss the expansion of horizon-unit temperature $\tau$ in \eqref{eq:horizon-unit-temperature},
\begin{equation}
\tau=\tau_0+\tau_2\epsilon^2+\tau_4\epsilon^4+\cdots\,,
 \qquad \tau_0=\frac{3}{4\pi}\,. 
 \label{eq:tau-expansion}
\end{equation}
Combining the expansions of \(\tau\) and \(z_h\) yields, up to $\epsO{2}$
\begin{align}
 \frac{T}{T_c}
 =1&+\left(\frac{\tau_2}{\tau_0}
 -\frac{z_{h2}}{z_{h0}}\right)\epsilon^2
 +\epsO{4}.
 \label{eq:T-ratio-expansion}
\end{align}
Parameterizing 
\be
\label{eq:Delta-T}
\Delta T  \equiv T_c-T
=t_2\epsilon^2+\epsO{4}, 
\ee 
we have 
\begin{equation}
 t_2=\frac{\tau_0z_{h2}-z_{h0}\tau_2}{z_{h0}^2}\,.
 \label{eq:t2-coefficient}
\end{equation}
Note that the index $0$ denotes the background solutions, which are known analytically.

At fixed coupling and vanishing source, the first law on either branch is
\(d\mathcal F_i=-s_i\,dT\), as shown in \eqref{eq:fixed-ensemble-free-energy-variation} of Appendix \ref{app:ther-bh}.  Therefore, at the same physical temperature, one obtains 
\begin{equation}
 d\Delta\mathcal F=-\Delta s\,dT,
 \qquad
 \Delta\mathcal F(T_c)=0,
 \label{eq:first-law-difference}
\end{equation}
where
\begin{equation}
 \Delta\mathcal F
 \equiv\mathcal F_{\mathrm{hairy}}-\mathcal F_{\mathrm{Sch}},
 \qquad
 \Delta s\equiv s_{\mathrm{hairy}}-s_{\mathrm{Sch}}.
\end{equation}
With the entropy normalization used here,
\begin{align}
 s_{\mathrm{hairy}}
 =\frac{4\pi}{z_h^2},~~~~~
 s_{\mathrm{Sch}}(T)
 =4\pi\left(\frac{4\pi T}{3}\right)^2.
\end{align}
Consequently,
\begin{equation}
 \Delta s=s_2\epsilon^2+\mathcal O(\epsilon^4)\,,
 \qquad
 s_2=-2s_0\frac{\tau_2}{\tau_0}\,,
 \qquad
 s_0=\frac{4\pi}{z_{h0}^2}\,.
 \label{eq:entropy-difference-leading}
\end{equation}
More generally, if the first nonzero temperature correction is
\begin{equation}
\label{eq:delta-T-general}
 \Delta T=t_{2r}\epsilon^{2r}
 +\mathcal O(\epsilon^{2r+2}),
 \end{equation}
then integration of Eq.~\eqref{eq:first-law-difference} gives
\begin{equation}
 \Delta\mathcal F
 =\frac{r}{r+1}s_2t_{2r}\epsilon^{2r+2}
 +\mathcal O(\epsilon^{2r+4}).
 \label{eq:free-energy-general-epsilon}
\end{equation}
Therefore, the scaling of free energy with respect to $\epsilon$ is completely determined by that of the temperature. In the remainder, we will focus on deriving the temperature scaling, from which the free-energy scaling follows via Eqs.~\eqref{eq:delta-T-general} and \eqref{eq:free-energy-general-epsilon}.

Note that the coefficient at $\mathcal{O}(\e^2)$ in $\Delta T$ is reduced to the two undetermined quantities, $\tau_2$ and $z_{h2}$. 
More explicitly, 
\begin{equation}
 \tau_2
 =\frac{-m^2\phi_1(1)^2/2-3\chi_2(1)}{8\pi}.
 \label{eq:tau2}
\end{equation}
From \eqref{eq:zh-from-B} we have $z_{h2}=-B_3$. Both quantities are completely determined once the perturbative functions $f_2$, $\chi_2$, and $\phi_3$ are obtained. 

At order \(\epsilon^2\), the metric equations are 
\bea 
\label{eq:eom-fchi2}
f_2'-\frac{3f_2}{u}
&=\frac{m^2\phi_1^2}{4u}
-\frac{u(u^3-1)}{4}\phi_1'^2\,, \\
\chi_2'&=\frac{1}{2}u \phi_1'^2\,.
\eea
Their integral forms are
\begin{align}
\begin{split}
 \chi_2(u)
 &=\frac12\int_0^u dy\,y\,\phi_1'(y)^2,
 \label{eq:chi2-integral}\\
 f_2(u)
 &=-u^3\int_u^1\frac{dy}{y^3}
 \left[
 \frac{m^2\phi_1(y)^2}{4y}
 +\frac{y f_0(y)}{4}\phi_1'(y)^2
 \right].
 \end{split}
\end{align}
With the boundary conditions for $f_2$ and $\chi_2$ in \eqref{eq:bc-fchi}, 
numerically
\be 
\chi_2(1)=0.091590\,.
\label{eq:T2}
\ee 
Thus we get the value of $\tau_2$ from \eqref{eq:tau2},
\be 
\label{eq:tau2-value}
\tau_2=0.0077078\,.
\ee
It depends only on the quadratic term in the scalar potential, which has been fixed to $m^2=-2$. The non-zero $\tau_2$, together with \eqref{eq:entropy-difference-leading}--\eqref{eq:free-energy-general-epsilon}, again shows that the scaling of free energy with respect to $\epsilon$ is completely determined by that of the temperature.

To solve the equation for $\phi_3$ and then determine $z_{h2}$, we introduce the formally self-adjoint
Sturm--Liouville differential expression
\begin{equation}
 \mathcal D y
 \equiv\left(\frac{f_0}{u^2}y'\right)'
 -\frac{m^2}{u^4}y.
 \label{eq:self-adjoint-operator}
\end{equation}
At order \(\epsilon^3\),
\begin{equation}
 \mathcal D\phi_3=R_3,
 \label{eq:phi3-equation}
\end{equation}
where
\begin{equation}
 R_3=
 \frac{4\lambda_4\phi_1^3-\frac12m^2\chi_2\phi_1}{u^4}
 -\left[
 \frac{f_2-\frac12\chi_2f_0}{u^2}\phi_1'
 \right]'.
 \label{eq:R3-compact}
\end{equation}
Imposing the boundary conditions \eqref{eq:phiu-bc} for UV AdS,
\begin{align}
 \phi_3(u)
 &=B_3u^2+\left(\frac18+2\lambda_4\right)u^3+\cdots,
 \qquad u\to0.
 \label{eq:phi3-UV}
 \end{align}
The regularity condition \eqref{eq:bc-fchi} then gives
 \begin{align}
 \phi_3'(1)
 &=-\frac{m^2}{3}\phi_3(1)
 -\left(\frac{4\lambda_4}{3}+\frac{m^4}{36}\right)
 \phi_1(1)^3.
 \label{eq:phi3-horizon}
\end{align}


Multiplying Eq.~\eqref{eq:phi3-equation} by the zero mode and applying the Wronskian identity gives
\begin{align}
 \int_0^1du\,\phi_1R_3
 &=\left[
 \frac{f_0}{u^2}
 \left(\phi_1\phi_3'-\phi_3\phi_1'\right)
 \right]\Bigg{|}_0^1
 =-B_3.
 \label{eq:phi3-solvability}
\end{align}
Using Eq.~\eqref{eq:zh-from-B},
\begin{equation}
 z_{h2}=\int_0^1du\,\phi_1R_3.
 \label{eq:zh2-solvability}
\end{equation} Plugging \eqref{eq:R3-compact} into the above relation \eqref{eq:zh2-solvability}, we have
\be
z_{h2}
=
4\lambda_4\int_0^1du\,\frac{\phi_1^4}{u^4}
+
\int_0^1du
\left[
-\frac{m^2}{2}\frac{\chi_2\phi_1^2}{u^4}
+\frac{f_2-\frac12\chi_2f_0}{u^2}(\phi_1')^2
\right], 
\ee
where one could further plug \eqref{eq:chi2-integral} into the above expression. 
This relation shows directly that \(z_{h2}\) is linear in \(\lambda_4\).
Using the linear onset scalar mode \eqref{eq:sol-phi1}, numerically, 
\begin{equation}
 z_{h2}(\lambda_4)=0.15069+2.0176\,\lambda_4\,.
 \label{eq:zh2-lambda4}
\end{equation}
The value of $z_{h2}$ receives contributions from the quartic interaction $\lambda_4 \phi^4$. Consequently, the scaling relations for temperature and free energy depend on $\lambda_4$.

For example, at \(\lambda_4=1/10\), we have 
\begin{equation}
 z_{h2}=0.35245,
 \end{equation}
and hence
\begin{align}
\begin{split}
 O
 =2.5811\,\epsilon+\mathcal O(\epsilon^3),~~~
 \Delta T
 =0.54065\,\epsilon^2+\mathcal O(\epsilon^4),~~~
 \Delta\mathcal F
 =-1.4613\,\epsilon^4+\mathcal O(\epsilon^6).
 \label{eq:ordinary-epsilon-scaling}
 \end{split}
\end{align}
Eliminating \(\epsilon\) gives
\begin{equation}
 O\simeq3.5103\,\Delta T^{1/2},
 \qquad
 \Delta\mathcal F\simeq-4.9993\,\Delta T^2.
 \label{eq:ordinary-temperature-scaling}
\end{equation}
This is the generic mean-field second-order scaling. Using \eqref{eq:ptO}, we obtain the relation between $\Delta p_t^{(1)}$  and $\Delta T$,
\be
\label{eq:scalrel1}
\Delta p_t^{(1)}
 =0.065849\,\Delta T
 +\mathcal O(\Delta T^2)\,.
\ee

\subsubsection{Tricriticality}
The temperature difference \eqref{eq:Delta-T} starts at $\mathcal{O}(\epsilon^2)$, with the coefficient $t_2$. It follows from \eqref{eq:t2-coefficient} and \eqref{eq:zh2-lambda4} that $t_2$ depends linearly on the quartic coupling \(\lambda_4\). Therefore, we can tune $\lambda_4$ to eliminate the leading $\mathcal{O}(\e^2)$, so that the temperature difference starts at order \(\epsilon^4\). This gives a unique value of $\lambda_4$,
\begin{equation}
 \lambda_4^t=-0.068490\,.\qquad
 \label{eq:lambda4-tricritical}
\end{equation}
The scaling of free energy difference now starts at order \(\epsilon^6\), as follows from the relation between $\Delta T$ and $\Delta \mathcal{F}$ \eqref{eq:free-energy-general-epsilon}. Note that the scaling of the condensate \eqref{eq:O-scaling} remains unchanged. The fine-tuned $\lambda_4^t$ realizes tricriticality, with 
\be 
\dF\sim \dT^{3/2}\,,\qquad O \sim \dT^{1/4}\,.
\ee 
The same mean-field tricritical scaling is well known from the statistical-mechanics literature  \cite{Riedel:1972tricritical,Wegner:1973logarithmic}. The present construction provides a holographic realization in which this boundary scaling is transmitted to the black-hole interior.

The scaling relations for continuous phase transitions in Sec.~\ref{sec:Generic second-order criticality} apply only when \(\lambda_4>\lambda_4^t\). For \(\lambda_4<\lambda_4^t\), the local branch has \(t_2<0\) and thus bends toward \(T>T_c\). It is thermodynamically disfavored since \(\Delta\mathcal F>0\). In this parameter regime, the system exhibits a first-order phase transition, which will be discussed in Sec. \ref{ss:1st} and Appendix \ref{app:first-order}. 


We now determine the scaling relations at the tricritical point.
At the fine-tuned $\lambda_4^t$, the leading scaling of $\dT$ is
\be
\Delta T =t_4\epsilon^4+ \mathcal{O}(\epsilon^6)\,,\qquad 
 t_4=\frac{\tau_0z_{h4}-z_{h0}\tau_4}{z_{h0}^2}.
\ee 
The coefficient $t_4$ is now reduced to the two quantities, $\tau_4$ and $z_{h4}$. 
More explicitly, let
\begin{equation}
 a_h=\phi_1(1),
 \qquad b_h=\phi_3(1),
 \qquad c_2=\chi_2(1),
 \qquad c_4=\chi_4(1).
\end{equation}
Expanding Eq.~\eqref{eq:horizon-unit-temperature} gives
\begin{equation}
 \tau_4=\frac{
 -(m^2 a_h b_h+\lambda_4^t a_h^4)
 +\frac14m^2 a_h^2 c_2
 -3c_4+\frac34c_2^2}{8\pi}.
 \label{eq:tau4}
\end{equation}
From \eqref{eq:zh-from-B} we have $z_{h4}=-B_5$. Both $\tau_4$ and $z_{h4}$ are completely determined once the perturbative functions $f_4$, $\chi_4$, and $\phi_5$ are obtained.

The fourth-order metric equations are
\begin{align}
 \chi_4'&=u\phi_1'\phi_3',
 \label{eq:chi4-equation}\\
 f_4'-\frac{3f_4}{u}
 &=\frac{m^2\phi_1\phi_3+\lambda_4^t\phi_1^4}{2u}
 +\frac{f_0\chi_4'+f_2\chi_2'}{2}.
 \label{eq:f4-equation}
\end{align}
Combining with the boundary conditions \eqref{eq:bc-fchi} gives
\be 
\chi_4(1)=-0.0032607\,.\qquad
\ee 
Then we have 
\be 
\tau_4=-1.2443\times10^{-4}\,.
\ee 
It depends only on the quartic term in the scalar potential, which has been fixed to $\lambda_4=\lambda_4^t$.

At order \(\epsilon^5\), the formally self-adjoint Sturm-Liouville differential equation for $\phi_5$ is
\begin{equation}
 \mathcal D\phi_5=R_5,
 \label{eq:phi5-equation}
\end{equation}
with
\begin{align}
 R_5={}&\frac{1}{u^4}\Bigg[
 12\lambda_4\phi_1^2\phi_3+6\lambda_6\phi_1^5
 -\frac{\chi_2}{2}
 \left(m^2\phi_3+4\lambda_4\phi_1^3\right)
 \nonumber\\
 &\hspace{2.4cm}
 +\left(-\frac{\chi_4}{2}+\frac{\chi_2^2}{8}\right)m^2\phi_1
 \Bigg]
 -\left(P_2\phi_3'+P_4\phi_1'\right)', 
 \label{eq:R5}
\end{align}
where
\begin{align}
 P_2=\frac{f_2-\frac12\chi_2f_0}{u^2},
~~~~~~
 P_4=\frac{
 f_4-\frac12\chi_2f_2
 +f_0\left(-\frac12\chi_4+\frac18\chi_2^2\right)}{u^2}.
 \label{eq:P4-definition}
\end{align}
The UV condition is \(\phi_5=B_5u^2+\cdots\), and horizon regularity gives
\begin{align}
 \phi_5'(1)={}&-\frac{m^2}{3}\phi_5(1)
 -\left(4\lambda_4+\frac{m^4}{12}\right)
 \phi_1(1)^2\phi_3(1)
-\left(2\lambda_6+\frac{m^2\lambda_4}{6}
 +\frac{m^6}{432}\right)\phi_1(1)^5.
 \label{eq:phi5-horizon}
\end{align}
The corresponding Wronskian identity is
\begin{equation}
 z_{h4}=-B_5=\int_0^1du\,\phi_1R_5\,,
 \label{eq:zh4-solvability}
\end{equation}
which gives
\begin{equation}
 \frac{\partial z_{h4}}{\partial\lambda_6}
 =6\int_0^1du\,\frac{\phi_1^6}{u^4}
 =0.38919>0\,.
 \label{eq:zh4-lambda6-derivative}
\end{equation}
From \eqref{eq:zh4-solvability} and \eqref{eq:R5}, numerically,
\begin{equation}
 z_{h4}(\lambda_6;\lambda_4^t)
 =-0.022326-0.56498\lambda_4^t
-3.3319(\lambda_4^t)^2+0.38919\lambda_6\,.
 \label{eq:zh4-couplings}
\end{equation}
The value of $z_{h4}$ receives contributions from the sextic interaction $\lambda_6 \phi^6$. Consequently, the scaling relations for temperature depend on $\lambda_6$.

For example, at \(\lambda_6=1\), we have
\be 
z_{h4}= 0.38993\,.
\ee 
Thus 
\begin{align}
 \Delta T
 =0.62046\,\epsilon^4+\mathcal O(\epsilon^6)\,,
 ~~~~~~
 \Delta\mathcal F
 =-2.2360\,\epsilon^6+\mathcal O(\epsilon^8)\,.
 \label{eq:tricritical-epsilon-scaling}
\end{align}
Therefore,
\begin{equation}
 O\simeq2.9082\,\Delta T^{1/4},
 \qquad
 \Delta\mathcal F\simeq-4.5752\,\Delta T^{3/2}.
 \label{eq:tricritical-temperature-scaling}
\end{equation}
This is the tricritical, or fourth-root scaling at the fine-tuned $\lambda_4=\lambda_4^t$. Using \eqref{eq:ptO}, we obtain the relation between $\Delta p_t^{(1)}$  and $\Delta T$,
\be
\label{eq:scalrel2}
 \Delta p_t^{(1)}
 =0.045197\,\Delta T^{1/2}
 +\mathcal O(\Delta T).
\ee

\subsubsection{Fourth-order multicriticality
}
The same construction also gives a sixth-root scaling if one tunes one more
coefficient in the scalar potential and removes both
the quadratic and quartic terms in the temperature expansion.  
  This requires
two independent tunings, \(t_2=t_4=0\), followed by an explicit calculation
of the first remaining coefficient \(t_6\).  We first determine the tuned
quartic and sextic couplings and then extend the perturbative hierarchy
through \(\phi_7\), \(f_6\), and \(\chi_6\).  The result below is
\(t_6>0\) for the potential used in this work, so this point is a genuine
local sixth-root multicritical point.

The first condition \(t_2=0\) gives \(\lambda_4=\lambda_4^t\), as analyzed above.  Once it is
imposed, \(t_4=0\) is equivalent to\footnote{The system 
exhibits a first-order phase transition for $\lambda_4=\lambda_4^t, \lambda_6<\lambda_6^t$.}
\begin{equation}
 \lambda_6=\lambda_6^t=-0.0024193\,,
 \label{eq:lambda6-tuned}
\end{equation}
providing a direct check of the second tuning. 
The fine-tuned $\lambda_6^t$ realizes fourth-order multicriticality, with 
\be 
\dF\sim \dT^{4/3}\,,\qquad O \sim \dT^{1/6}\,.
\ee 
On the slice
\(\lambda_4=\lambda_4^t\), values \(\lambda_6<\lambda_6^t\) give \(t_4<0\)
and a locally disfavored branch above \(T_c\).  This local near-critical statement alone
does not establish the existence of a finite-amplitude first-order coexistence line.

For the temperature coefficient $t_6$, let
\begin{equation}
 a_h=\phi_1(1),\qquad b_h=\phi_3(1),\qquad
 d_h=\phi_5(1),\qquad c_j=\chi_j(1),
\end{equation}
and define
\begin{align}
 w_2={}&\frac{m^2}{2}a_h^2,\\
 w_4={}&m^2a_hb_h+\lambda_4a_h^4,\\
 w_6={}&m^2a_hd_h+\frac{m^2}{2}b_h^2
 +4\lambda_4a_h^3b_h+\lambda_6a_h^6.
\end{align}
Expansion of the horizon temperature then yields
\begin{equation}
 \tau_6=\frac{1}{8\pi}\left[
 -w_6+\frac{c_2}{2}w_4
 +\left(\frac{c_4}{2}-\frac{c_2^2}{8}\right)w_2
 -3c_6+\frac32c_2c_4-\frac{c_2^3}{8}
 \right]\,,
 \label{eq:tau6}
\end{equation}
which further determines the value of $t_6$,
\begin{equation}
 t_6=\frac{\tau_0z_{h6}-z_{h0}\tau_6}{z_{h0}^2}.
 \label{eq:t6}
\end{equation}
From \eqref{eq:zh-from-B} we have $z_{h6}=-B_7$. Both $\tau_6$ and $z_{h6}$ are completely determined once the perturbative functions $f_6$, $\chi_6$, and $\phi_7$ are obtained.

We define
\begin{align}
 E_2=-\frac{\chi_2}{2},~~~~~
 E_4=-\frac{\chi_4}{2}+\frac{\chi_2^2}{8},~~~~~
 E_6=-\frac{\chi_6}{2}
 +\frac{\chi_2\chi_4}{4}-\frac{\chi_2^3}{48},
 \label{eq:E246}
\end{align}
so that \(e^{-\chi/2}=1+E_2\epsilon^2+E_4\epsilon^4
+E_6\epsilon^6+\cdots\).  The sixth-order metric equations are
\begin{align}
 \chi_6'
&=u\left(\phi_1'\phi_5'+\frac12\phi_3'^2\right),
 \label{eq:chi6-equation}\\
 f_6'-\frac{3f_6}{u}
 &=\frac{1}{2u}\left[
 m^2\phi_1\phi_5+\frac{m^2}{2}\phi_3^2
+4\lambda_4\phi_1^3\phi_3+\lambda_6\phi_1^6\right]+\frac12\left(f_0\chi_6'+f_2\chi_4'+f_4\chi_2'\right).
 \label{eq:f6-equation}
\end{align}
The corresponding coefficient in \(e^{-\chi/2}f/u^2\) is
\begin{equation}
 P_6=\frac{f_6+E_2f_4+E_4f_2+E_6f_0}{u^2}.
 \label{eq:P6-definition}
\end{equation}
Combining with the boundary conditions \eqref{eq:bc-fchi} gives
\be 
\chi_6(1)=1.4694\times 10^{-4}\,.\qquad
\ee 
Then we have 
\be 
\tau_6=4.0174\times10^{-6}\,.
\ee 
It depends only on the sextic term in the scalar potential, which has been fixed to $\lambda_6=\lambda_6^t$.

At order \(\epsilon^7\), write
\begin{equation}
 \mathcal D\phi_7=R_7,
 \label{eq:phi7-equation}
\end{equation}
where
\begin{equation}
 R_7=
 \frac{\widetilde Q_7+E_2Q_5+E_4Q_3+E_6m^2\phi_1}{u^4}
 -\left(P_2\phi_5'+P_4\phi_3'+P_6\phi_1'\right)',
 \label{eq:R7}
\end{equation}
and
\begin{align}
 Q_3={}&m^2\phi_3+4\lambda_4\phi_1^3,\\
 Q_5={}&m^2\phi_5+12\lambda_4\phi_1^2\phi_3
 +6\lambda_6\phi_1^5,\\
 \widetilde Q_7={}&12\lambda_4
 \left(\phi_1^2\phi_5+\phi_1\phi_3^2\right)
 +30\lambda_6\phi_1^4\phi_3+8g_8\phi_1^7.
 \label{eq:Q357}
\end{align}
Although the full profile \(\phi_7\) can be reconstructed, it is not needed
to determine the horizon-radius coefficient.  The same Green identity used
at lower orders gives directly
\begin{equation}
 z_{h6}=-B_7=\int_0^1du\,\phi_1R_7.
 \label{eq:zh6-solvability}
\end{equation}
At the tuned couplings $\lambda_6=\lambda_6^t$, the dependence on the actual octic coefficient $g_8=\lambda\lambda_8$ is
\begin{align}
 z_{h6}(g_8)
 &=1.2608\times10^{-5}
 +0.13559\,g_8\,.
 \label{eq:zh6-g8}
\end{align}
The slopes provide an analytic and numerical check:
\begin{equation}
 \frac{\partial z_{h6}}{\partial g_8}
 =8\int_0^1du\,\frac{\phi_1^8}{u^4}
 =0.13559\,.
 \label{eq:zh6-g8-slopes}
\end{equation}
The value of $z_{h6}$ receives contributions from the eighth-order  interaction $g_8 \phi^8$. Consequently, the scaling relations for temperature depend on $g_8$.

For the exponential potential \eqref{eq:potential}, we choose \(g_8=\lambda\lambda_8=1/10\).
Therefore
\begin{equation}
 z_{h6}=0.0013685\,.
 \label{eq:t6-original-potential}
\end{equation}
Equivalently,
\begin{equation}
 \Delta T
 =0.0021661\,\epsilon^6+\mathcal O(\epsilon^8).
 \label{eq:sixth-temperature-epsilon}
\end{equation}
The branch thus lies below \(T_c\), and since \(s_2<0\), its leading
free-energy difference is also negative:
\begin{equation}
 \Delta\mathcal F
 =\frac34s_2t_6\epsilon^8+\mathcal O(\epsilon^{10})
 =-0.0087819\,\epsilon^8+\mathcal O(\epsilon^{10}).
\end{equation}
Eliminating \(\epsilon\) gives the explicit sixth-root laws
\begin{align}
\begin{split}
 O
 \simeq7.1755\,\Delta T^{1/6},~~~
 \Delta p_t^{(1)}
 \simeq0.27515\,\Delta T^{1/3},~~~
 \Delta\mathcal F
 \simeq-31.335\,\Delta T^{4/3}.
 \label{eq:sixth-root-scaling}
 \end{split}
\end{align}
Using \eqref{eq:ptO}, we obtain the relation between $\Delta p_t^{(1)}$  and $\Delta T$,
\be
\label{eq:scalrel3}
 \Delta p_t^{(1)}
=0.27515\,\Delta T^{1/3}
 +\mathcal{O}(\Delta T^{2/3}).
\ee

The nonzero positive value of \(t_6\) establishes the local
sixth-root scaling.  This perturbative conclusion does not exclude a
disconnected, globally lower-free-energy branch away from the bifurcation
point. More generally, Eq.~\eqref{eq:t6} shows that the
physical below-\(T_c\) branch requires
\(g_8>-4.4902\times10^{-5}\).  At equality \(t_6\) also vanishes, and one
would have to compute \(t_8\) before assigning a higher multicritical
scaling.

\subsubsection{Multicriticality and thermodynamic scaling relations}

The ordinary critical point, the tricritical point, and the fourth-order multicritical point discussed above are successive members of a single perturbative hierarchy. Each additional tuning removes the next even power of \(\epsilon\) in \eqref{eq:tau-expansion} from the leading temperature difference and thereby changes the critical scaling. 

Higher-order multicritical points, obtained by successively tuning lower-order interactions to vanish, have long been studied within the renormalization-group framework \cite{Wegner:1975higher}. In the present holographic setting, this hierarchy is realized by successively eliminating the lower-order coefficients in the temperature expansion, leading to the scaling $O\sim\Delta T^{1/(2r)}$. 
We now summarize this pattern in general and connect it with the corresponding Landau description. 

The structure of the higher-order coefficients follows directly from
perturbation theory.  A coupling multiplying \(\phi^{2j}\) first appears in
the scalar equation at order \(\epsilon^{2j-1}\), and therefore first
contributes to \(z_{h,2j-2}\).  Assigning the weight
\begin{equation}
 w(\lambda_{2j})=j-1,
 \end{equation}
one finds that \(z_{h,2r}\) is a finite polynomial in monomials of total
weight no greater than \(r\):
\begin{equation}
 z_{h,2r}
 =\sum_{\{m_j\}}C_{\{m_j\}}
 \prod_{j=2}^{r+1}\lambda_{2j}^{m_j},
 \qquad
 \sum_{j=2}^{r+1}(j-1)m_j\le r.
 \label{eq:zh2n-weight-rule}
\end{equation}
For example,
\begin{align}
\begin{split}
 z_{h2}
 &=\mathcal Z_{(0)}^{(2)}
 +\mathcal Z_{(1)}^{(2)}\lambda_4,\\
 z_{h4}
 &=\mathcal Z_{(0,0)}^{(4)}
 +\mathcal Z_{(1,0)}^{(4)}\lambda_4
 +\mathcal Z_{(2,0)}^{(4)}\lambda_4^2
 +\mathcal Z_{(0,1)}^{(4)}\lambda_6,
  \label{eq:zh-pattern-examples}
 \\
 z_{h6}
 &=\mathcal Z_{(0,0,0)}^{(6)}
 +\mathcal Z_{(1,0,0)}^{(6)}\lambda_4
 +\mathcal Z_{(2,0,0)}^{(6)}\lambda_4^2
 +\mathcal Z_{(3,0,0)}^{(6)}\lambda_4^3
 +\mathcal Z_{(0,1,0)}^{(6)}\lambda_6
 +\mathcal Z_{(1,1,0)}^{(6)}\lambda_4\lambda_6
 +\mathcal Z_{(0,0,1)}^{(6)}g_8.
 \end{split}
\end{align}

The pattern is clear from the perturbative construction.  If
\begin{equation}
\Delta T=t_2\epsilon^2+t_4\epsilon^4+\cdots+t_{2r}\epsilon^{2r}+\cdots\,,
\end{equation} 
then the scaling \(O\sim\Delta T^{1/(2r)}\) requires
\begin{equation}
 t_2=t_4=\cdots=t_{2r-2}=0,
 \qquad t_{2r}>0.
 \label{eq:general-tuning-condition}
\end{equation}
Thus the ordinary second-order transition is generic, the fourth-root scaling
requires tuning one coupling, and the sixth-root scaling requires tuning two
couplings.  
The tuned values are
determined recursively by the solvability conditions of the higher-order
inhomogeneous scalar equations, and the allowed dependence on the potential
couplings is organized by the weight rule \eqref{eq:zh2n-weight-rule}.
This is why higher critical scalings require higher-order terms in the
potential.  
A scaling $O\sim\Delta T^{1/(2r)}$ requires enough even powers
\begin{equation}
 V(\phi)=-6-\phi^2+\lambda_4\phi^4+\lambda_6\phi^6
 +\cdots+\lambda_{2r+2}\phi^{2r+2}+\cdots 
\end{equation}
to tune the $r-1$ coupling constants.  
In the present model \eqref{eq:potential} the super-exponential term
contains an infinite tower of higher even powers, but near the critical point
only the first few terms in its small-field expansion enter 
at any fixed order of multicriticality.

The same exponents follow from an analytic mean-field Landau description, with
the source coupled linearly to the order parameter.  Suppose the first nonzero
stabilizing interaction is of order \(O^{2m}\):
\begin{equation}
 \Delta\mathcal F_{\mathrm{eff}}
 =-a\,\Delta T\,O^2
 +u_mO^{2m}+\cdots,
 \qquad a>0,
 \qquad u_m>0.
 \label{eq:Landau-general}
\end{equation}
Minimization gives
\begin{equation}
 O\sim\Delta T^{1/(2m-2)},
 \qquad
 \Delta\mathcal F\sim-\Delta T^{m/(m-1)}.
\end{equation}
Writing \(|\Delta\mathcal F|\sim\Delta T^{2-\alpha_{\mathrm{crit}}}\), one obtains\footnote{
 Turning on a source $J$, we have an additional term $-JO$ in \eqref{eq:Landau-general}. At \(T=T_c\), we obtain \(J\sim O^{\delta_{\mathrm{crit}}}\), and therefore $ \delta_{\mathrm{crit}}=2m-1.$}
\begin{equation}
 \beta_{\mathrm{crit}}=\frac{1}{2m-2},
 \qquad
 \alpha_{\mathrm{crit}}=\frac{m-2}{m-1}. 
\end{equation}
Identifying \(m=r+1\), we obtain 
\begin{equation}
 \beta_{\mathrm{crit}}=\frac{1}{2r},
 \qquad
 \alpha_{\mathrm{crit}}=\frac{r-1}{r},
 \qquad
 \delta_{\mathrm{crit}}=2r+1.
 \label{eq:general-critical-exponents}
\end{equation}

Finally, the small-\(\epsilon\) expansion is not universally applicable all the way to the
singularity.  It is valid in the exterior, near the horizon, and on the first Kasner epoch before the scalar reaches the steep
super-exponential wall.  At later Kasner epochs, this expansion might break  down. Nevertheless, this deep-interior
breakdown does not affect the boundary critical exponents derived above.  It
only limits how far a fixed local Kasner epoch can be followed before the
potential wall cuts it off.  Before that cutoff, \(p_t^{(1)}\) is analytic in
\(\epsilon\) and obeys Eq.~\eqref{eq:ptO}.

\subsection{Critical scaling of subsequent Kasner epochs}
\label{sec:subsequent-kasner-epochs}

The previous subsection established the critical scaling of the temporal
Kasner exponent in the \emph{first} local epoch, as in \eqref{eq:scalrel1}, \eqref{eq:scalrel2}, \eqref{eq:scalrel3} for different orders of phase transitions.  That result follows by
matching the linear threshold profile from the exterior into the interior
before the first bounce.  Here we ask how this boundary critical behavior is
transmitted to the Kasner exponents in subsequent epochs, which are generated
by nonlinear bounces off the super-exponential potential.  It is important to
distinguish a fixed later epoch, such as \(n=2\) or \(n=3\), from the
late-epoch limit \(n\to\infty\): the former requires regularity of the
finite-bounce matching map, whereas the latter is controlled by a separate
large-\(n\) asymptotic solution.

As found in previous subsection, for the $r$-th order of multicriticality,
\begin{equation}
 \Delta T=t_{2r}\epsilon^{2r}
 +\mathcal O(\epsilon^{2r+2})\,,
 \qquad r=1\,,2\,,3\,,
 \label{eq:DeltaT-r-hierarchy}
\end{equation}
for the ordinary critical, tricritical, and fourth-order multicritical
points, respectively.  Equations~\eqref{eq:v-order-parameter} and
\eqref{eq:pt1-v1} then give the established first-epoch result
\begin{equation}
 \Delta p_t^{(1)}
 =\frac{C_v^2}{9\,t_{2r}^{1/r}}\,
 \Delta T^{1/r}
 +\mathcal O\!\left(\Delta T^{2/r}\right)\,.
 \label{eq:first-epoch-r-scaling}
\end{equation}
We also expect that $\Delta p_t^{(n)}\to0$ as $\Delta T\to 0$, since all epochs satisfy $0<\Delta p_t^{(n+1)}<\Delta p_t^{(n)}\,.$

For each fixed \(n\), assume that the \(n\)-th plateau can be continuously tracked for all sufficiently small nonzero \(\epsilon\), and that the finite nonzero limit \(C_{v,n}=\lim_{\epsilon\to0}v_n(\epsilon)/\epsilon\) exists. We have\footnote{We distinguish the symbols $\mathcal{O}$ and $o$ as follows. In the limit $x\to0$, $y_1(x)=\mathcal O(y_2(x))$ means that $y_1(x)/y_2(x)$ remains bounded, whereas $y_1(x)=o(y_2(x))$ means that $y_1(x)/y_2(x)\to0$. Thus $o(y_2)$ denotes a correction parametrically smaller than $y_2$.}  
\begin{equation}
 v_n(\epsilon)=C_{v,n}\epsilon+ o(\epsilon)\,,
 \qquad n\ \text{fixed}\,.
 \label{eq:fixed-epoch-matching}
\end{equation}
This is a matched-asymptotic assumption and can be checked numerically. 

For every local Kasner epoch, the exact relation between the scalar velocity
and the Kasner exponent is
\begin{equation}
 \Delta p_t^{(n)}
 \equiv p_t^{(n)}-p_t^c
 =\frac{4v_n^2}{3(v_n^2+12)}
 =\frac{v_n^2}{9}+\mathcal O(v_n^4)\,.
 \label{eq:pn-vn}
\end{equation}
{Combining Eqs.~\eqref{eq:DeltaT-r-hierarchy},
\eqref{eq:fixed-epoch-matching}, and \eqref{eq:pn-vn} gives
\begin{equation}
 \Delta p_t^{(n)}
 =\frac{C_{v,n}^2}{9\,t_{2r}^{1/r}}\,\Delta T^{1/r}\big(1
 +o(1)\big)\,,
 \qquad n\ \text{fixed}\,.
 \label{eq:fixed-later-epoch-scaling}
\end{equation}
Thus a regularly matched fixed epoch inherits the same temperature exponent
as the first epoch: \(1\), \(1/2\), and \(1/3\) at the ordinary critical,
tricritical, and fourth-order multicritical points, respectively.  Its
coefficient is nevertheless independent data: \(C_{v,n}\) must be obtained 
from a numerical fit, and could not be inferred
from \(C_v=C_{v,1}\) alone.
}

For the super-exponential potential \eqref{eq:potential}, a complementary result applies in 
the late-epoch limit.  The analysis of Ref.~\cite{Hartnoll:2022snh} gives
\begin{equation}
 |v_n(T)|
 =b(T)\,\left[n-n_0(T)\right]^{-1/6}\,
 \left(1+o_{n\to\infty}(1)\right)\,,
 \qquad n-n_0\gg1\,,
 \label{eq:late-epoch-velocity}
\end{equation}
where \(b>0\) and \(n_0\) are branch-dependent integration constants.  
 We assume that a finite critical limit of the shift $n_0(T)$ satisfies
\begin{align}
 n_0(T)
 &=n_c+A\,\Delta T^\gamma
 +o\!\left(\Delta T^\gamma\right)\,,
 &\gamma&>0\,,
 \label{eq:n0-critical-ansatz}\\
 b(T)
 &=B|\epsilon|\left[1+\mathcal O(\epsilon^2)\right]\,,
 &B&>0\,.
 \label{eq:b-critical-ansatz}
\end{align}
Under these assumptions, in the joint critical and late-epoch regime one
finds
\begin{equation}
 \Delta p_t^{(n)}
 =\frac{B^2}{9\,t_{2r}^{1/r}}\,
 (n-n_c)^{-1/3}\Delta T^{1/r}
 \left[
 1+\mathcal O\!\left(
 \Delta T^{\min(\gamma,1/r)}
 \right)+o_{n\to\infty}(1)
 \right]\,.
 \label{eq:late-epoch-temperature-scaling}
\end{equation}
Therefore, at large $n$, we expect that 
\begin{equation}
 \frac{\Delta p_t^{(n)}}{\Delta p_t^{(1)}}
 =\frac{B^2}{C_v^2}(n-n_c)^{-1/3}[1+o_T(1)] \,.
 \label{eq:first-to-late-general}
\end{equation}

\subsection{Possible loss of the first Kasner regime at lower temperatures}
\label{ss:first-epoch-switch}

The analysis above focuses on the first local Kasner regime in the
vicinity of the transition temperature.  As the temperature is lowered away
from \(T_c\), however, the interior evolution changes qualitatively.  The
scalar reaches the steep super-exponential part of the potential at an
earlier stage of its evolution, and the associated force truncates the first
kinetic-dominated interval before it can develop into a distinct Kasner
plateau. As the temperature is lowered, the first Kasner regime identified at higher temperatures may cease to exist as a distinct local epoch. In the following we briefly estimate this special temperature below which the first Kasner epoch is lost. 

To estimate when this loss occurs, it is convenient to introduce the logarithmic radial coordinate
\begin{equation}
 \rho\equiv\log\frac{z}{z_h}\,,
 \qquad
 \dot X\equiv\frac{dX}{d\rho}\,.
 \label{eq:first-loss-rho}
\end{equation}
Within the \(n\)-th kinetic-dominated epoch, the leading Kasner behavior is
\begin{equation}
 \dot\phi\simeq v_n\,,
 \qquad
 \ddot\phi\simeq0\,,
 \qquad
 |f|\propto
 e^{(3+v_n^2/4)\rho}\,.
 \label{eq:first-loss-kasner-condition}
\end{equation}
Thus a distinct Kasner plateau requires an extended radial interval over
which the scalar velocity is approximately constant.

In terms of \(\rho\), the scalar equation~\eqref{eq:eom-u-phi-compact} can be written as
\begin{equation}
 \ddot\phi
 =\frac{1}{f}\left(
 \partial_\phi V-\frac{\dot\phi}{2}V
 \right).
 \label{eq:first-loss-rho-equation}
\end{equation}
For the super-exponential term
\begin{equation}
 V_{\rm se}(\phi)=\lambda e^{\lambda_8\phi^8}\,,
\end{equation}
its contribution to the scalar acceleration is
\begin{equation}
 \left(\ddot\phi\right)_{\rm se}
 =
 \frac{1}{f}
 \left(8\lambda_8\phi^7-\dot\phi/2\right)V_{\rm se}\,.
 \label{eq:first-loss-superexp-acceleration}
\end{equation}
Since the existence of a Kasner plateau requires the scalar velocity to remain nearly constant, the relevant quantity is the change of $\dot\phi$ induced by the super-exponential interaction relative to $\dot\phi$ itself. This motivates the dimensionless ratio
\begin{equation}
 \mathcal R_{\rm se}
 \equiv
 \left|
 \frac{(\ddot\phi)_{\rm se}}{\dot\phi}
 \right|
 =
 \left|
 \frac{8\lambda_8\phi^7-\dot\phi/2}{\dot\phi}
 \right|
 \frac{|V_{\rm se}|}{|f|}\,.
 \label{eq:first-loss-force-ratio}
\end{equation}
When
\(R_{\rm se}\ll1,\)
the super-exponential interaction only weakly perturbs the scalar velocity, so that a distinct Kasner plateau can develop. In contrast, when
\(R_{\rm se}=O(1),\)
the scalar velocity changes significantly over a logarithmic radial interval of order unity, preventing the formation of an extended kinetic-dominated plateau.
\footnote{The loss of the first Kasner epoch at low temperature is not universal. As the temperature decreases, the horizon value of the scalar field $\phi_h$ generally increases along the branches considered here. If the polynomial part of the potential remains the leading correction near the first Kasner regime, the super-exponential term does not eliminate this regime, which can then persist smoothly to $T=0$. We find this behavior, for example, 
for $\lambda_4=\lambda_4^t$, $\lambda_6=1$, and $\lambda=\lambda_8=1/10$. For other parameter choices, however, the first Kasner epoch may instead terminate  at a finite temperature $T_{\rm loss}$.}

 This criterion
depends not only on the magnitude of \(V_{\rm se}\), but also
on the factor \(8\lambda_8\phi^7-\dot\phi/2\), which measures the steepness of the
potential wall.  It is consequently more appropriate than the simpler
estimate \(|V_{\rm se}|\sim|f|\).  
Before the first bounce, we approximate the solution by
\begin{equation}
 \phi(\rho)\simeq\phi_h+2c_h\rho\,,
 \qquad
 |f(\rho)|\simeq f_K e^{(3+c_h^2)\rho}\,,
 \label{eq:first-switch-kasner-approx}
\end{equation}
where horizon regularity gives
\begin{equation}
 c_h=\frac{\partial_\phi V(\phi_h)}{V(\phi_h)}\,.
 \label{eq:first-switch-ch}
\end{equation}
Here \(f_K\) is the normalization obtained by matching through the short
post-horizon region into the first kinetic-dominated interval.  Treating
\(c\simeq c_h\) as constant and \(f_K \simeq 1\) 
is the main approximation in the estimate.

The loss of a numerically extracted plateau also depends on the
plateau-identification prescription.  
We adopt the criterion that an epoch cannot be clearly extracted if its radial interval satisfies $\Delta \rho<\rho_p$, where $\rho_p$ is a small parameter chosen by hand. Therefore, the condition for the ``loss'' of the first epoch is estimated by requiring the
super-exponential force to become order one at $\rho_p$:
\begin{equation}
 \mathcal R_{\rm se}(\rho_p)\simeq1.
 \label{eq:first-switch-at-rhostar}
\end{equation}
Substituting Eq.~\eqref{eq:first-switch-kasner-approx} gives
\begin{align}
 0 
 &\simeq
 \log\frac{|\lambda|}{f_K}
 +\lambda_8\phi_p^8
 -(3+c_h^2)\rho_p
 +\log\left|
 \frac{8\lambda_8\phi_p^7-c_h}{2c_h}
 \right|\,,
 \label{eq:first-switch-G}\\
 \phi_p
 &\equiv\phi_h+2c_h\rho_p \,.
 \label{eq:first-switch-phistar}
\end{align} 
Once we specify the numerical criterion \(\rho_p\), the equations give a special horizon value \(\phi_h^{\rm loss}\), and thus determine a special temperature $T_{\rm loss}$, at which the first Kasner epoch appears to be lost. This estimation method allows us to obtain an approximate temperature $T_{\rm loss}$, without performing a scan over the entire phase diagram.

A quantitative determination of $T_{\rm loss}$ requires a separate analysis; here we only show the existence and physical interpretation of the phenomenon. 
We will compare this semi-analytical result with the numerical result in Sec. \ref{ss:2nd}. 


\subsection{Low-temperature scaling from the IR fixed point}
\label{ss:lowT}

We close the analytical discussion with the opposite asymptotic limit,
\(T\to0\).  
The powers obtained below are not universal critical exponents: they are fixed by the IR AdS\(_4\) geometry of a given zero-temperature branch.  We first derive the scaling of the horizon data,
which is robust, and then state separately the additional condition required
to transfer that scaling to a specified local Kasner epoch.

Suppose that the zero-temperature domain wall approaches a nondegenerate IR
AdS\(_4\) extremum with
\begin{equation}
 \partial_\phi V(\phi_\ast)=0,
 \qquad
 V(\phi_\ast)<0,
 \qquad
 f_\ast=-\frac{V(\phi_\ast)}{6}>0.
 \label{eq:lowT-ir-fixed-point}
\end{equation}
The radius of the IR AdS$_4$ and the effective mass of the scalar field are
\begin{align}
    L^2_\text{IR} = \frac{1}{f_*}\,,~~~ m^2_\text{IR} = \partial^2_{\phi}V(\phi_*)\,.
\end{align}

The linearized scalar perturbation around the IR AdS$_4$ fixed point admits two independent asymptotic modes 
\begin{equation}
 \phi^{(0)}(z)-\phi_\ast
 =\mathcal A_{\rm IR} \,z^{\delta_{\rm IR}^-}
 +\mathcal B_{\rm IR} \,z^{\delta_{\rm IR}^+},
 \qquad z\to\infty\,,
\end{equation}
where 
\begin{align}
    \delta_{\rm IR}^{\pm} = \frac{3}{2}\pm \sqrt{\frac{9}{4}+m^2_{\rm  IR}L^2_{\rm IR}}\,.
\end{align}
For a nondegenerate IR AdS$_4$ extremum satisfying  $\partial_\phi^2V(\phi_*)>0$, we have $\delta_{\rm IR}^-<0$ and $\delta_{\rm IR}^+>3$. Consequently, the mode proportional to
\(z^{\delta_{\rm IR}^-}\)
vanishes as
\(z\rightarrow\infty\),
whereas the mode proportional to
\(z^{\delta_{\rm IR}^+}\)
grows and therefore does not approach the IR fixed point. A zero-temperature domain-wall solution terminating at the IR fixed point therefore requires the growing mode to be absent, implying
\(\mathcal B_{\rm IR}=0\). The remaining solution is therefore governed by the leading irrelevant deformation of the IR fixed point.

According to the standard AdS/CFT dictionary, the larger exponent determines the conformal dimension of the dual IR operator $\Delta_{\mathcal O}^{\rm IR}=\delta_{\rm IR}^+.$ It is therefore convenient to introduce the positive irrelevant falloff exponent
\be
\delta_{\rm IR}
 \equiv \Delta_{\mathcal O}^{\rm IR}-3=-\delta_{\rm IR}^-=
\frac12
\left(
\sqrt{9+4m_{\rm IR}^2L_{\rm IR}^2}-3
\right)\,.
\ee
The scalar field thus approaches the IR fixed point as
\begin{equation}
\phi^{(0)}(z)-\phi_*
=
A_{\rm IR}z^{-\delta_{\rm IR}}
+
o\!\left(z^{-\delta_{\rm IR}}\right),
\qquad
z\rightarrow\infty.
 \label{eq:lowT-zeroT-tail}
\end{equation}

The above analysis applies to the zero-temperature domain-wall solution, whose IR region extends to
\(z\rightarrow\infty\).
At small but finite temperature, the domain wall is smoothly deformed into a black-hole solution.
The IR AdS$_4$ throat is then truncated by a regular event horizon located at
\(z=z_h\),
while the geometry outside the horizon remains arbitrarily close to the zero-temperature solution in the limit
\(T\rightarrow0\).
Since the scalar field has already relaxed to the IR fixed point before reaching the horizon, the near-horizon geometry is governed by the planar AdS$_4$ Schwarzschild solution associated with the IR vacuum. Introducing the dimensionless radial coordinate
\(u=z/z_h\), the leading throat geometry is
\begin{equation}
 f(u)=f_\ast(1-u^3)+o(1),
 \qquad
 \chi(u)=\chi_\ast+o(1),
 \label{eq:lowT-throat-geometry}
\end{equation}
and hence
\begin{equation}
 T=
 \frac{3f_\ast e^{-\chi_\ast/2}}{4\pi z_h}\,
 [1+o(1)].
 \label{eq:lowT-Tzh}
\end{equation}

We now extend the above zero-temperature analysis to small but nonzero temperature.
As discussed above, the finite-temperature solution is obtained by smoothly truncating the infinitely long IR AdS$_4$ throat with a regular event horizon at $z=z_h$.
In the low-temperature limit, the overlap region \(1\ll z\ll z_h\), equivalently $1/z_h\ll u\ll 1$
is simultaneously described by the zero-temperature IR solution and the finite-temperature throat geometry.

Matching the scalar field to the asymptotic behavior in Eq.~\eqref{eq:lowT-zeroT-tail}, we write
\begin{equation}
 \phi(z;T)-\phi_\ast
 =
 \mathcal A_{\rm IR}\, z_h^{-\delta_{\rm IR}}\eta(u)
 +o\!\left(z_h^{-\delta_{\rm IR}}\right),
 \label{eq:lowT-eta-ansatz}
\end{equation}
where the dimensionless function $\eta(u)$ describes the radial profile inside the finite-temperature IR throat. Substituting this ansatz into the linearized scalar equation, one obtains
\begin{equation}
 \left(\frac{1-u^3}{u^2}\eta'\right)'
 =
 \mu_{\rm IR}\frac{\eta}{u^4},
 \qquad
 \mu_{\rm IR}
 \equiv\frac{\partial_\phi^2V(\phi_\ast)}{f_\ast}
 =\delta_{\rm IR}(\delta_{\rm IR}+3), 
 \label{eq:lowT-eta-eq}
\end{equation}
where the prime denotes differentiation with respect to $u$.  

The solution is determined by matching to the zero-temperature IR falloff and imposing the regularity condition at the horizon
\begin{equation}
 \eta(u)=u^{-\delta_{\rm IR}}\,[1+o(1)]
 \quad (u\to0),
 \qquad
 \eta'(1)=-\frac{\mu_{\rm IR}}{3}\eta(1).
 \label{eq:lowT-eta-boundary-conditions}
\end{equation}
These two conditions uniquely determine the horizon value \(\eta_h\equiv\eta(1)\). Here \(u\to0\) is understood in the matching sense \(z_h\to\infty\), \(u\to0\), with fixed \(z=uz_h\gg 1\), rather than as the asymptotic AdS boundary.

Evaluating Eq.~\eqref{eq:lowT-eta-ansatz} at the horizon and using Eq.~\eqref{eq:lowT-Tzh} immediately yields 
\begin{equation}
 \phi_h-\phi_\ast
 =
 C_\phi T^{\delta_{\rm IR}}
 +o\!\left(T^{\delta_{\rm IR}}\right), 
~~~~~~~ 
 C_\phi
 =
 \eta_h\mathcal A_{\rm IR}
 \left(
 \frac{4\pi e^{\chi_\ast/2}}{3f_\ast}
 \right)^{\delta_{\rm IR}}.
 \label{eq:lowT-Cphi}
\end{equation}
The exponent $\delta_{\rm IR}$ is determined entirely by the local IR fixed-point data and is therefore universal, whereas the prefactor $C_\phi$ also depends on the global normalization of the domain-wall solution through $\mathcal A_{\rm IR}$ and the boundary time normalization encoded in $\chi_*$.

To relate the IR scalar scaling to the Kasner exponents inside the horizon, we first determine the scaling of the scalar velocity at the horizon, which provides the initial data for the interior evolution. Horizon regularity gives the 
scalar velocity
\begin{equation}
 v_h
 \equiv z_h\phi'(z_h)
 =\frac{2 \partial_\phi V(\phi_h)}{V(\phi_h)}.
 \label{eq:lowT-vh-def}
\end{equation}
Expanding around the IR fixed point $\phi_*$ and using Eq.~\eqref{eq:lowT-Cphi}, we immediately obtain
\begin{equation}
 v_h
 =
 C_{v,h}T^{\delta_{\rm IR}}
 +o\!\left(T^{\delta_{\rm IR}}\right),
 \qquad
 C_{v,h}
 =
 \frac{2 \partial_\phi^2 V(\phi_\ast)}{V(\phi_\ast)}C_\phi.
 \label{eq:lowT-vh-scaling}
\end{equation}

We now follow the scalar evolution into the $n$-th Kasner plateau. We assume that, for each fixed $n$, the corresponding Kasner plateau persists continuously in the low-temperature limit, so that
\begin{equation}
 \mathcal M_n
 \equiv
 \lim_{T\to0}
 \frac{v_n(T)}{v_h(T)}
\end{equation}
where $\mathcal M_n$ is finite and nonzero.
It then follows immediately that
\begin{equation}
 v_n(T)
 =
 \mathcal M_n v_h(T)+o\!\left(v_h(T)\right)=
 \widetilde C_nT^{\delta_{\rm IR}}
 +o\!\left(T^{\delta_{\rm IR}}\right),
 \qquad
 \widetilde C_n
 \equiv
 \mathcal M_n C_{v,h}.
 \label{eq:lowT-epoch-matching}
\end{equation}

Since the Kasner exponents are analytic functions of the scalar velocity $v_n$, substituting Eq.~\eqref{eq:lowT-epoch-matching} into the exact Kasner relations immediately yields
\begin{align}
 p_t^{(n)}-\ptsch
 &=
 \frac{4v_n^2}{3(v_n^2+12)}
 =
 \frac{\widetilde C_n^2}{9}\, 
 T^{2\delta_{\rm IR}}
\, \left[1+o(1)\right],
 \label{eq:lowT-pt-scaling}\\
 p_x^{(n)}-\pxsch
 &=
 -\frac{2v_n^2}{3(v_n^2+12)}
 =
 -\frac{\widetilde C_n^2}{18}\, 
 T^{2\delta_{\rm IR}}
\, \left[1+o(1)\right],
 \label{eq:lowT-px-scaling}\\
  p_\phi^{(n)}- \pphisch
 &=
 \frac{4\sqrt2\,v_n}{12+v_n^2}
 =
 \frac{\sqrt2}{3}\widetilde C_n\, T^{\delta_{\rm IR}}
\, \left[1+o(1)\right].
 \label{eq:lowT-pphi-scaling}
\end{align}
Therefore, the deviations of the Kasner exponents from the Schwarzschild values obey
\begin{equation}
 p_t^{(n)}-\ptsch
 \propto T^{2\delta_{\rm IR}}\,,
 \qquad
 p_x^{(n)}-\pxsch
 \propto T^{2\delta_{\rm IR}}\,,
 \qquad
 p_\phi^{(n)}-\pphisch \propto T^{\delta_{\rm IR}}\,,
 \label{eq:lowT-conditional-powers}
\end{equation}
where the scaling follows from the assumption \eqref{eq:lowT-epoch-matching}, whose validity can be tested numerically.

We now evaluate the corresponding IR quantities for the potential \eqref{eq:potential}. For
\(\lambda_4=\lambda=\lambda_8=1/10\) and \(\lambda_6=0\), 
we obtain 
\begin{equation}
 \phi_\ast=1.3381\,,
 \qquad
 f_\ast=1.2151\,,
 \qquad
 \partial_\phi^2 V(\phi_\ast)=19.692\,, \qquad
 \delta_{\rm IR}=2.7961\,.
 \label{eq:lowT-superexp-exponents}
\end{equation}
Solving 
\eqref{eq:lowT-eta-eq}--\eqref{eq:lowT-eta-boundary-conditions} gives
\begin{equation}
 \mu_{\rm IR}=16.207\,,
 \qquad
 \eta_h=0.53479\,.
 \label{eq:lowT-transfer-numbers}
\end{equation}
while the remaining normalization is fixed by the zero-temperature domain-wall solution.
For \(\kappa=-1\), we obtain
\begin{equation}
 C_\phi=-75.048\,,
 \qquad
 C_{v,h}=405.42
 \label{eq:lowT-corrected-amplitudes}
\end{equation}
The 
regular 
solution for $\eta(u)$ can be continued through the horizon.  Its linear large-\(u\) behavior is
\begin{equation}
 \eta(u)=K_{\rm IR}\log u+\mathcal O(1)\,,
 \qquad
 K_{\rm IR}=-0.10821\,.
 \label{eq:lowT-KIR}
\end{equation}
If it is continuously connected to the first Kasner plateau,
the transfer coefficient is
\begin{equation}
 \mathcal M_{\rm IR}
 =
 -\frac{3K_{\rm IR}}{\mu_{\rm IR}\eta_h}
 =0.037456\,,
 \label{eq:lowT-MIR}
\end{equation}
leading to the analytic prediction
\begin{equation}
 v_1 
 \simeq15.186\,T^{2.7961}\,,
 \qquad
 p_t^{(1)} -\ptsch
 \simeq25.622\,T^{5.5922}\,.
 \label{eq:lowT-superexp-linear-channel}
\end{equation}


To illustrate the effect of the super-exponential interaction, we repeat the same analysis for the polynomial model obtained by setting
\(\lambda=0\)
in Eq.~\eqref{eq:potential}.
For \(\lambda_4=1/10\) and \(\lambda_6=0\), we obtain
\begin{equation}
 \phi_\ast=\sqrt5\,,
 \qquad
 f_\ast=\frac{17}{12}\,,
 \qquad
\partial_\phi^2 V(\phi_\ast)=4\,,
 \qquad
 \delta_{\rm IR}=0.75245\,.
 \label{eq:lowT-polynomial-data}
\end{equation}
The corresponding transfer solution gives
\begin{equation}
 \eta_h^{\rm poly}=0.92664\,,
 \qquad
 K_{\rm IR}^{\rm poly}=-0.62730\,.
 \label{eq:lowT-polynomial-transfer}
\end{equation}
Consequently,
\begin{equation}
 C_\phi^{\rm poly}=-7.4945\,,
 \qquad
 C_{v,h}^{\rm poly}=7.0537\,.
 \label{eq:lowT-polynomial-horizon-amplitudes}
\end{equation}
For any fixed epoch satisfying Eq.~\eqref{eq:lowT-epoch-matching}, the full
coefficient can therefore be written as
\begin{equation}
 p_t^{(n)}-\ptsch
 =5.5282\,\mathcal M_n^2\,
 T^{1.5049}
 +o\!\left(T^{1.5049}\right).
 \label{eq:lowT-no-superexp-general-coefficient}
\end{equation}
For the strict linear IR-throat channel,
\begin{equation}
 \mathcal M_{\rm IR}^{\rm poly}=0.71928\,,
 \qquad
 v_1 ^{\rm poly}
\simeq5.0735\,T^{0.75245}\,,
 \label{eq:lowT-polynomial-linear-velocity}
\end{equation}
and hence
\begin{equation}
 p_t^{(1)}-\ptsch
\simeq 2.8601\,T^{1.5049}\,.
 \label{eq:lowT-no-superexp-slope}
\end{equation}
The comparison clearly shows the role of the super-exponential interaction.
Although it affects the UV potential only at high order, it substantially modifies the IR fixed point by increasing the local curvature of the potential.
Consequently, the irrelevant exponent increases from
\(
\delta_{\rm IR}=0.75245
\)
to
\(
2.7961,
\)
leading to a much faster approach of the Kasner exponents to their Schwarzschild values at low temperature.
These analytic predictions \eqref{eq:lowT-superexp-linear-channel} and \eqref{eq:lowT-no-superexp-slope} will be tested numerically 
in Sec. \ref{ss:num-low-temperature}.

\section{Numerical analysis}
\label{sec:num}

The previous section derived the near-critical scaling of the thermodynamic 
observables and local Kasner exponents, as well as their low-temperature scaling.  We now solve the full nonlinear
boundary-value problem and test those predictions numerically.  Our analysis
contains three aspects.  First, we verify the thermodynamic and
Kasner scaling laws at the ordinary critical and tricritical points.  Second,  
we compare these two models in 
the extremely low-temperature regime governed by their respective IR fixed
points. Finally, we also numerically study a first-order phase transition model.

As in the previous section, numerically we work in unit 
\(\kappa=-1\). The detailed numerical procedure can be found in Appendix \ref{app:numerical-procedure}. 
Unless stated otherwise, the super-exponential parameters in \eqref{eq:potential} are fixed to
$
 \lambda=\lambda_8=\frac{1}{10}.
$
\subsection{Second-order phase transition}
\label{ss:2nd}

\begin{figure}[h!]
\begin{center}
\includegraphics[width=0.44\textwidth]{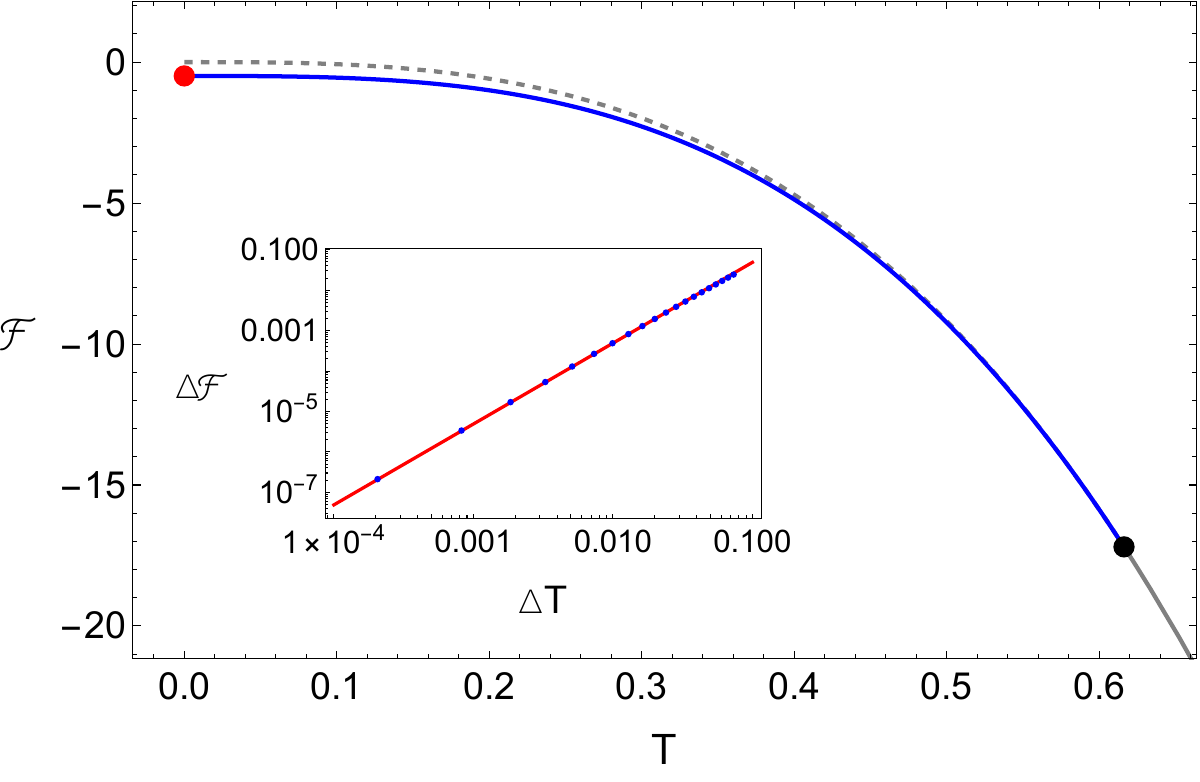}\,~~~~~
\includegraphics[width=0.43\textwidth]{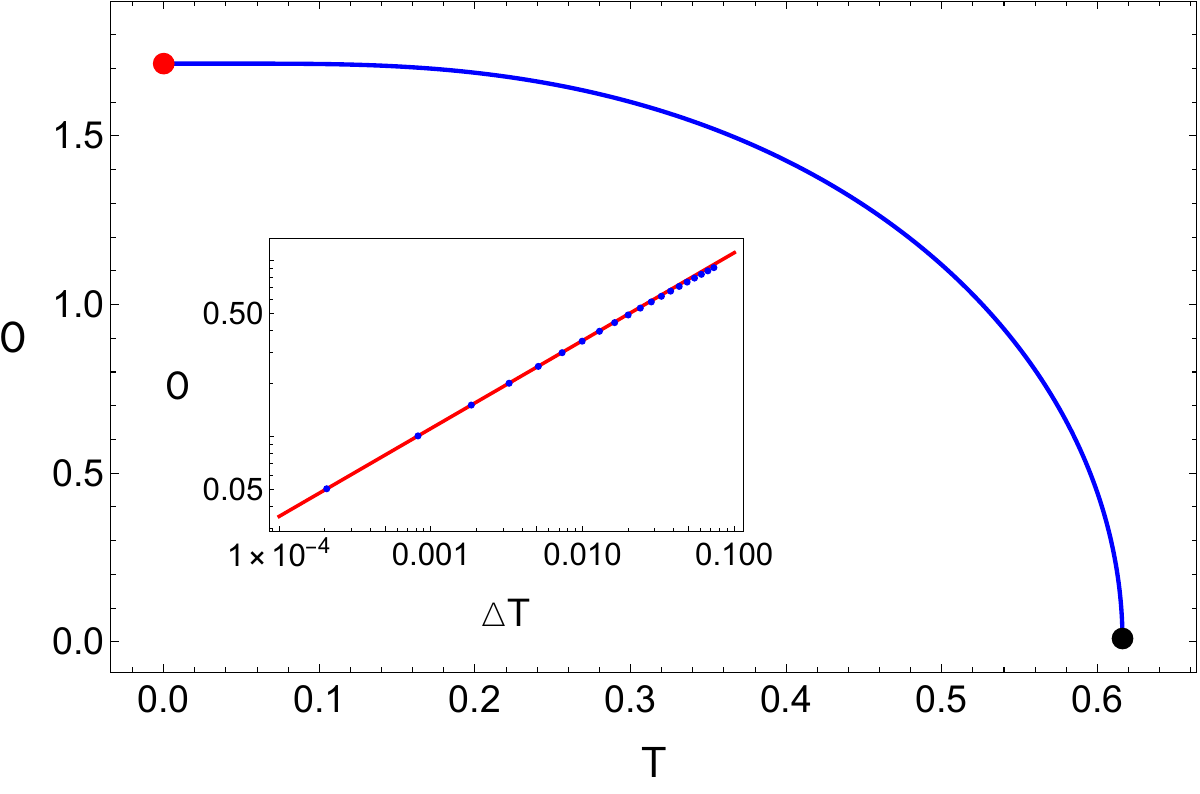}
\end{center}
\vspace{-0.7cm}
\caption{\small Thermodynamics and condensate at the generic second-order critical point at $\lambda_4=\lambda=\lambda_8=1/10, \lambda_6=0$.
{\bf Left:} free-energy densities of the Schwarzschild ({\em gray}) and hairy ({\em blue}) black holes. 
{\bf Right:} The order parameter $O /(-\kappa)$ as a function of $T/(-\kappa)$. In both panels, the black dot marks the critical point of the phase transition, while the red dot denotes the corresponding zero-temperature domain wall solution. The insets are log-log plots: \(\Delta\mathcal F\) versus \(\Delta T=T_c-T\) on the left, and \(O\) versus
\(\Delta T\) on the right. We work in unit $\kappa=-1$.
}
 \label{fig:num-second-order-thermodynamics}
\end{figure}

For the second-order phase transition, we choose 
\begin{equation}
 \lambda_4=\frac{1}{10}\,,
 \qquad
 \lambda_6=0\,,
 \label{eq:num-second-order-parameters}
\end{equation}
as a representative example.
Under the sourceless boundary condition $\kappa\alpha-\beta=0$, the hairy black hole solution can only be found below the critical temperature,
\begin{equation}
 T_c^{\rm num}\simeq0.61618\,,
 \end{equation}
which is in agreement with the perturbative result
\(T_c^{\rm an}=0.6161818601\) obtained in \eqref{eq:tc-ana}.  Below \(T_c\), the condensate is nonzero and
\(\Delta\mathcal F<0\), as shown in Fig. \ref{fig:num-second-order-thermodynamics}, indicating that the hairy solution is thermodynamically preferred.

 Close to the transition temperature $T_c\,,$ numerically we find
\begin{align}
 O
 &\simeq3.5071\,\Delta T^{0.49991}\,,~~~~~~
 \Delta\mathcal F
\simeq-4.9821\,\Delta T^{1.9996}\,.
 \label{eq:num-second-order-free-energy}
\end{align}
The analytical predictions in \eqref{eq:ordinary-temperature-scaling} are
\(3.5103\,\Delta T^{1/2}\) and
\(-4.9993\,\Delta T^2\), respectively.  Thus the fitted powers reproduce the
mean-field values, while the fitted amplitudes differ from the analytical ones by a relative deviation of order $10^{-3}$.\footnote{The deviations can be further reduced by considering data closer to $T_c\,,$ where the perturbative expansion becomes more accurate. In the following, we omit the detailed error estimates and simply state that the numerical results agree well with the analytical predictions. }

\begin{figure}[h!]
\begin{center}
\includegraphics[
width=0.45\textwidth]{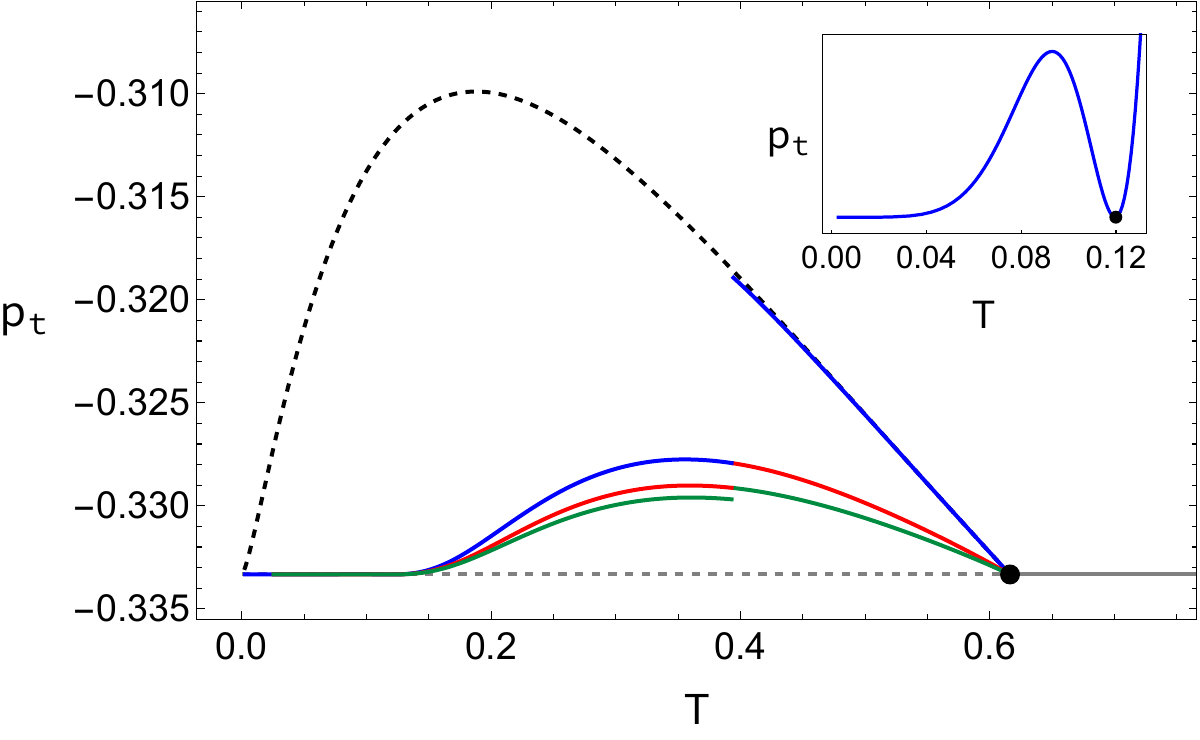}~~~
\includegraphics[
width=0.453\textwidth]{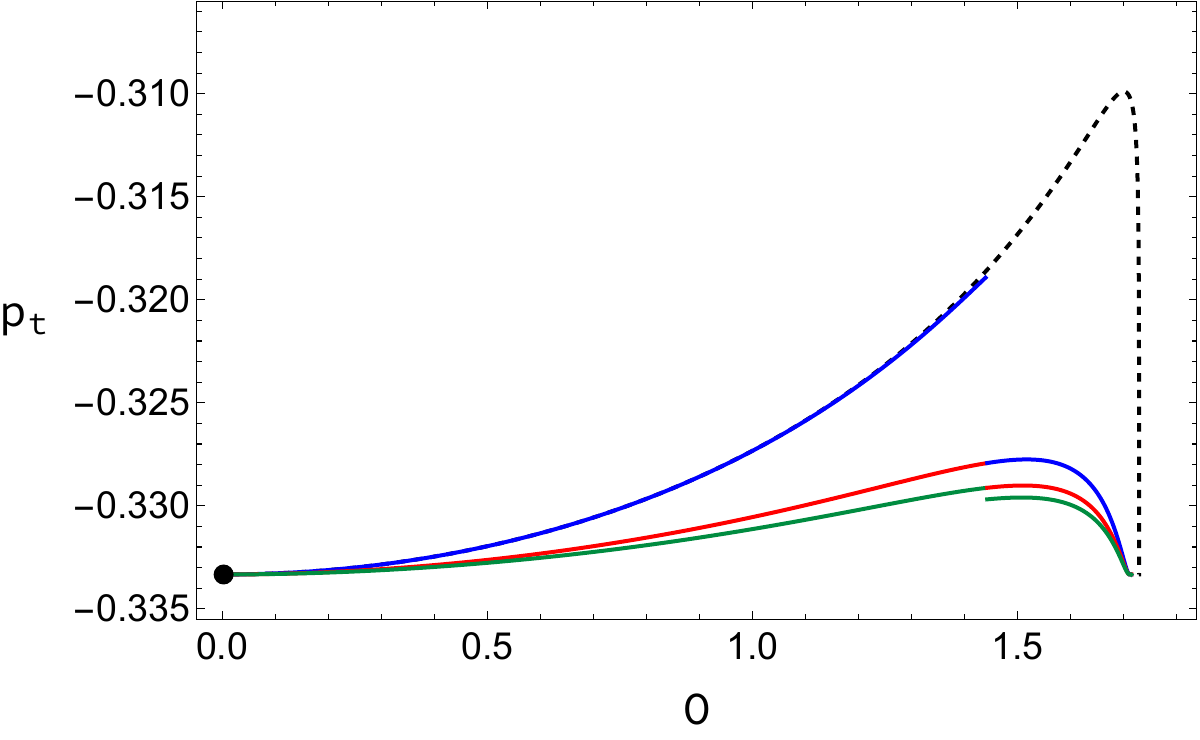}
\end{center}
\vspace{-0.6cm}
\caption{\small The Kasner exponent $p_t$ as a function of the temperature $T$ ({\bf left}) and condensate $O$ ({\bf right}) for $\lambda_4=\lambda=\lambda_8=1/10, \lambda_6=0$.  In the left plot, the gray line is the
 Schwarzschild value \(p_t=-1/3\); the first ({\em blue}), second ({\em red}), and third ({\em green}) epochs are shown
 separately, while the polynomial model without the super-exponential term (i.e. set $\lambda=0$) ({\em black dashed}) is
 included for comparison. Interestingly, $p_t\simeq p_t^{\text{Sch}}$ at a special temperature  $T=0.11965$, as shown in the inset of the left panel, and the interior of the hairy black hole becomes Schwarzschild-like. Details of the interior are discussed in 
 Appendix \ref{app:screened-kasner}. 
 We work in unit $\kappa=-1$. }
 \label{fig:num-second-order-kasner}
\end{figure}

\begin{figure}[h!]
 \centering
 \includegraphics[width=0.43\textwidth]{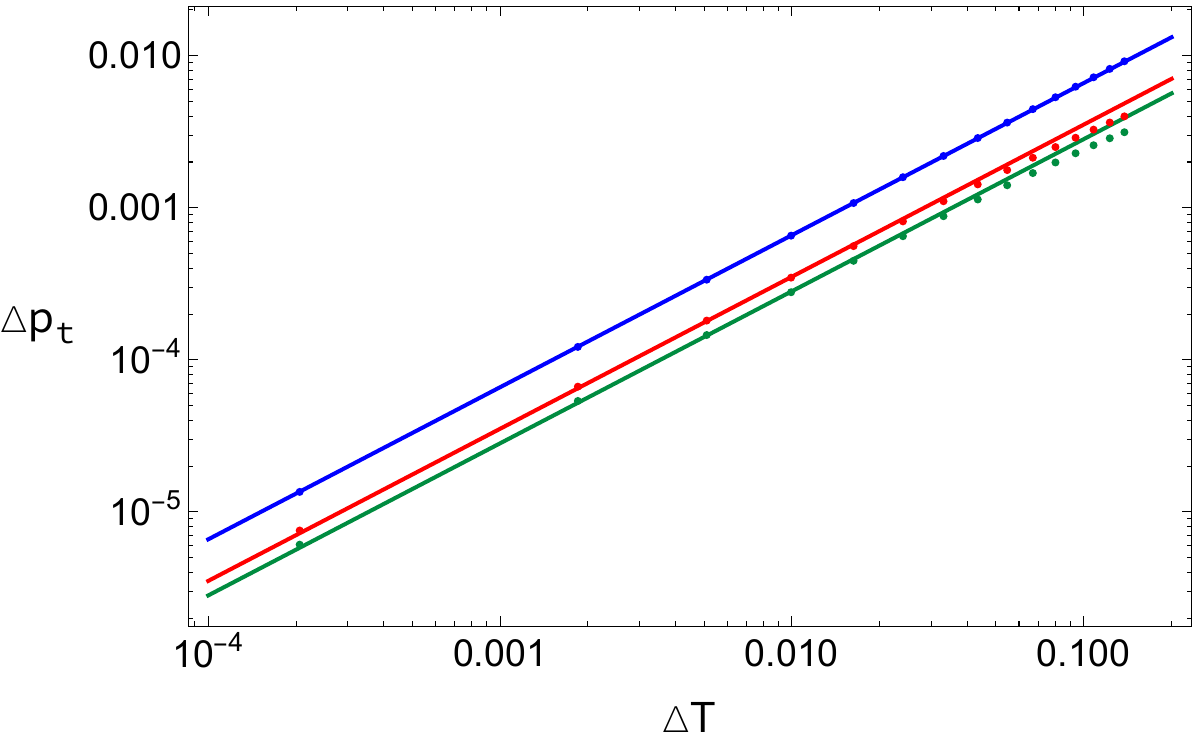}~~~~~
 \includegraphics[width=0.44\textwidth]{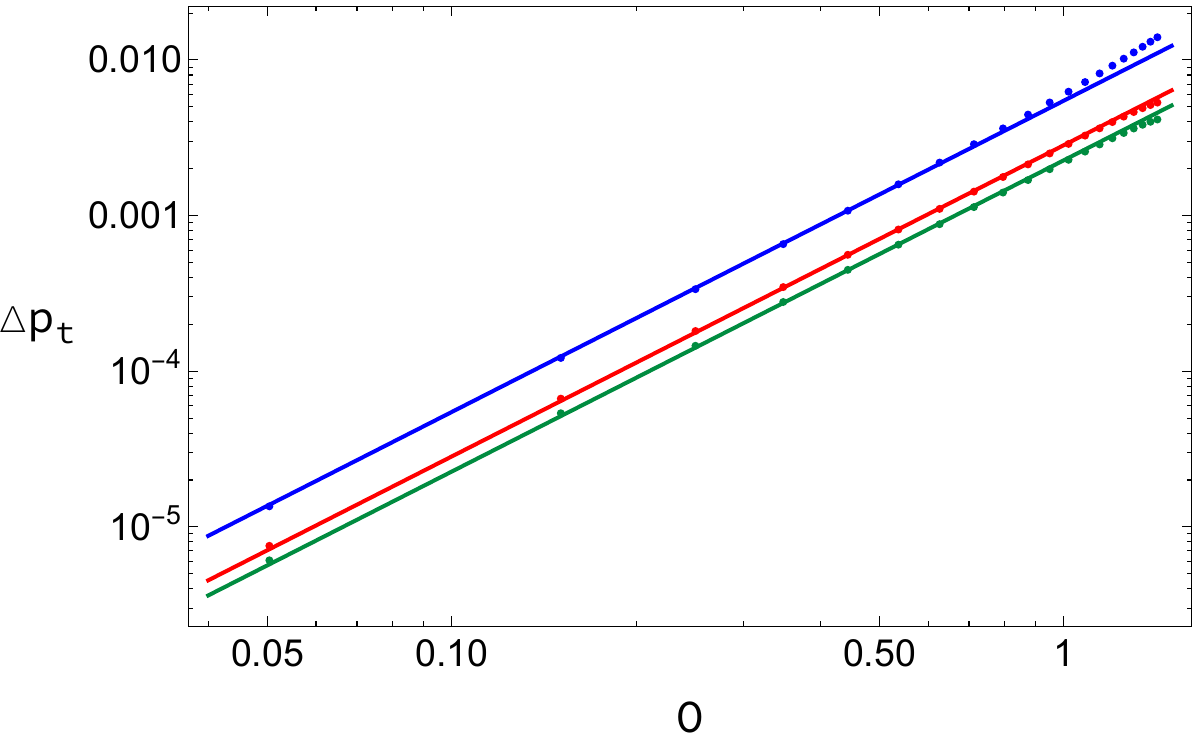}
 \caption{\small The first three continuously tracked Kasner epochs in the ordinary critical phase transition for $\lambda_4=\lambda=\lambda_8=1/10, \lambda_6=0$.  {\bf Left}: \(\Delta p_t^{(n)}\) versus
 \(\Delta T\);   {\bf Right}: \(\Delta p_t^{(n)}\) versus \(O\).  In both panels different colors denote different epochs: $n=1$ ({\em blue}), $2$ ({\em red}), $3$ ({\em green}).
 All epochs have the predicted critical powers. 
 }
 \label{fig:num-subsequent-epochs}
\end{figure}

Next we study the Kasner exponent across the second-order phase transition above. The Kasner exponent as a function of $T$ can be found in Fig. \ref{fig:num-second-order-kasner}. 
We find the first Kasner exponent numerically satisfies
\be 
\dpt^{(1)} \simeq 0.065875 \,\Delta T^{\,1.0000}\,,~~~\dpt^{(1)} \simeq 0.0053551\,O^{\,2.0006}\,,
\ee 
as in the left plot of Fig. \ref{fig:num-subsequent-epochs}. 
These relations match well with the analytical results Eqs. \eqref{eq:scalrel1} and \eqref{eq:ptO}. 

The second and third Kasner epochs close to the transition temperature obey
\begin{align}
 \Delta p_t^{(2)}
 &\simeq0.035305\,\Delta T^{0.99546}\,,~~~~~~~
 \Delta p_t^{(2)}
 \simeq0.0029032\,O^{1.9914}\,,
 \label{eq:num-second-epoch-O}\\
  \Delta p_t^{(3)}
 &\simeq 0.028345\,\Delta T^{0.99483}\,,~~~~~~~
 \Delta p_t^{(3)}
 \simeq 0.0023345\,O^{1.9902}\,.
 \label{eq:num-3-epoch-O}
\end{align}
The numerical results are shown in Fig. \ref{fig:num-subsequent-epochs}. 
The fitted powers agree with the fixed-epoch prediction
\eqref{eq:fixed-later-epoch-scaling}: a regular finite sequence of interior
bounces preserves the powers \(\Delta T^1\) and \(O^2\).
The amplitudes could not be fixed by the first epoch.  
Note that this is a test of fixed-epoch matching at small value of $n$, not of the large-\(n\) analytic scaling law.

The preceding fits concern epochs that can be tracked continuously in a neighborhood of \(T_c\). Away from the critical region, the
super-exponential force can truncate the first kinetic-dominated interval before a sufficiently long plateau forms, as seen in the blue curve in the left plot of Fig. \ref{fig:num-second-order-kasner}. The earliest plateau at low temperature then belongs to a later epoch at high  temperature.  The resulting jump at $T_{\rm loss}$ in the extracted curve is 
not a discontinuity of the bulk solution or a new thermodynamic phase transition, as discussed in Sec. \ref{ss:first-epoch-switch}.

We adopt the criterion that a Kasner epoch can not be clearly extracted if its radial interval satisfies $\Delta \rho<\rho_p$. 
Numerically we find that we can take
\begin{equation}
 \rho_p\simeq\log 45\,.
 \label{eq:num-switch-rho-window}
\end{equation} 
For a leading estimate we take the order-one matching value
\(f_K\simeq1\) in \eqref{eq:first-switch-kasner-approx}. 
With the 
plateau choice, Eq.~\eqref{eq:first-switch-G} and \eqref{eq:first-switch-phistar} then give
$
 \phi_h^{\rm loss}\simeq 0.534.
$
At this horizon value \(\phi_h\), the corresponding temperature is 
\begin{equation}
 T_{\rm loss}\simeq0.395\,.
 \label{eq:num-switch-temperature-window}
\end{equation}
This matches well with the numerical value
\(T_{\rm loss}^{\rm num}\simeq0.394\) in Fig. \ref{fig:num-second-order-kasner}.

\subsubsection{The scaling at extremely low temperature}
\label{ss:num-low-temperature}

Now we focus on the system at extremely low temperature.  
For the model with 
the super-exponential interaction with $\lambda_4=\lambda=\lambda_8=1/10, \lambda_6=0$, the result from Sec. \ref{ss:lowT} is
\bea
  p_t^{(1)} - p_t^{\text{Sch}} &\simeq25.622\,T^{5.5922}\,.
 \label{eq:num-lowT-superexp}
\eea
For the polynomial model obtained by setting $\lambda=0$,  
$\lambda_4=1/10$ and $\lambda_6=0$, it takes 
\bea
  p_t^{(1)} - p_t^{\text{Sch}} 
 &\simeq2.8601\,T^{1.5049}\,.
 \label{eq:num-lowT-polynomial}
\eea

\begin{figure}[h!]
 \centering
 \includegraphics[width=0.445\textwidth]{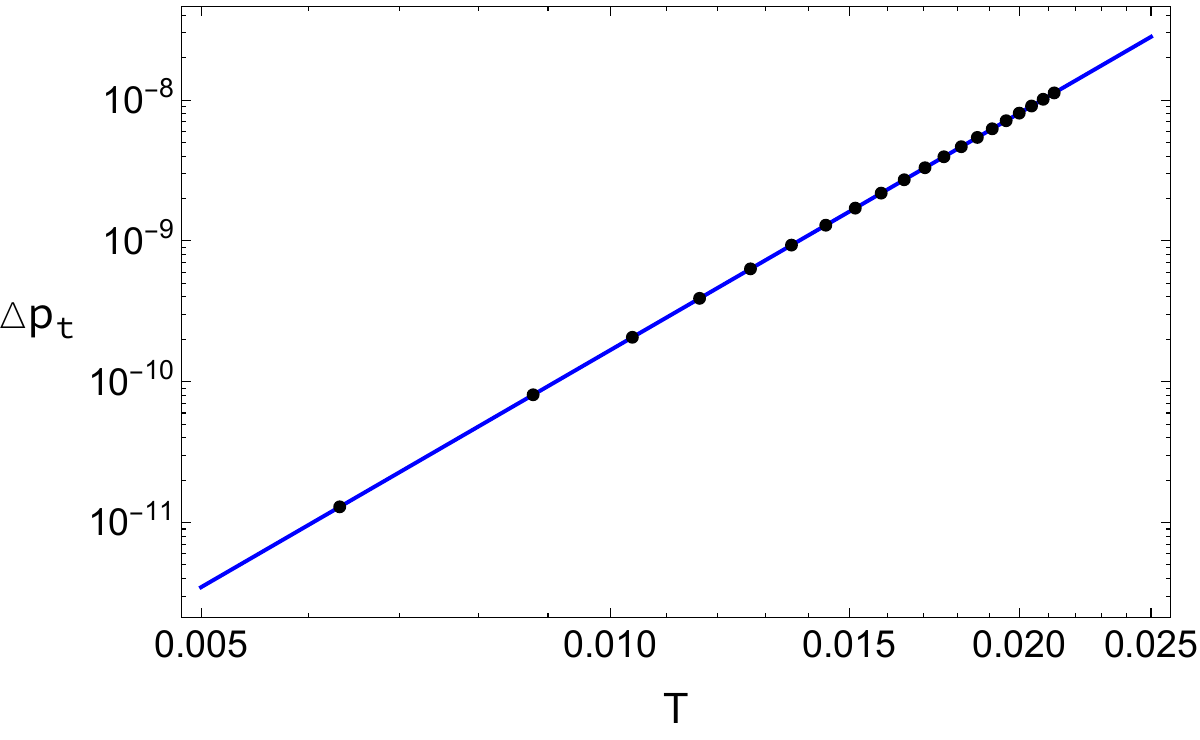}~~~~~
 \includegraphics[width=0.468\textwidth]{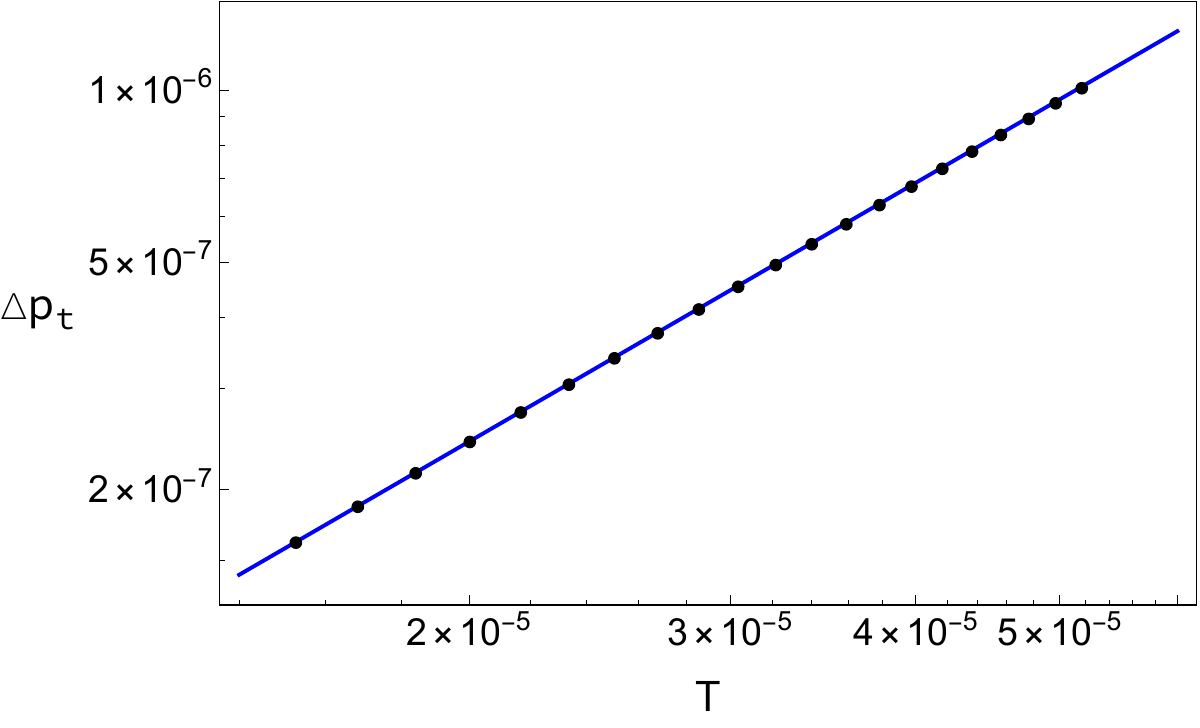}
 \caption{\small
 Log-log plot of \(\Delta p_t=p_t^{(1)}-\ptsch\) at extremely low-temperature of the super-exponential model with $\lambda_4=\lambda=\lambda_8=1/10, \lambda_6=0$ ({\bf left}), and
 polynomial model with $\lambda=0$, $\lambda_4=1/10$ and $\lambda_6=0$ ({\bf right}). The black dots are numerical data, while the blue lines are the analytical results.  In both models, the numerical fitted amplitude and exponent agree with the analytic values to the first three significant digits. 
 }
 \label{fig:num-low-temperature-comparison}
\end{figure}

The larger exponent in Eq.~\eqref{eq:num-lowT-superexp} indicates that the
super-exponential model approaches the Schwarzschild value \(p_t^{\text{Sch}} =-1/3\) much more rapidly. The  Fig.~\ref{fig:num-low-temperature-comparison} shows the log-log plot of $p_t^{(1)}-\ptsch$ as functions of $T$.  In both models, the data approach straight lines as the temperature is lowered, indicating the scaling
\begin{align}
 p_t^{(1)}-\ptsch
 &\simeq
 25.641\,T^{5.5922}\,,  {~~~~~~~\rm super\text{-}exponential~ model}\,;
 \label{eq:num-lowT-fit1}
 \\
 p_t^{(1)}-\ptsch
 &\simeq
2.8669\,T^{1.5051}\,,  {~~~~~~~\rm polynomial~ model}\,.
 \label{eq:num-lowT-fit2}
\end{align}
The black dots are the numerical data, while the blue lines represent the analytical results \eqref{eq:num-lowT-superexp} and \eqref{eq:num-lowT-polynomial}. It can be seen that the numerical fitting matches well with analytical result. 

The analytical prediction is that the scaling exponent is label-independent for the interior Kasner epochs; equivalently, later epochs share the same scaling form with different amplitudes. Due to numerical challenges, however, we are able to validate this only for the first Kasner exponent at low temperature.

\subsection{Phase transition at the tricritical point}
\label{ss:2nd2}

\begin{figure}[h!]
\begin{center}
\includegraphics[width=0.455\textwidth]{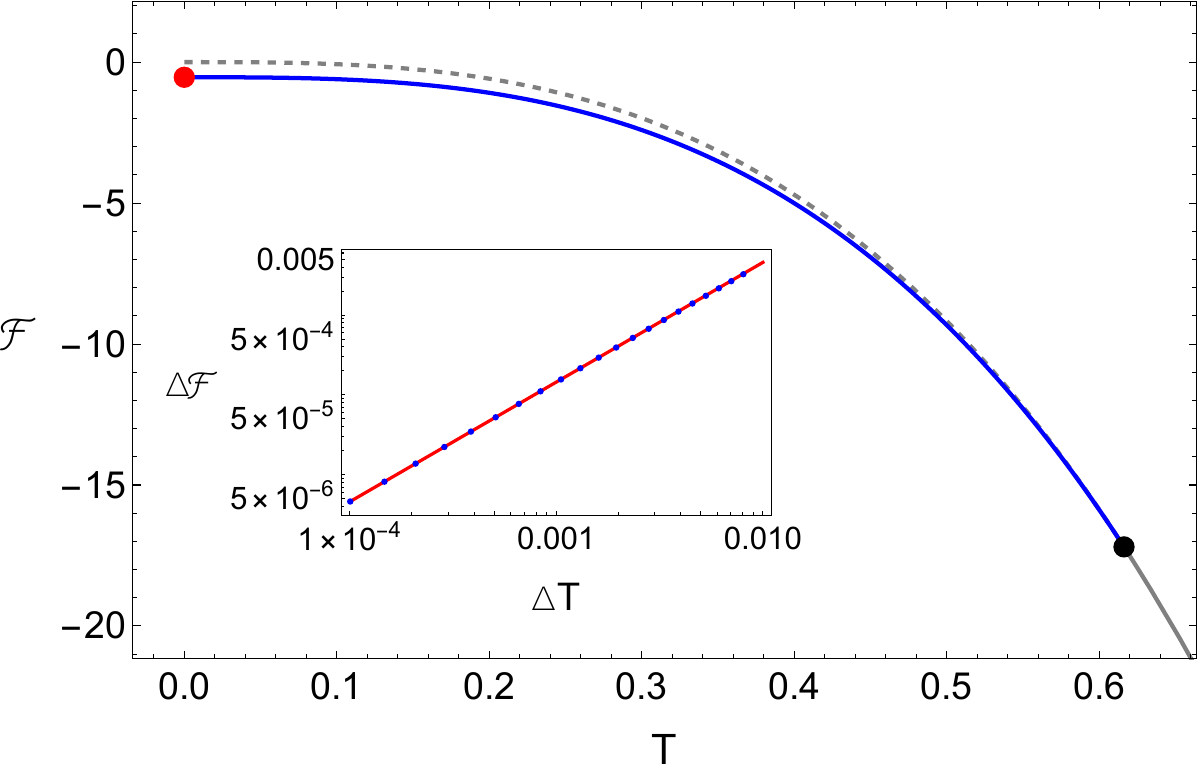}\,~~~~
\includegraphics[width=0.45\textwidth]{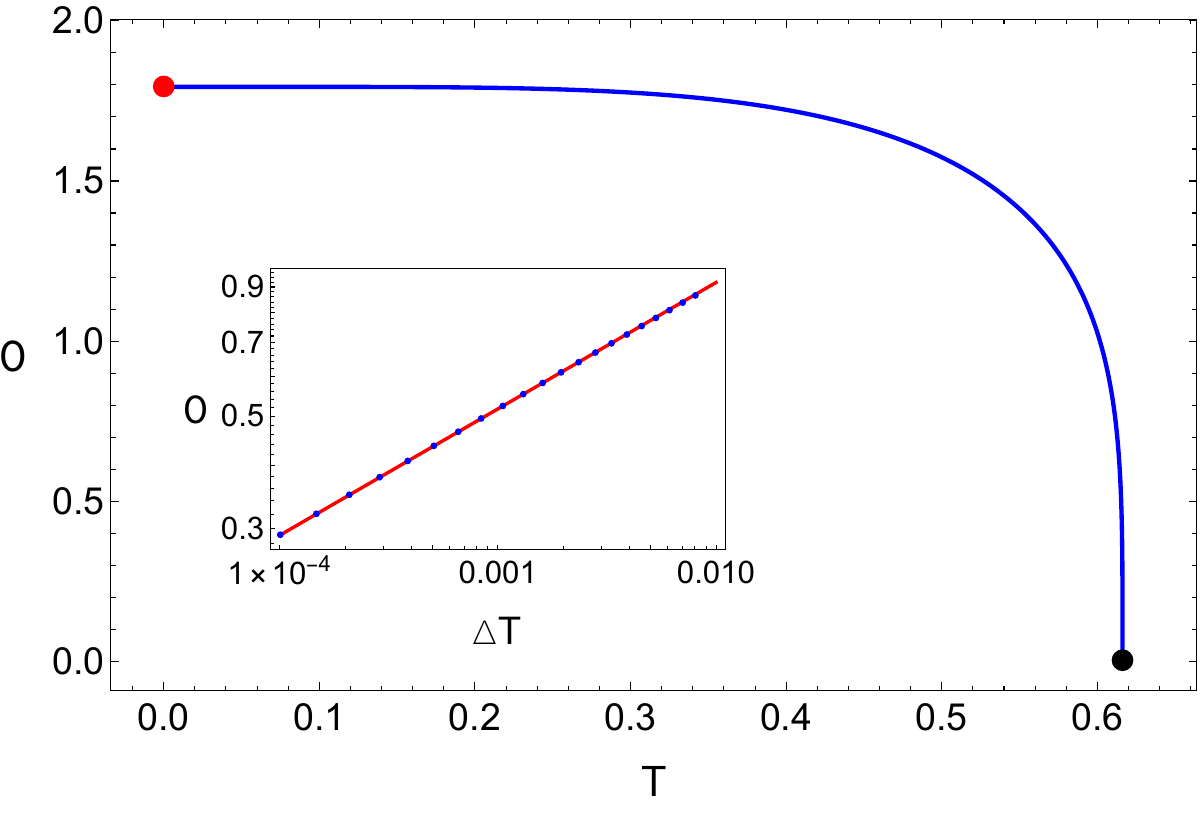}
\end{center}
\vspace{-0.7cm}
\caption{\small Free energy ({\bf left}) of the hairy ({\em blue}) and Schwarzschild ({\em dashed black}) solutions, and condensate ({\bf right}), as functions of temperature at tricritical point with $\lambda_4=\lambda_4^t\,,\lambda_6=1\,.$ The insets are log-log plots. The black and red dots are at transition temperature and zero temperature, respectively.   We work in unit $\kappa=-1$.}
\label{fig:pt-tri}
\end{figure}

We next numerically study the phase transition at the tricritical point by tuning 
\begin{equation}
 \lambda_4=\lambda_4^t=-0.068490\,,
 \qquad
 \lambda_6=1\,.
 \label{eq:num-tricritical-parameters}
\end{equation}
The free energy and condensate as functions of $T/(-\kappa)$ can be found in Fig. \ref{fig:pt-tri}. The critical temperature
remains \(T_c\simeq0.61618\).  
Close to the transition temperature,  the numerical 
fits are
\begin{align}
 O
 \simeq2.9028\,\Delta T^{0.24985}\,,
~~~~~~
 \Delta\mathcal F
 \simeq-4.5877\,\Delta T^{1.5002}\,.
 \label{eq:num-tricritical-free-energy}
\end{align}
They agree well with the analytic laws
\(2.9082\,\Delta T^{1/4}\) and
\(-4.5752\,\Delta T^{3/2}\) in \eqref{eq:tricritical-temperature-scaling}. 

\begin{figure}[h!]
 \centering
 \includegraphics[width=0.44\textwidth]{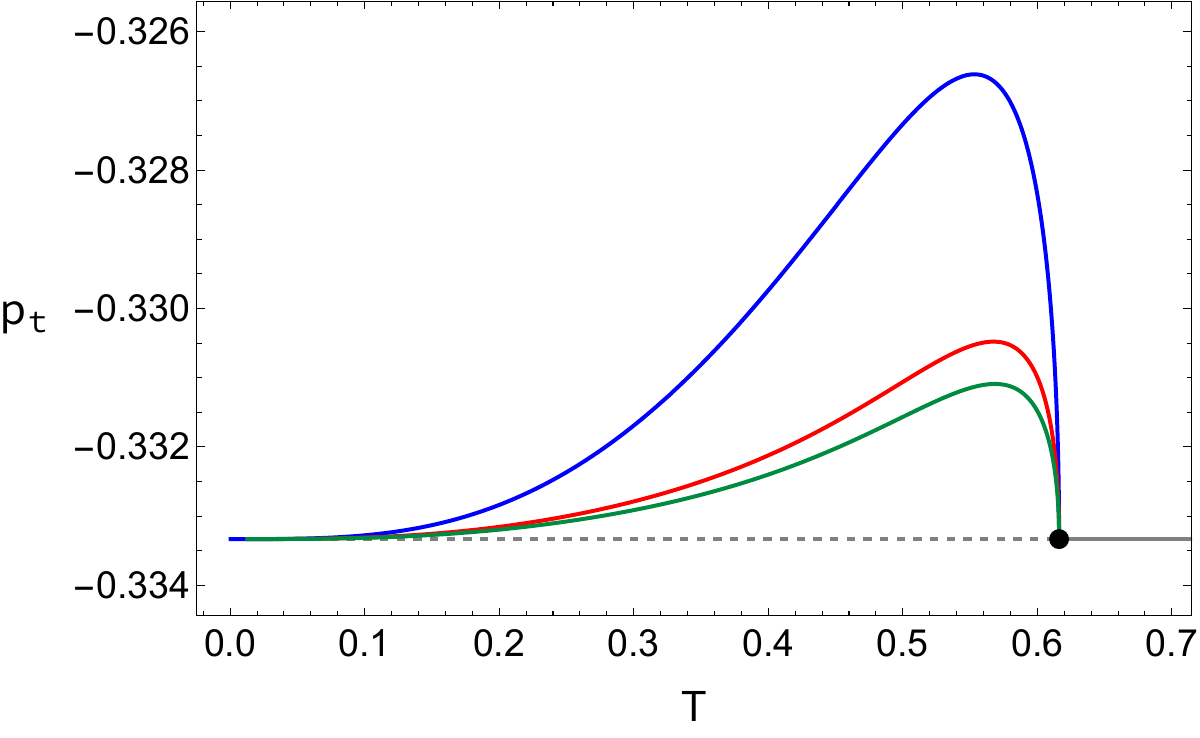}~~~~
 \includegraphics[width=0.445\textwidth]{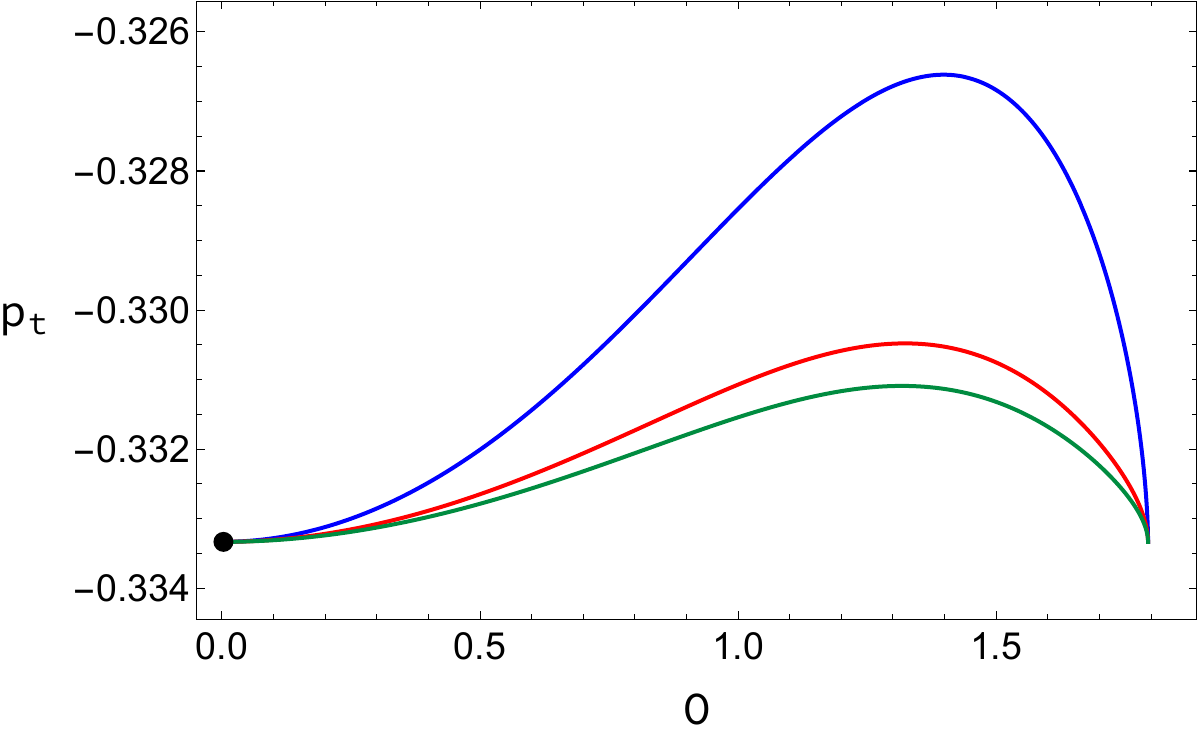}
 \caption{\small First three Kasner epochs at the tricritical point.  {\bf Left}:
 \( p_t^{(n)}\) versus \(T\).
 {\bf Right}: \( p_t^{(n)}\) versus \(O\). In both panels different colors denote different epochs: $n=1$ ({\em blue}), $2$ ({\em red}), $3$ ({\em green}).  We work in unit $\kappa=-1$.}
 \label{fig:num-tricritical-kasner}
\end{figure}

\begin{figure}[h!]
 \centering
 \includegraphics[width=0.44\textwidth]{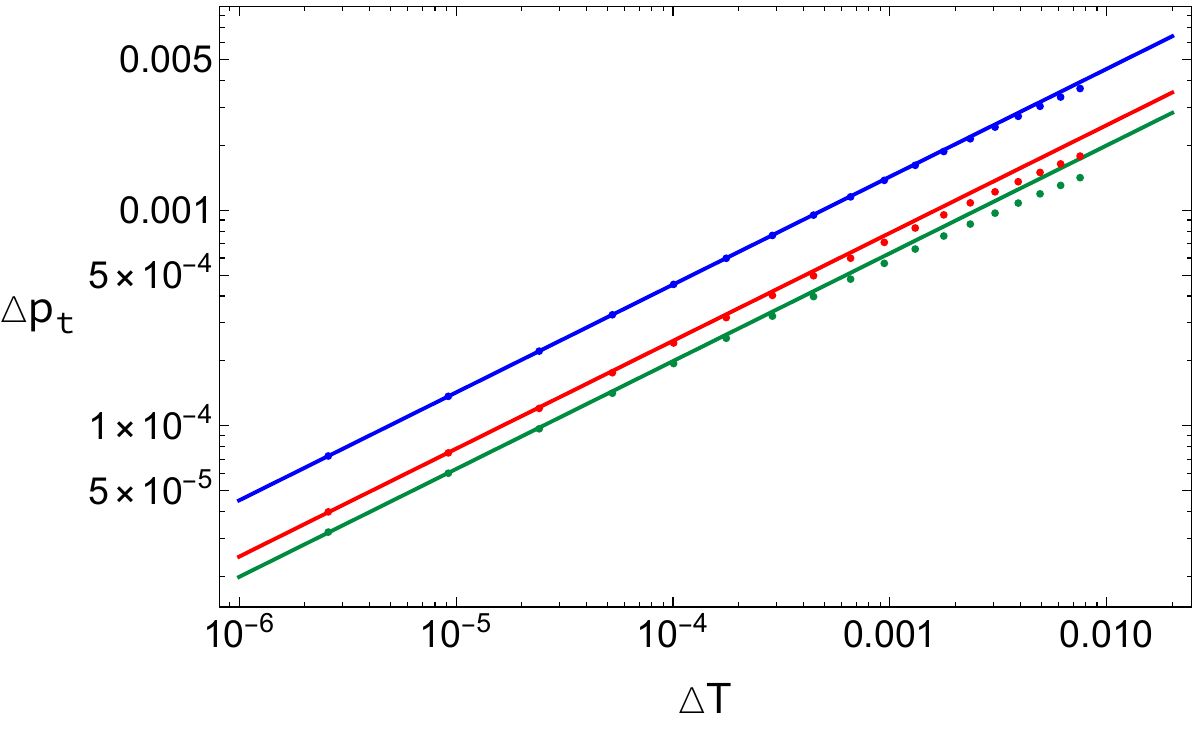}~~~~
 \includegraphics[width=0.445\textwidth]{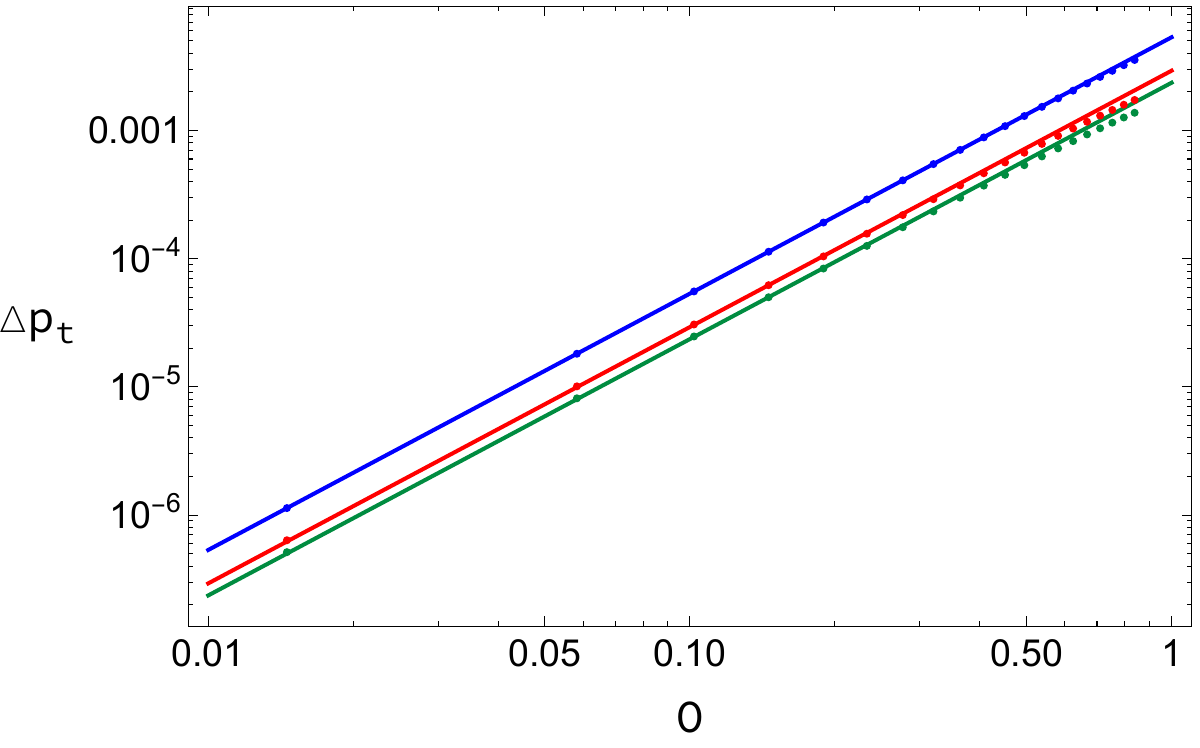}
 \caption{\small The zoom-in plot of the first three continuously tracked Kasner epochs at the tricritical point.  {\bf Left}: \(\Delta p_t^{(n)}\) versus
 \(\Delta T\);   {\bf Right}: \(\Delta p_t^{(n)}\) versus \(O\).  Different colors denote different epochs: $n=1$ ({\em blue}), $2$ ({\em red}), $3$ ({\em green}).}
 \label{fig:Kasner-scaling-tricri-pt}
\end{figure}

For the first three Kasner epochs near the tricritical point, as shown in Fig. \ref{fig:num-tricritical-kasner} and  Fig. \ref{fig:Kasner-scaling-tricri-pt}, we find
\begin{align}
 \Delta p_t^{(1)}
 \simeq0.045024\,\Delta T^{0.49970}\,,
~~~~~~
 \Delta p_t^{(1)}
 \simeq0.0053430\,O^{1.9999}\,,\\
  \Delta p_t^{(2)}
 \simeq 0.023829\,\Delta T^{0.49647}\,,
~~~~~~
 \Delta p_t^{(2)}
 \simeq 0.0028671\,O^{1.9870}\,,\\
   \Delta p_t^{(3)}
 \simeq 0.019099\,\Delta T^{0.49603}\,,
~~~~~~
 \Delta p_t^{(3)}
 \simeq 0.0023024\,O^{1.9852}\,.
 \label{eq:num-tricritical-pt-O}
\end{align}
For the first Kasner epoch, the numerical results are again in good agreement with the analytical predictions at tricriticality given in \eqref{eq:scalrel2} and \eqref{eq:ptO}. For the second and third Kasner epochs, our analytical analysis only predicts the scaling exponents in \eqref{eq:fixed-later-epoch-scaling}, which agree well with the numerical results.

The comparison between the two continuous transitions is summarized in
Table~\ref{tab:num-analytic-comparison}.  The important structural result is 
that the relation \(\Delta p_t^{(1)}\propto O^2\) and its
coefficient are common to both critical points because they share the same linear threshold mode.  What changes at tricriticality is the relation between \(O\) and \(\Delta T\).  Consequently, the temperature power
of \(\Delta p_t^{(1)}\) changes from \(1\) to \(1/2\).  The ordinary
transition therefore produces a finite slope discontinuity in \(p_t(T)\), whereas the tricritical scaling implies a divergent one-sided derivative in the strict asymptotic limit. In both transitions, later Kasner epochs inherit the same temperature exponent as the first epoch, with epoch-dependent coefficients. 

\begin{table}[h!]
 \centering
 \small
 \setlength{\tabcolsep}{4pt}
 \begin{tabular}{c|cc|cc}
 \hline
 &\multicolumn{2}{c|}{ordinary critical point}
 &\multicolumn{2}{c}{tricritical point}\\
 quantity & analytic & numerical & analytic & numerical\\
 \hline
 \(O\)
 &\(3.5103\,\Delta T^{1/2}\)
 &\(3.5071\,\Delta T^{0.49991}\)
 &\(2.9082\,\Delta T^{1/4}\)
 &\(2.9028\,\Delta T^{0.24985}\)\\
 \(\Delta\mathcal F\)
 &\(-4.9993\,\Delta T^2\)
 &\(-4.9821\,\Delta T^{1.9996}\)
 &\(-4.5752\,\Delta T^{3/2}\)
 &\(-4.5877\,\Delta T^{1.5002}\)\\
 \(\Delta p_t^{(1)}(\Delta T)\)
 &\(0.065849\,\Delta T\)
 &\(0.065875\,\Delta T^{1.0000}\)
 &\(0.045197\,\Delta T^{1/2}\)
 &\(0.045024\,\Delta T^{0.49970}\)\\
 \(\Delta p_t^{(1)}(O)\)
 &\(0.0053440\, O^2\)
 &\(0.0053551\, O^{2.0006}\)
 &\(0.0053440\, O^2\)
 &\(0.0053430\, O^{1.9999}\)\\
 \hline
 \end{tabular}
 \caption{\small Direct comparison of the analytical and numerical near-critical relations.}
 \label{tab:num-analytic-comparison}
\end{table}

The same numerical strategy can test the fourth-order multicritical point, where the analytic prediction is
\(\Delta p_t^{(1)}\propto\Delta T^{1/3}\).  We do not perform an independent numerical scan around this tuned point, and therefore do not quote a numerical fit for this case.

\subsubsection{The scaling at extremely low temperature}

We now study the extremely low-temperature behavior at the tricritical point.
Here the polynomial couplings are fixed by \(\lambda_4=\lambda_4^t,\,\lambda_6=1\).
For the model with the super-exponential interaction,
\(\lambda=\lambda_8=1/10\), following the low-temperature analysis in
Sec.~\ref{ss:lowT}, we obtain 
\begin{equation}
 p_t^{(1)}-\ptsch
 \simeq 0.13085\,T^{3.3506}\,.
 \label{eq:num-lowT-tri-superexp}
\end{equation}
For comparison, we also consider the polynomial model obtained by setting
\(\lambda=0\), while keeping the same tricritical values
\(\lambda_4=\lambda_4^t\) and \(\lambda_6=1\).  In this case one finds
\begin{equation}
 p_t^{(1)}-\ptsch
 \simeq 0.13153\,T^{3.3397}\,.
 \label{eq:num-lowT-tri-polynomial}
\end{equation}

\begin{figure}[h!]
 \centering
 \includegraphics[width=0.44\textwidth]{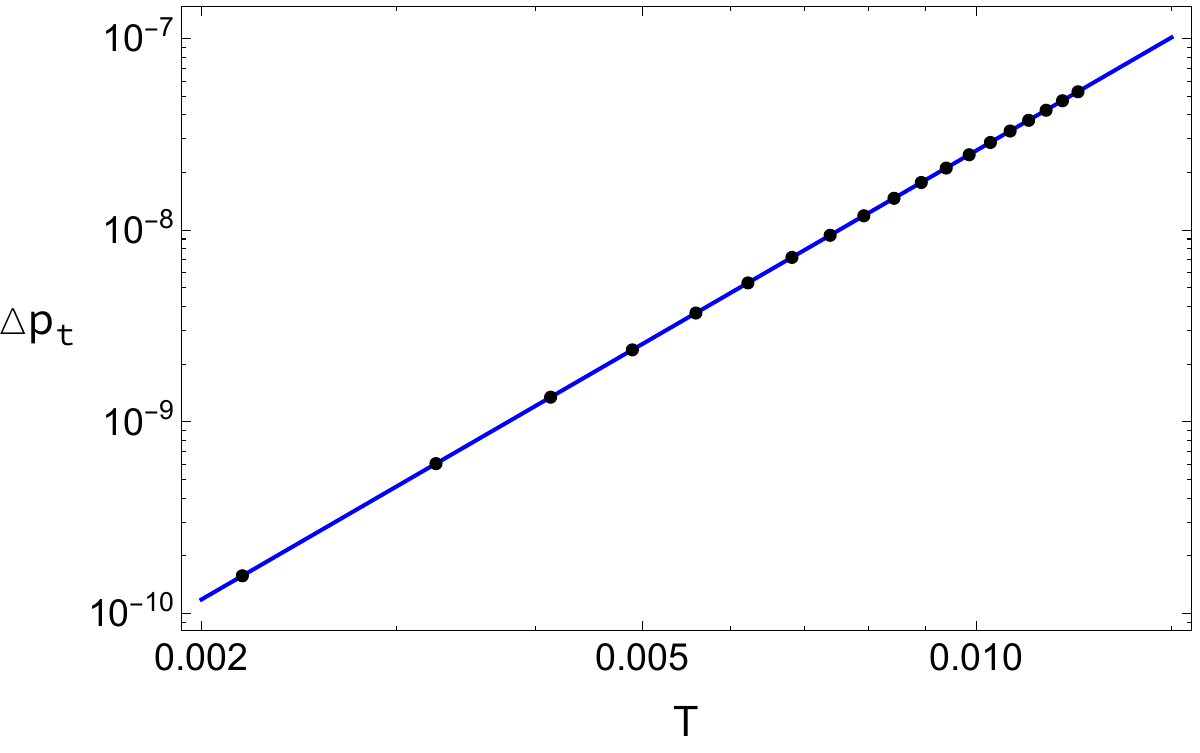}~~~~
 \includegraphics[width=0.445\textwidth]{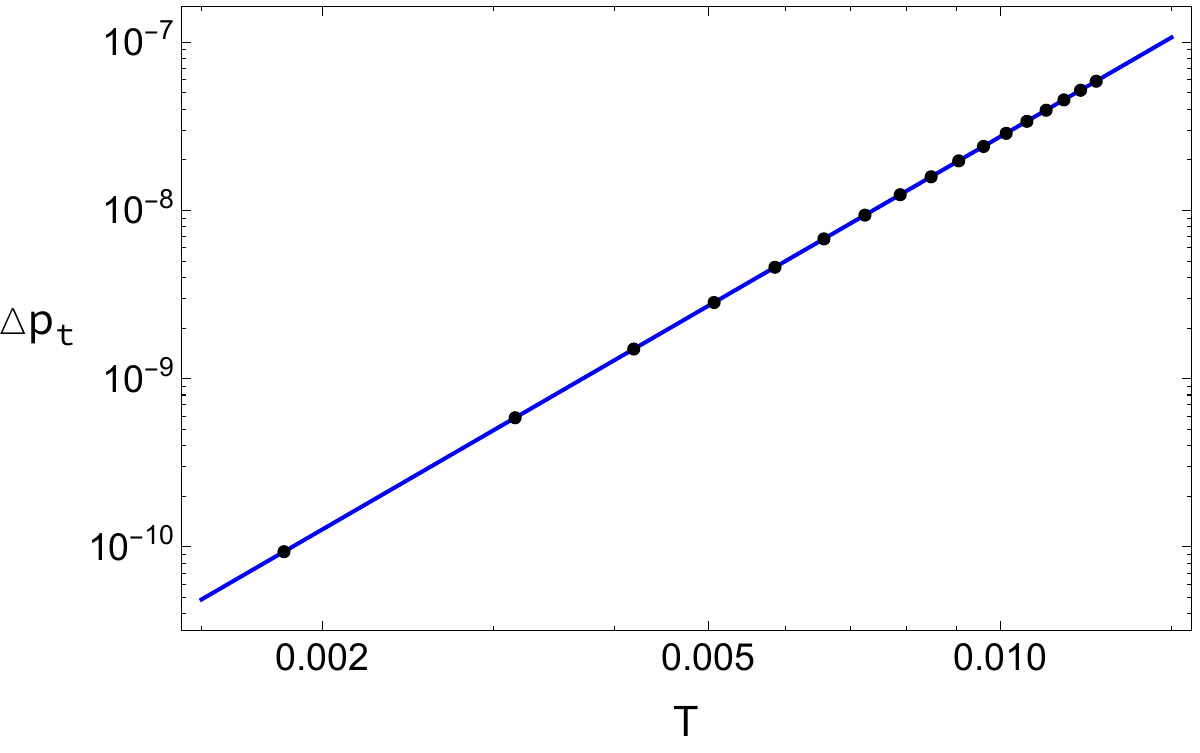}
 \caption{\small Log-log plot of \(\Delta p_t=p_t^{(1)} -\ptsch\) at extremely low-temperature of the super-exponential with $\lambda=\lambda_8=1/10, \lambda_4=\lambda_4^t$ and $\lambda_6=1$ ({\bf left}), and
 polynomial models with $\lambda=0$, $\lambda_4=\lambda_4^t$ and $\lambda_6=1$ ({\bf right}) at the tricritical point. The black dots are numerical data, while the blue lines
are the analytical results.  In both models, the numerical fitted coefficient and exponent agree with the analytic values to
the first three significant digits.}
 \label{fig:tri-scaling-lowT}
\end{figure}

From Fig. \ref{fig:tri-scaling-lowT}, the corresponding numerical fits take the form
\begin{align}
 p_t^{(1)}-\ptsch
 &\simeq
 0.13081\,T^{3.3505}\,,  {~~~~~~~\rm super\text{-}exponential~ model}\,;\\
 p_t^{(1)}-\ptsch
 &\simeq
 0.13148\,T^{3.3397}\,,  {~~~~~~~\rm polynomial~ model}\,.
 \label{eq:num-lowT-tri-poly-fit}
\end{align}
The fitted amplitudes and exponents agree with the analytical predictions in Eqs.~\eqref{eq:num-lowT-tri-superexp} and \eqref{eq:num-lowT-tri-polynomial} to the first three significant digits, providing a direct numerical check that the
low-temperature Kasner scaling at the tricritical point is again controlled by the irrelevant exponent of the zero-temperature IR fixed point.
As in the ordinary transition discussed in the previous subsection, extracting the corresponding exponents for later Kasner epochs remains challenging.

The two exponents in Eqs.~\eqref{eq:num-lowT-tri-superexp} and
\eqref{eq:num-lowT-tri-polynomial} are close to each other.  This is because,
for the tricritical parameters considered here, the super-exponential term
only slightly changes the location and curvature of the IR fixed point.  As a
result, both models approach the Schwarzschild value \(\ptsch=-1/3\) with nearly
the same power law.

\subsection{First-order phase transition }
\label{ss:1st}

\begin{figure}[h!]
 \centering
 \includegraphics[width=0.325\textwidth]{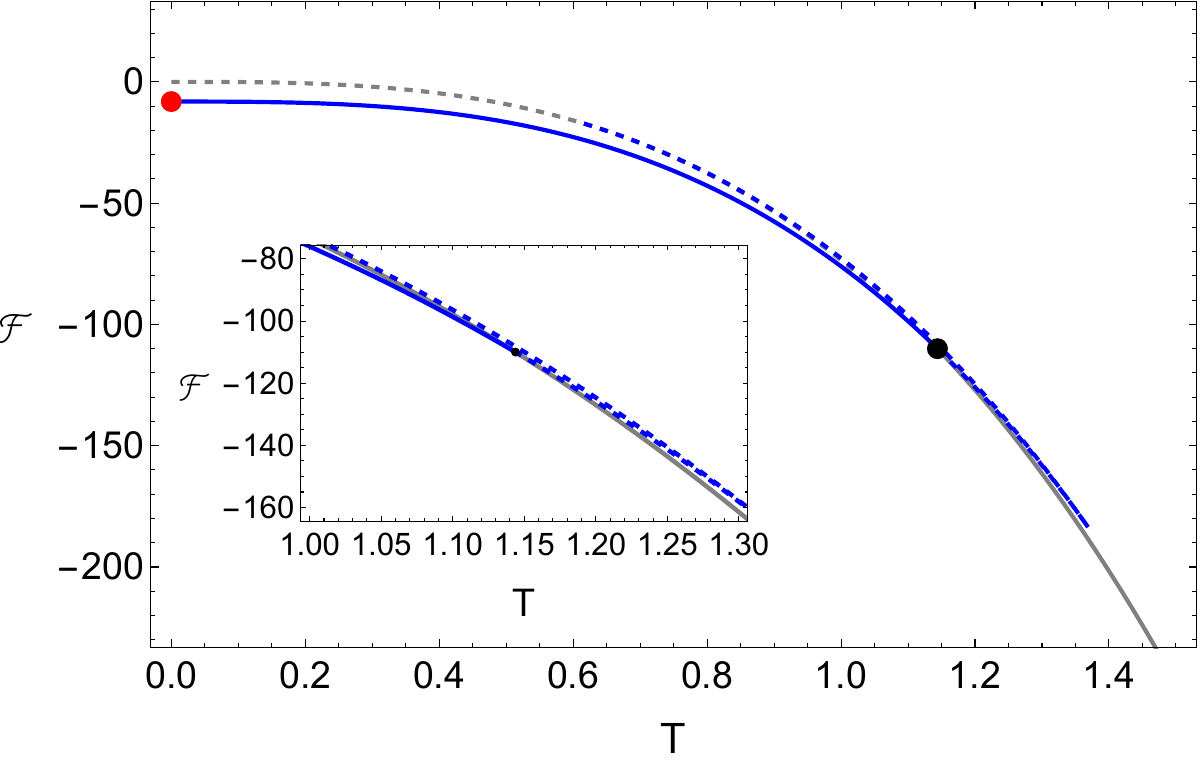}~
\includegraphics[width=0.3\textwidth]{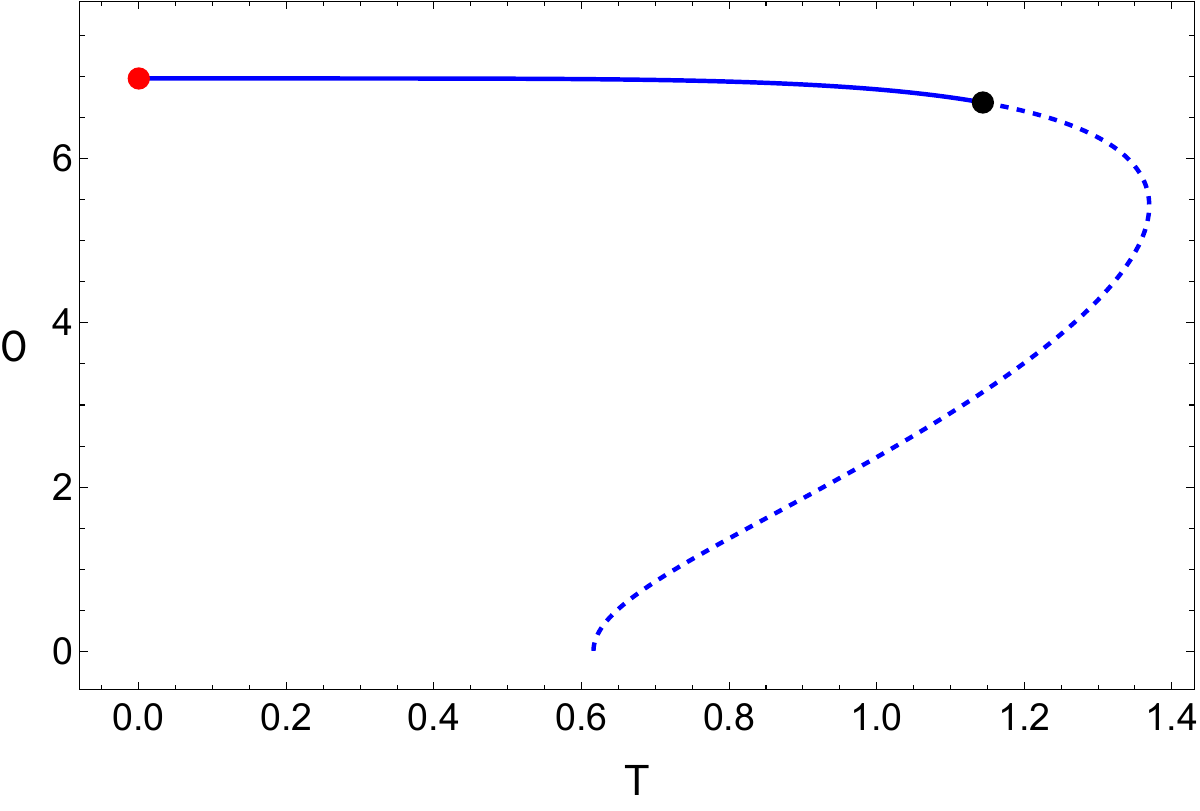}~
 \includegraphics[width=0.33\textwidth]{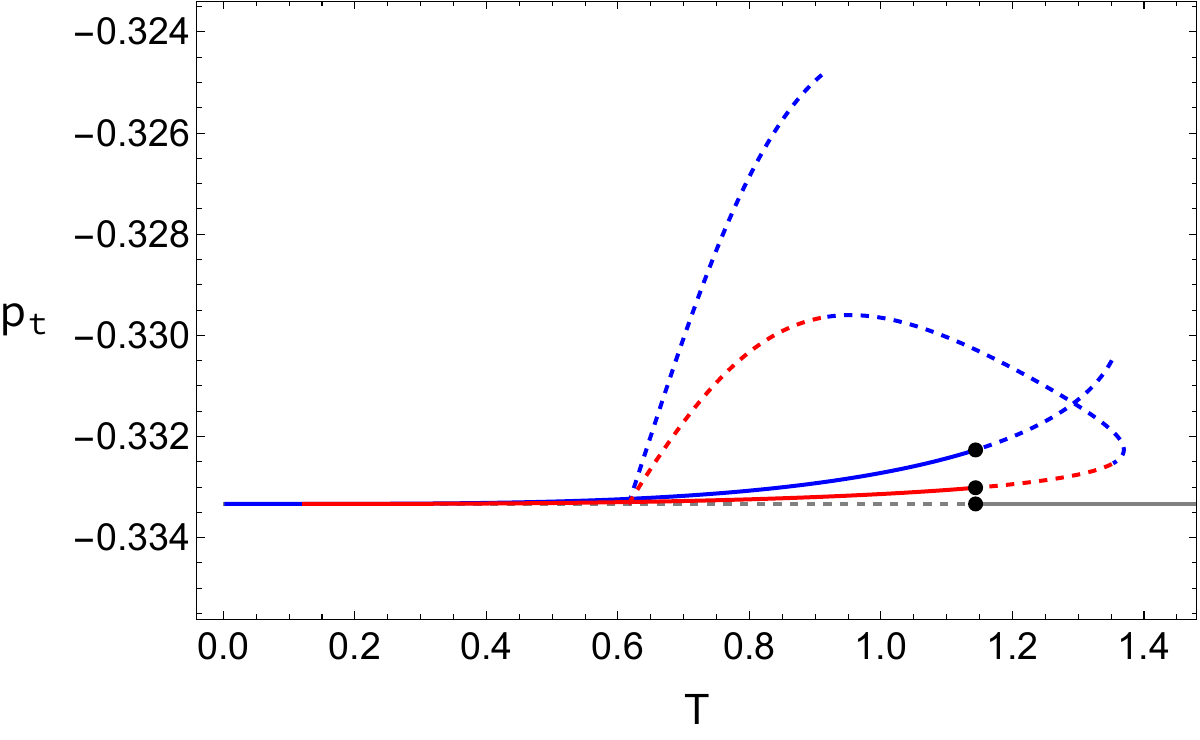}
 \caption{\small Phase diagram of a first-order transition for
$\lambda_4=-7/20$, $\lambda_6=1.$ The solid and dashed lines represent the stable and unstable black holes, respectively.
{\bf Left:} free energy of the Schwarzschild branch (\emph{gray}) and the hairy branch (\emph{blue}),
including metastable segments; their global crossing determines the
first-order transition temperature \(T_1\). 
{\bf Middle:} condensate \(O\) as a function of temperature, showing a finite jump at \(T_1\). 
{\bf Right:} Kasner
exponent \(p_t\) for the corresponding black-hole interiors.  The blue and red lines represent the first and second Kasner epochs, respectively. The solid and dashed lines are for stable and unstable solutions, respectively. 
The discontinuity
at \(T_1\) is caused by thermodynamic branch selection and is
distinct from finite-temperature epoch relabeling. 
We work in unit $\kappa=-1$.
}
 \label{fig:num-first-order-comparison}
\end{figure}

For \(\lambda_4<\lambda_4^t\) 
, the system exhibits a first-order phase transition, as discussed in detail in Appendix \ref{app:first-order}. 
The numerical scan follows the full nonlinear branches in the same fixed \(\kappa=-1\) ensemble, and locates a coexistence temperature \(T_1\) satisfying
\begin{equation}
 \mathcal F_{\rm hairy}(T_1)=\mathcal F_{\rm Sch}(T_1)\,,
 \qquad
 O_{\rm hairy}(T_1)\neq0\,,
 \qquad
 \Delta s(T_1)\neq0\,.
 \label{eq:num-first-order-criteria}
\end{equation}
A representative phase diagram displayed in
Fig.~\ref{fig:num-first-order-comparison} contains both stable
and metastable portions of every branch until the global free-energy crossing is identified.

At coexistence, the thermodynamically selected solution switches between two
distinct bulk geometries.  The condensate, horizon scalar, and any
continuously tracked local Kasner data therefore have finite jumps,
\begin{equation}
 \left[O\right]_{T_1}\neq0,
 \qquad
 \left[\Delta p_t^{(n)}\right]_{T_1}\neq0
 \quad\text{in general}.
 \label{eq:num-first-order-jumps}
\end{equation}

This discontinuity of $p_t$ has also been observed before in the literature \cite{Liu:2021hap, Zhao:2026mkx}. 
Unlike the critical and tricritical cases, the first-order transition itself occurs at a finite value of the condensate and does not involve a vanishing Kasner deviation at the transition temperature.

\subsubsection{The scaling at extremely low temperature}

We finally examine the extremely low-temperature behavior in the first-order
case.  The polynomial couplings are chosen as
\(\lambda_4=-7/20,\, \lambda_6=1\).
For the model with the super-exponential interaction,
\(\lambda=\lambda_8=1/10\), following the low-temperature analysis in Sec.~\ref{ss:lowT}, we obtain 
\begin{equation}
 p_t^{(1)}-\ptsch
 \simeq 6.5589\times10^{-4}\,T^{3.7730}\,.
 \label{eq:num-lowT-first-superexp}
\end{equation}
For the polynomial model obtained by setting \(\lambda=0\), while keeping
\(\lambda_4=-7/20\) and \(\lambda_6=1\), we obtain
\begin{equation}
 p_t^{(1)}-\ptsch
 \simeq 6.6263\times 10^{-4}\,T^{3.7582}\,.
 \label{eq:num-lowT-first-polynomial}
\end{equation}

\begin{figure}[h!]
 \centering
 \includegraphics[width=0.44\textwidth]{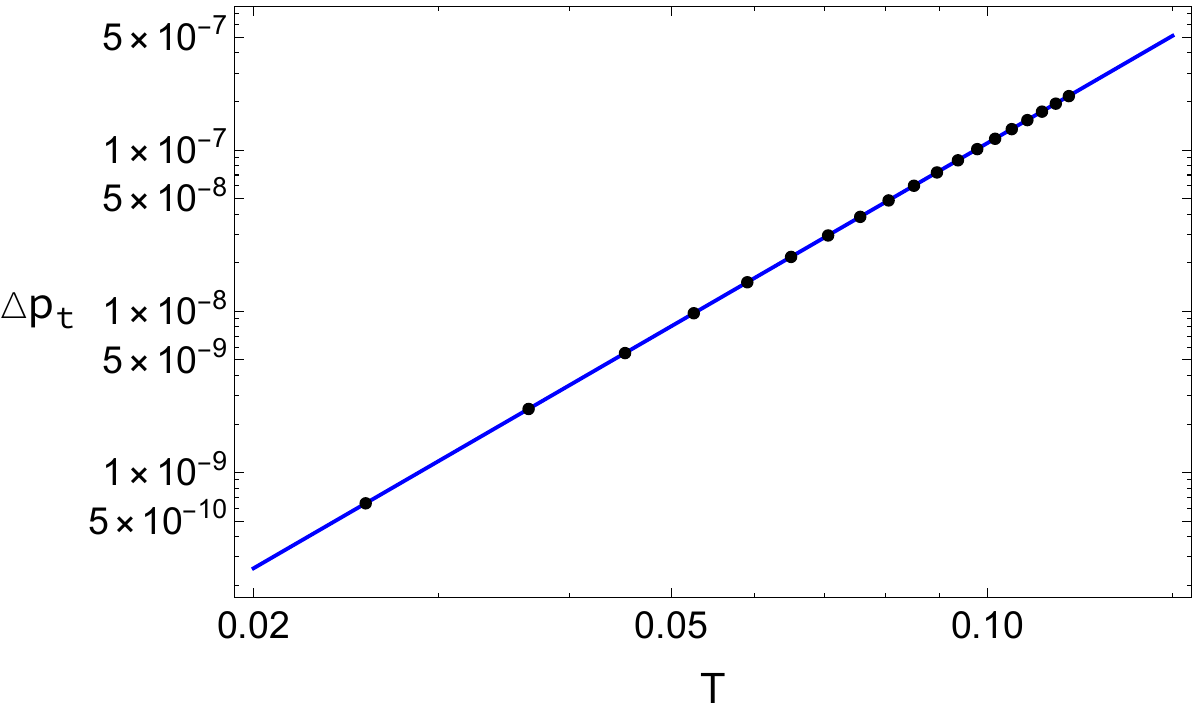}~~~~
 \includegraphics[width=0.445\textwidth]{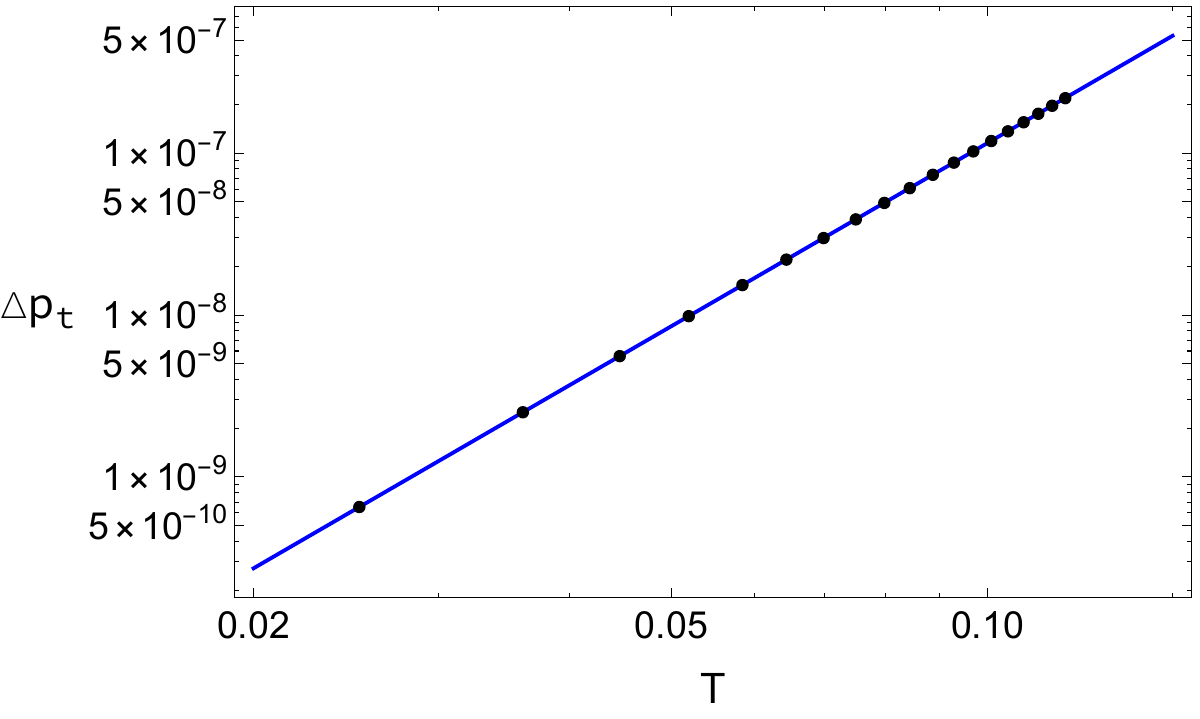}
 \caption{\small Log-log plot of \(\Delta p_t=p_t^{(1)} -\ptsch\) at extremely low-temperature of the super-exponential with $\lambda=\lambda_8=1/10, \lambda_4=-7/20$ and $\lambda_6=1$ ({\bf left}), and
 polynomial models with $\lambda=0$, $\lambda_4=-7/20$ and $\lambda_6=1$ ({\bf right}) for first-order phase transition. The black dots are numerical data, while the blue lines
are the analytical results.  In both models, the numerical fitted coefficient and exponent agree with the analytic values to
the first three significant digits.}
 \label{fig:1st-scaling-lowT}
\end{figure}

From Fig. \ref{fig:1st-scaling-lowT}, the corresponding numerical fits are
\begin{align}
 p_t^{(1)}-\ptsch
 &\simeq
 6.5561\times 10^{-4}\,T^{3.7730},
  {~~~~~~~\rm super\text{-}exponential~ model}; 
 \label{eq:num-lowT-first-superexp-fit}\\
 p_t^{(1)}-\ptsch
 &\simeq
 6.6237\times 10^{-4}\,T^{3.7581}, {~~~~~~~\rm polynomial~ model}.\label{eq:num-lowT-first-poly-fit}
\end{align}
The fitted coefficients and exponents agree well with the analytic predictions in
Eqs.~\eqref{eq:num-lowT-first-superexp} and
\eqref{eq:num-lowT-first-polynomial}. 
The close agreement between the super-exponential and
polynomial exponents is due to the fact that, for the parameters used here, the
two zero-temperature IR fixed points have similar irrelevant dimensions. 

\section{Conclusion and discussion}
\label{sec:cd}

We have investigated how boundary criticality is encoded in the spacelike
interiors of planar black holes in four-dimensional Einstein-scalar gravity. 
The model includes polynomial couplings that control the order of the boundary phase transition and a super-exponential interaction that generates bounces in the interior. The full scalar potential also determines the low-temperature infrared geometry. Our main result is that the critical and multicritical scaling of the boundary order parameter is encoded in local 
Kasner epochs behind the horizon. This result suggests that other holographic phase transitions may likewise leave characteristic imprints on the geometry behind the horizon.

Near a continuous transition, we show analytically that the Kasner exponent in the first local epoch satisfies
\begin{equation}
 p_t^{(1)}-\ptsch\propto O^2,
\end{equation} provided that this epoch can
be tracked continuously to the critical solution.  The condensate scaling
therefore fixes the interior temperature scaling: the Kasner deviation ($p_t^{(1)}-\ptsch$) is
linear in \(T_c-T\) at the ordinary critical point of second-order phase transition, proportional to its square
root (\(\sqrt{T_c-T}\)) at the tricritical point, and proportional to its cubic root (\((T_c-T)^{1/3}\)) at a fourth-order multicritical point. The ordinary and
tricritical predictions, including their amplitudes, agree well with our numerical
results. A fixed later Kasner epoch inherits the same critical power, while its amplitude is epoch dependent. 
Additionally, the holographic realization of tricriticality and higher multicritical scaling is interesting in its own right. It would be interesting to explore whether this holographic realization can be connected to multicritical phenomena in experimentally accessible physical systems \cite{Riedel:1972tricritical,Wegner:1975higher}.

We have also clarified an apparent discontinuity in the extracted first Kasner exponent found in an ordinary critical phase transition model, shown in Fig. \ref{fig:num-second-order-kasner}. This behavior occurs when the super-exponential force terminates the first kinetic plateau before it can be identified by the chosen extraction criterion. The next plateau is then labeled as the first epoch. The resulting jump is not a discontinuity of the underlying black-hole solution. 

At zero temperature the hairy black hole approaches an AdS$_4$-to-AdS$_4$ domain wall. In the extremely low-temperature limit, the irrelevant scaling dimension of the IR fixed point controls  how rapidly the horizon data and the first Kasner
geometry approach their limiting values. We have also numerically confirmed this scaling for the first Kasner epoch at low temperature, in the cases of ordinary critical, tricritical, and first-order phase transitions. 

Several questions remain open. 
Our analytical construction predicts the tuned coupling
$\lambda_6^t$ required for sixth-root multicritical scaling, and a
detailed numerical verification would provide a nontrivial test of this
prediction. It would also be interesting to identify boundary observables sensitive to individual bounces \cite{Caceres:2022smh, Caceres:2023zhl, Li:2026bof}, to determine how spatial inhomogeneities modify the homogeneous epoch structure \cite{Liu:2022rsy} and to study time-dependent holographic phase transitions. Further directions include generalizing the analytical method here to other holographic phase transitions \cite{Hartnoll:2020rwq, Hartnoll:2020fhc, Gao:2023zbd,
Zhang:2025tsa, Xu:2025edz, Li:2025mhy, Arean:2026jie}, and exploring phase transitions in analytical hairy black hole systems \cite{Arean:2024pzo, Xiong:2026npi}. 
Finally, the sequence of Kasner epochs generated by scalar-field bounces in the
present homogeneous setting is closely related to the cosmological-billiard
picture developed in
Refs.~\cite{Damour:2000hv,Damour:2000wm,Damour:2002et}.
In more general gravitational systems, additional gravitational and matter
degrees of freedom can lead to chaotic BKL dynamics, with an irregular sequence
of Kasner epochs generated by successive wall reflections
\cite{BKL-book,Hartnoll:2020fhc,Cai:2023igv,Caceres:2026mug}.
It would be particularly interesting to investigate how a boundary phase
transition is encoded in such a chaotic black-hole interior. Because individual
epochs may depend sensitively on the exterior data, the robust imprint of
criticality may reside not in a particular Kasner plateau, but in statistical
properties of the full epoch sequence. Such a generalization would test whether
information about boundary criticality survives the chaotic approach to the
singularity.

\subsection*{Acknowledgments}
We thank Shan-Ming Ruan, Hao-Tian Sun, Ya-Wen Sun, Xin-Meng Wu, You-Jie Zeng, Shao-Dong Zhou and Yu Zhou for useful discussions. This work was supported by the National Natural Science Foundation of China Grants No. 12375041, 12447134 and 12575046. H-D.~L.~also acknowledges the support from 
the 
Postdoctoral Innovation Project of Shandong Province SDCX-ZG-202503036.

\vspace{0.95cm}
\appendix
\addtocontents{toc}{\protect\setcounter{tocdepth}{0}}

\section{Thermodynamics of black holes}
\label{app:ther-bh}
In this appendix we derive the free energy density in the double-trace ensemble. The calculation procedure is the same as the system without the super-exponential potential in \cite{Liu:2021hap}. 
The
renormalized action can be obtained by adding a proper boundary term. For alternative quantization with the double-trace
deformation, the renormalized action is
\be
\label{LorRenAction}
\begin{split}
 S_{\rm ren} &=
 \int_{\mathcal M} d^4x\sqrt{-g}
 \left[R-\frac{1}{2}g^{ab}\partial_a\phi\partial_b\phi-V(\phi)\right]
 +\int_{\partial\mathcal M}d^3x\sqrt{-\gamma}\,2K \\
 &\quad
 +\int_{\partial\mathcal M}d^3x\sqrt{-\gamma}
 \left[-4+\phi n^a\partial_a\phi
 +\frac{1}{2}\left(1+\kappa\varepsilon\right)\phi^2
 \right] \,.
\end{split}
\ee
Here the cutoff surface is at $z=\varepsilon$, with $\varepsilon\to0$,
$\gamma_{\mu\nu}$ is the induced metric, $K$ is the extrinsic curvature and
$n^a$ is the outward-pointing unit normal. 

Before extracting the near-boundary coefficients, it is useful to write the
regulated free energy directly in terms of the radial functions.  With
Euclidean period $1/T$ and spatial volume $V_2=\int dxdy$, we define the density of free energy,
\begin{equation}
 \mathcal{F}(\varepsilon)=
 \frac{T}{V_2}
 S^E_{\rm ren}
\Big{|}_{z\geq\varepsilon}\,.
 \label{eq:free-energy-regulated}
\end{equation}
On shell, using $R-\frac12(\nabla\phi)^2-V=V$ obtained from the trace of the Einstein equations, the bulk term can be integrated once by the equation for $f$,
\begin{equation}
 \frac{e^{-\chi/2}V}{z^4}
 =2\left(\frac{e^{-\chi/2}f}{z^3}\right)' .
\end{equation}
The horizon contribution vanishes because $f(z_h)=0$, and the regulated free
energy is therefore a cutoff-boundary expression,
\begin{equation}
\begin{split}
 \mathcal F(\varepsilon)
 =
 \left.
 \frac{e^{-\chi/2}}{z^3}
 \bigg[
 -4f+zf'-zf\chi'
 +4\sqrt f
 +zf\phi\phi'
 -\frac12\sqrt f(1+\kappa z)\phi^2
 \bigg]
 \right|_{z=\varepsilon}.
\end{split}
\label{eq:free-energy-functions}
\end{equation}
The first two terms are the integrated bulk contribution and the
Gibbons-Hawking term, while the second line contains the scalar and
gravitational counterterms.  Thus the on-shell action depends only on the fields and their first derivatives at the cutoff surface.

Substituting the near-boundary expansion \eqref{eq:nbexp} into
\eqref{eq:free-energy-functions}, the cutoff-dependent terms cancel and the
Euclidean on-shell action reduces to
\be
\label{On-shellEucAction}
 S_{\rm on-shell}^E
 =\frac{V_2}{T}\left(m_T-\frac{\kappa\alpha^2}{2}\right)\,,
\ee
so the free energy density may be written in terms of the boundary data as
\be
\label{free-energy}
 \mathcal{F}\equiv \frac{F}{V_2}
 =\frac{T S_{\rm on-shell}^E}{V_2}
 =m_T-\frac{\kappa\alpha^2}{2}\,.
\ee
Evaluating the conserved charge $Q$ in \eqref{conserved-charge} both at the boundary and at the horizon, with the expansions \eqref{eq:nbexp} and \eqref{eq:nhexpansion}, gives
\be 
\label{eq:Q-bdry-horizon}
m_T=
\frac{2}{3}\al \beta -
\frac{4\pi T}{3 z_h^2}=
\frac{2}{3}\kappa \al^2 -
\frac{4\pi T}{3 z_h^2}
\ee 
The free energy \eqref{free-energy} is then written as
\be 
\label{eq:F}
\mathcal{F}=\frac{\kappa \alpha^2}{6}-\frac{4\pi T}{3 z_h^2}\,.
\ee

The energy density from the renormalized Brown-York tensor is
\be
 \mathcal{E}=
 \lim_{\epsilon\to0}\sqrt{-\gamma}\,T^0_{~0}
 =-2m_T+2\alpha\beta-\frac{\kappa\alpha^2}{2}\,.
 \label{eq:energy-density}
\ee
The entropy density is $s=4\pi/z_h^2$. 
{The radially conserved charge relation \eqref{eq:Q-bdry-horizon} is now
\be
 -Ts=-\frac{4\pi T}{z_h^2}
 =3m_T-2\alpha\beta\,.
 \label{eq:minus-Ts}
\ee
Combining \eqref{eq:energy-density}, \eqref{eq:minus-Ts} and \eqref{free-energy} gives
\begin{equation}
\label{eq:etsF}
 \mathcal{E}-Ts=m_T-\frac{\kappa\alpha^2}{2}=\mathcal{F}\,,
\end{equation}
which reproduces the thermodynamic identity \eqref{fmT-thermo-relation}.
}

We finally consider the variation of the on-shell action. Let
\(\beta_E=1/T\) denote the Euclidean period. For a smooth family of regular solutions at
fixed spatial volume, the renormalized boundary variation is
\begin{equation}
 \delta S_E^{\rm on-shell}
 =
 V_2\mathcal E\,\delta\beta_E
 +\beta_EV_2
 \left(
 -\alpha\,\delta J
 -\frac{\alpha^2}{2}\,\delta\kappa
 \right),
 \qquad
 J=\kappa\alpha-\beta.
 \label{eq:IE-general-variation}
\end{equation}
Using \eqref{free-energy}, this becomes
\begin{equation}
 \beta_E\,\delta\mathcal F
 =
 (\mathcal E-\mathcal F)\delta\beta_E
 +\beta_E
 \left(
 -\alpha\,\delta J
 -\frac{\alpha^2}{2}\,\delta\kappa
 \right).
\end{equation}
Using equations~\eqref{eq:etsF} and 
\(\delta\beta_E/\beta_E=-\delta T/T\), we obtain 
\begin{equation}
 \delta\mathcal F
 =
 -s\,\delta T-\alpha\,\delta J
 -\frac{\alpha^2}{2}\,\delta\kappa.
 \label{eq:general-free-energy-variation}
\end{equation}
Since, in the ensemble we studied, \(J=0\) and \(\kappa\)
are held fixed, we have 
\begin{equation}
\delta\mathcal F=-s\,\delta T.
 \label{eq:fixed-ensemble-free-energy-variation}
\end{equation}

\section{First-order behavior below the tricritical coupling}
\label{app:first-order}
This appendix explains why the continuous transition found for \(\lambda_4>\lambda_4^{t}\)
is replaced by a first-order transition when
\(\lambda_4<\lambda_4^{t}\)
in the fixed \(\kappa=-1\) ensemble. The local perturbative expansion around the critical solution shows how the hairy branch changes as the tricritical coupling is crossed, while the global free-energy comparison determines which branch is thermodynamically preferred.

The perturbative expansion construction in Sec.~\ref{sec:analy} near the Schwarzschild solution takes
the form
\begin{equation}
  O
 =A_1\epsilon+\mathcal O(\epsilon^3),
 \qquad
 \Delta T
 =t_2(\lambda_4)\epsilon^2
 +t_4(\lambda_4,\lambda_6)\epsilon^4+\cdots,
 \qquad
 \Delta T\equiv T_c-T .
 \label{eq:first-order-eps}
\end{equation}
For \(\lambda_4>\lambda_4^t\), one has \(t_2>0\), so the hairy branch emerges
continuously below \(T_c\). 
At fixed temperature, the entropy difference
\begin{equation}
\Delta s\equiv s_{\rm hairy}-s_{\rm Sch}
\end{equation}
admits the expansion
\begin{equation}
\Delta s=s_2\epsilon^2+O(\epsilon^4).
\end{equation}
Expanding both branches around the critical solution shows that the horizon-radius correction cancels in the entropy difference, yielding
\begin{equation}
s_2=-2s_0\frac{\tau_2}{\tau_0}.
\end{equation}
Since the perturbative solution gives $\tau_2>0$, one obtains
$
s_2<0.
$ 
At fixed \(\kappa\), the first law gives
\begin{equation}
 d\Delta\mathcal F=-\Delta s\,dT,
 \qquad
 \Delta s=s_2\epsilon^2+\mathcal O(\epsilon^4),
\end{equation}
and hence
\begin{equation}
 \Delta\mathcal F
 =\frac{s_2t_2}{2}\epsilon^4+\mathcal O(\epsilon^6).
 \label{eq:first-order-free-energy-local}
\end{equation}
Hence, for \(\lambda_4>\lambda_4^t\), \(t_2>0, s_2<0\) imply \(\Delta \mathcal F<0\), so the continuously emerging hairy branch is thermodynamically preferred. At $\lambda_4=\lambda_4^{t}$, the coefficient $t_2$ vanishes and the transition becomes tricritical. For $\lambda_4<\lambda_4^{t}$, one instead finds $t_2<0$, so the perturbative hairy branch bends toward $T>T_c$ and is locally disfavored.

The perturbative expansion obtained above can be translated into the corresponding Landau effective theory. Writing
\begin{equation}
 \Delta\mathcal F_{\rm eff}
 =a(T-T_c) O^2
 +u_4\, O^4
 +c\, O^6+\cdots,
 \qquad
 a,c>0 ,
 \label{eq:first-order-landau}
\end{equation}
the equilibrium condition
$
\partial_O\Delta \mathcal F_{\rm eff}=0
$ 
implies
\begin{equation}
T_c-T=\frac{2u_4}{a}O^2+O(O^4).
\end{equation}
On the other hand, from the bulk perturbative relation  \eqref{eq:first-order-eps}, we have 
\begin{equation}
T_c-T
=
\frac{t_2(\lambda_4)}{A_1^2}O^2+\cdots.
\end{equation}
Matching these two expansions immediately gives
\begin{equation}
u_4
=
b(\lambda_4-\lambda_4^t),
\qquad
b>0,
\end{equation}
where we used the perturbative result
$t_2(\lambda_4)\propto\lambda_4-\lambda_4^t$.

Therefore, for \(\lambda_4<\lambda_4^t\), the quartic coefficient \(u_4\) becomes negative,
while the positive sixth-order term stabilizes the free energy.  The finite-condensate
minimum becomes degenerate with the Schwarzschild minimum at
\begin{equation}
  O_\ast^2=-\frac{u_4}{2c},
 \qquad
 T_\ast-T_c=\frac{u_4^2}{4ac}>0 ,
 \label{eq:first-order-landau-crossing}
\end{equation}  
which is the 
Landau mechanism for a first-order transition. The perturbative expansion therefore predicts that the quartic Landau coefficient changes sign precisely at the tricritical coupling. Once $\lambda_4<\lambda_4^t$, the preferred state switches discontinuously to a finite condensate, providing the local Landau description of the first-order transition.

A similar statement applies to \(\lambda_6\) for the case 
\(\lambda_4=\lambda_4^t\).  Near
\(\lambda_6=\lambda_6^t\), the quartic term is absent and the leading terms
beyond the quadratic one are proportional to
\( O^6\) and
\( O^8\).  If
\(\lambda_6<\lambda_6^t\) makes the sixth-order coefficient negative while
the eighth-order coefficient remains positive, the same mechanism predicts a
first-order transition.  

However, the actual transition temperature is determined by the full nonlinear hairy and Schwarzschild solutions through the free-energy crossing
\begin{equation}
 \Delta\mathcal F(T_\ast)
 \equiv
 \mathcal F_{\rm hairy}(T_\ast)
 -\mathcal F_{\rm Sch}(T_\ast)
 =0 ,
 \qquad
  O(T_\ast)\neq0 ,
 \label{eq:first-order-crossing}
\end{equation}
with both phases evaluated at the same \(\kappa=-1\).  After the physical
rescaling, the free energies are
\begin{equation}
 \mathcal F_{\rm hairy}
 =m_T-\frac{\kappa\alpha^2}{2}
 =\frac{\kappa\alpha^2}{6}
 -\frac{4\pi T}{3z_h^2},
 \qquad
 \mathcal F_{\rm Sch}
 =-\left(\frac{4\pi T}{3}\right)^3 .
 \label{eq:first-order-free-energies}
\end{equation}
The entropy discontinuity is
\begin{equation}
 \Delta s_\ast
 =-\left.\frac{d\Delta\mathcal F}{dT}\right|_{T=T_\ast},
\end{equation}
and the magnitude of the latent heat is \(T_\ast|\Delta s_\ast|\).  A 
multivalued condensate curve is therefore suggestive but not decisive, as can be seen e.g. in Fig. \ref{fig:num-first-order-comparison}.  A 
finite-condensate free-energy crossing among physical solutions is the
criterion that establishes the first-order transition expected for
\(\lambda_4<\lambda_4^t\).

\section{Numerical procedure to obtain the solution}
\label{app:numerical-procedure}

In this appendix, we describe the numerical procedure for starting from the solution with 
$z_h=1$ and then applying a scaling transformation to obtain the solution with 
$\kappa=-1$. The precise procedure is as follows. 

We first construct a representative solution in horizon units by setting
\begin{equation}
 z_h=1,
 \qquad
 \chi_h=0,
\end{equation}
in the regular horizon expansion \eqref{eq:nhexpansion}.  The horizon scalar
\(\phi_h\) is the only shooting parameter, and the condition
\(V(\phi_h)<0\) is required for a positive nonextremal temperature.  After
integrating to the AdS boundary, we extract \(\alpha\), \(\beta\), \(m_T\),
and \(\chi_0\) from the UV expansion \eqref{eq:nbexp}.  Shifting
\(\chi\to\chi-\chi_0\) fixes the boundary speed of light and gives the
representative temperature
\begin{equation}
 T_{\rm rep}
 =-\frac{V(\phi_h)}{8\pi}e^{\chi_0/2}.
 \label{eq:num-representative-temperature}
\end{equation}

The representative solution must then be mapped to the fixed physical
ensemble.  Under a planar rescaling by \(\ell>0\),
\begin{equation}
 \alpha\to\frac{\alpha}{\ell},
 \qquad
 \beta\to\frac{\beta}{\ell^2},
 \qquad
 m_T\to\frac{m_T}{\ell^3},
 \qquad
 T\to\frac{T_{\rm rep}}{\ell},
 \qquad
 z_h\to\ell.
 \label{eq:num-planar-rescaling}
\end{equation}
For a hairy representative, \(\alpha\neq0\), the source-free condition
\(\beta_{\rm phys}=-\alpha_{\rm phys}\) fixes
\begin{equation}
 \ell=-\frac{\beta}{\alpha}>0.
 \label{eq:num-fixed-kappa-scale}
\end{equation}
A representative solution with \(\ell\leq0\) does not belong to the chosen
fixed-\(\kappa\) ensemble and is discarded.  The physical observables are
then
\begin{align}
 T&=\frac{T_{\rm rep}}{\ell},
 &O&=\frac{\alpha}{\ell},
 &s&=\frac{4\pi}{\ell^2},
 \label{eq:num-physical-data-a}\\
 \mathcal F_{\rm hairy}
 &=\frac{m_T}{\ell^3}
 +\frac12\left(\frac{\alpha}{\ell}\right)^2,
 &
 \mathcal F_{\rm Sch}(T)
 &=-\left(\frac{4\pi T}{3}\right)^3.
 \label{eq:num-physical-data-b}
\end{align}

The same solution is integrated inward from the horizon.  To keep the
notation consistent with the Kasner relations, we extract the scalar
velocity
\begin{equation}
 v_{\rm eff}(z)
 \equiv\frac{d\phi}{d\log z}=z\phi'(z).
 \label{eq:num-effective-velocity}
\end{equation}
 On a kinetic-dominated plateau, 
\(v_{\rm eff}\simeq v_n\), and
\begin{equation}
 p_t^{(n)}=\frac{v_n^2-4}{v_n^2+12}.
 \label{eq:num-pt-from-v}
\end{equation}
The velocity is invariant under the planar rescaling, so the plateau may be
identified before or after imposing \(\kappa=-1\).

\section{A special hairy black hole with Schwarzschild-like interior}
\label{app:screened-kasner}

In this appendix we describe a special numerical solution for which the exterior geometry is a hairy black hole, while the Kasner regime inside the horizon is Schwarzschild-like.  This provides a useful example
showing that the existence of scalar hair outside the horizon does not by itself guarantee a nonzero scalar velocity in the interior.

For a local kinetic-dominated Kasner regime, from \eqref{eq:nth-kasner-epoch} and \eqref{eq:rel2} the scalar velocity is
$
 v\equiv z\phi'(z),
$
and the temporal Kasner exponent is
$
 p_t=\frac{v^2-4}{v^2+12}.
$
Thus \(v=0\) gives
\begin{equation}
 p_t=-\frac13,
\end{equation}
which is the same value as the Schwarzschild-AdS black brane.  More importantly, when the scalar velocity vanishes in the interior, the scalar is locally constant and the super-exponential wall does not play any nontrivial role.  Therefore there is no scalar bounce, and the interior is not divided into several Kasner epochs.  The whole interior is described by a single Schwarzschild-like Kasner epoch, with
\begin{equation}
\label{eq:interior-sch}
 p_t=-\frac13,\qquad
 p_x=p_y=\frac23,\qquad
 p_\phi=0.
\end{equation}
In this sense the exterior scalar hair is screened from the interior Kasner dynamics: The solution is therefore an exterior hairy black hole with a Schwarzschild-like interior.

\begin{figure}[h!]
\begin{center}
\includegraphics[
width=0.45\textwidth]{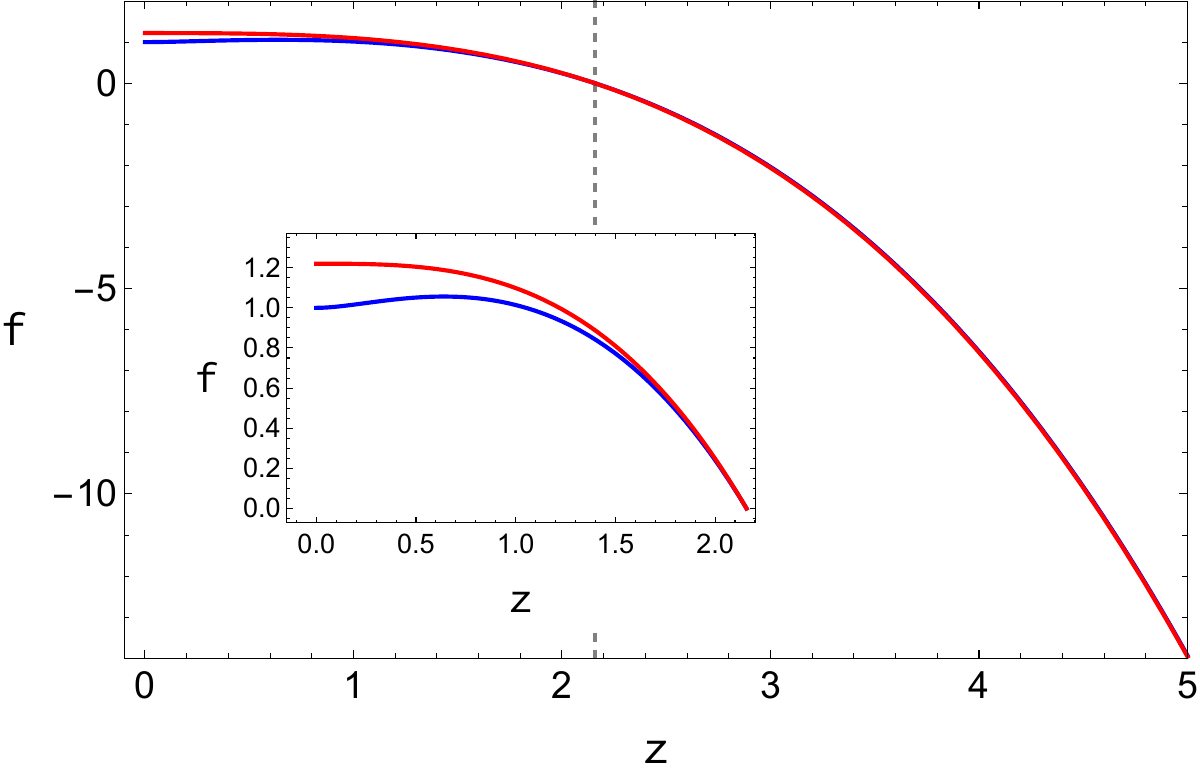}~~~
\includegraphics[
width=0.436\textwidth]{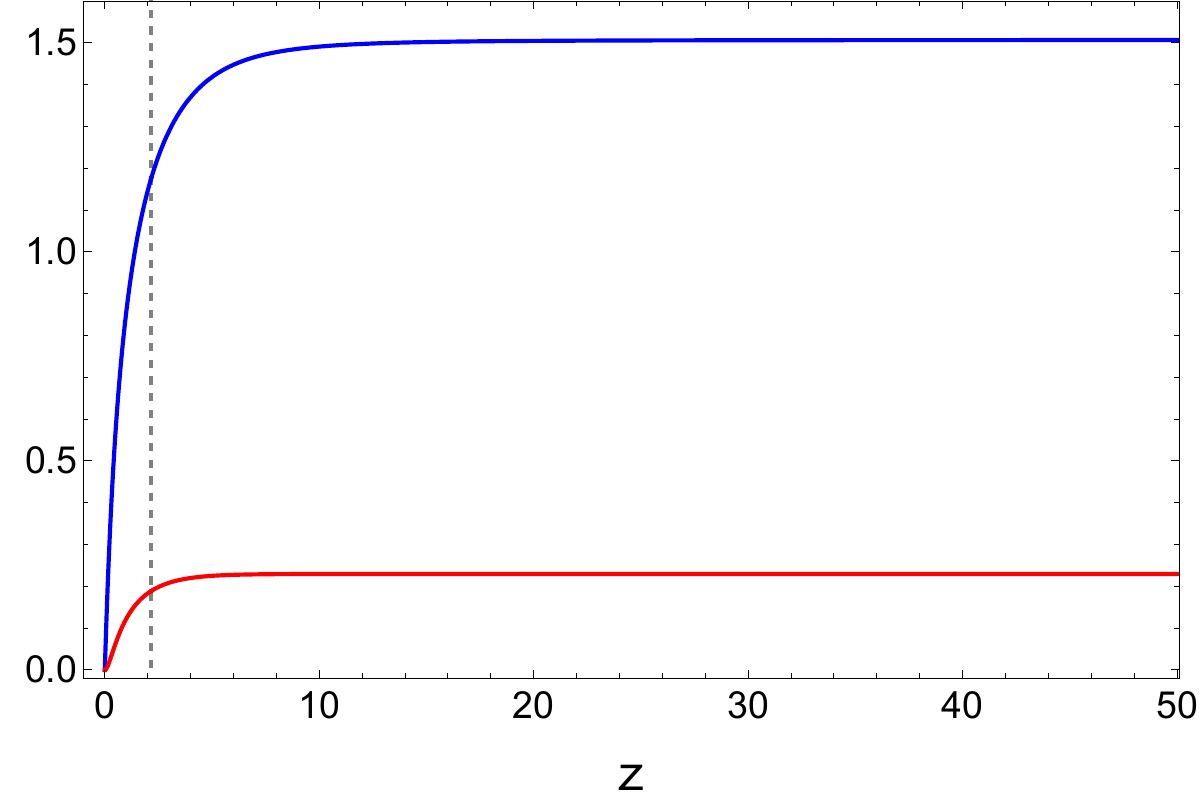}
\end{center}
\vspace{-0.6cm}
\caption{\small A special hairy black-hole solution with a Schwarzschild-like
interior, with the parameter $\lambda_4=\lambda=\lambda_8=1/10, \lambda_6=0$ and $T=0.11965, \kappa=-1$.  The dashed vertical line denotes the event horizon.  {\bf Left:} The 
metric field \(f(z)\).  The blue curve is the numerical hairy black-hole
solution, while the red curve is the Schwarzschild-AdS black brane with the
same horizon position. The inset shows the region near the horizon. Beyond the horizon, the numerical solution agrees with the Schwarzschild-AdS form extremely well; we have checked this agreement numerically up to \(z/z_h\sim10^{100}\).   {\bf Right:} 
The scalar field \(\phi(z)\) ({\em blue}) 
and metric field \(\chi(z)\) ({\em red}). 
Although the exterior solution has nonzero scalar hair and a nonzero boundary
condensate, the scalar becomes constant behind the horizon and \(\chi(z)\)
also approaches a constant.  Hence 
\(v\simeq0\), no scalar bounce occurs, and the whole interior is
described by a single Schwarzschild Kasner epoch \eqref{eq:interior-sch}.}
\label{fig:screened-kasner}
\end{figure}

The numerical solution shown in Fig.~\ref{fig:screened-kasner} has
nonzero scalar hair in the exterior and nonzero boundary condensate.  However,
after integrating through the horizon, the first interior plateau satisfies
\begin{equation}
 v_1\simeq0,
 \qquad
 p_t^{(1)}+\frac13\simeq0.
\end{equation}
Equivalently, the metric function in the corresponding interior interval is
well approximated by
\begin{equation}
 f(z)\simeq f_\ast^{\rm loc}
 \left(1-\frac{z^3}{z_h^3}\right),
 \label{eq:app-screened-f}
\end{equation}
with an effective normalization \(f_\ast^{\rm loc}\), while $\chi, \phi$ are constant.  In the example displayed
below, \(f_\ast^{\rm loc}\simeq1.2159\).  The normalization differs from the
Schwarzschild value because the exterior solution and the time normalization
are not those of the vacuum black brane, but the exponent is nevertheless the
Schwarzschild one.

This behavior can be understood from the local Kasner data.  The scalar field
outside the horizon fixes the boundary condensate, but the first interior
Kasner exponent depends on the logarithmic scalar velocity \(v=z\phi'\) in the
kinetic-dominated region.  A nontrivial exterior profile may evolve through
the horizon in such a way that this velocity crosses zero before the first
Kasner plateau is formed.  At this point the scalar is locally almost
constant, the kinetic contribution is subleading, and the metric approaches
the Schwarzschild Kasner exponent.  We could therefore refer to this solution as ``a
screened Kasner solution": the scalar hair is present in the exterior, but its
leading imprint on the first interior Kasner exponent is screened.

The screening is not expected to be protected by a symmetry.  It should occur
only at isolated points, or along codimension-one loci by tuning $\lambda_4$, in the space of
hairy black-hole solutions.  Small changes of the horizon scalar generally
produce a nonzero \(v_1\), and hence move the first Kasner exponent away from
\(-1/3\).  Nevertheless, the solution is useful because it shows that the map
from exterior order parameter to interior Kasner data is not simply determined
by the magnitude of the condensate; it depends on the full radial evolution of
the scalar field between the boundary, the horizon, and the interior
Kasner epoch.

\end{document}